\documentclass[5p,preprint]{elsarticle}
\usepackage[textsize=footnotesize]{todonotes}
\usepackage{lscape}
\usepackage{array,graphicx}
\DeclareGraphicsExtensions{.pdf,.jpeg,.png,.eps,.jpg}
\usepackage{url}
\usepackage{hyperref}
\usepackage{comment}
\usepackage{tikz}
\usetikzlibrary{calc,positioning, arrows.meta, shapes.geometric}
\def\checkmark{\tikz\fill[scale=0.4](0,.35) -- (.25,0) -- (1,.7) -- (.25,.15) -- cycle;}
\usepackage{soul}
\usepackage{booktabs}
\usepackage{pifont}
\usepackage{tablefootnote}
\usepackage{soul}
\usepackage{adjustbox}
\usepackage{tablefootnote}
\usepackage[utf8]{inputenc}
\usepackage{textgreek}
\usepackage{balance}
\usepackage{wrapfig}
\usepackage{fontawesome5}
\usepackage{tabularray}
\usepackage{array}
\usepackage{hhline}
\usepackage{wasysym}
\usepackage[dvipsnames]{xcolor}

\usepackage{msc}
\usepackage{pgfplots}
\usepackage{listings}
\makeatletter
\def\lst@makecaption{%
  \def\@captype{table}%
  \@makecaption
}
\makeatother
\usepackage{xspace}
\newcommand{\ea}{\emph{et al.}\xspace}

\newcommand*\rot{\rotatebox{90}} 
\newcommand*\TRot{\rotatebox{90}} 

\pgfplotsset{compat=1.18}
\begin{document}

\begin{frontmatter}

\title{A Survey of Internet Censorship and its Measurement: Methodology, Trends, and Challenges}

\author[uulmaddress]{Steffen Wendzel}
\ead{steffen.wendzel@uni-ulm.de}

\author[uulmaddress]{Simon Volpert}
\ead{simon.volpert@uni-ulm.de}

\author[uulmaddress]{Sebastian Zillien}
\ead{sebastian.zillien@uni-ulm.de}

\author[uulmaddress]{Julia Lenz}
\ead{julia.lenz@uni-ulm.de}

\author[hagenaddress]{Philip Rünz}
\ead{philip.ruenz@studium.fernuni-hagen.de}

\author[cnraddress]{Luca Caviglione}
\ead{luca.caviglione@cnr.it}

\address[uulmaddress]{Institute of Information Resource Management (IRM), Dep.\ Engineering, Computer Science and Psychology, Ulm University, Germany}
\address[hagenaddress]{Faculty of Mathematics \& Computer Science, University of Hagen, Germany}
\address[cnraddress]{National Research Council of Italy, Genoa, Italy}

\begin{abstract}
Internet censorship limits the access of nodes residing within a specific network environment to the public Internet, and vice versa. During the last decade, techniques for conducting Internet censorship have been developed further. Consequently, methodology for \textit{measuring} Internet censorship had been improved as well.

In this paper, we firstly provide a survey of {network-level} Internet censorship techniques. Secondly, we survey censorship measurement methodology. {We further cover the censorship of circumvention tools and its measurement, as well as} available datasets. In cases where it is beneficial, we bridge the terminology and taxonomy of Internet censorship with related domains, namely traffic obfuscation and information hiding. We further extend the technical perspective with recent trends and challenges, including human aspects of Internet censorship.
\end{abstract}


\end{frontmatter}

\newenvironment{bluebox}
  {\begin{lrbox}{0}\begin{tabular}{@{}l@{}}}
  {\end{tabular}\end{lrbox}%
   \setlength{\fboxsep}{12pt}
  {\centering\fcolorbox{black}{GreenYellow}{\usebox{0}}\par}}

\begin{bluebox}
    This is a pre-print. The final version of this paper\\
    appeared in \emph{Computers \& Security}:\\
    \url{https://doi.org/10.1016/j.cose.2025.104732}
\end{bluebox}

\section{Introduction}
The Internet enables the interconnectivity of numerous users and the interaction with millions of websites and online services. Sometimes, content or user-interaction is considered inappropriate or undesired by entities controlling fractions of the Internet. Such entities can be state governments, regional governments and any form of corporate and non-governmental organizations. The desire to control -- or: censor -- content of the Internet for its users has been studied by a plethora of researchers. Internet censorship of countries has already been discussed in the mid-1990's \cite{ang1996censorship} and early forms of Internet censorship have also been used in an \emph{idealistic vision of self-governance} by the Internet community~\cite{murdoch2013introduction}. Since then, due to the increasing number of Internet users, social media platforms and the ever-growing amount of online content, Internet censorship became a widespread and persistent phenomenon~\cite{FOTN24}.

While there are already some surveys on Internet censorship, none of these studies up-to-date techniques of both, Internet censorship \emph{and} its measurement. Further, no previous work draws clear links between Internet censorship and the related domains of network traffic obfuscation and network information hiding through covert channels and steganography. Additionally, previous surveys did not cover related human aspects.

{To close this gap, we provide a survey on network-level Internet censorship techniques, their measurement and selected related topics. While we cover some human aspects, surveying political and juridical aspects of censorship (e.g., content removal by authorities) is outside the scope of our paper. We then only provide links to such topics when it aids the understanding.}

The \textbf{key contributions of this article} are the following:

\begin{enumerate}
    \item We perform survey on \emph{Internet censorship methodology}, i.e., how it is technically realized {on the network level}.
    \item We {also survey network-level} censorship \emph{measurement}.
    \item We discuss {the censorship techniques applied to circumvention tools and the measurement of the related censorship techniques}.
    \item {We survey} measurement \emph{datasets} and show how they overlap and where they have singularities.
    \item Whenever reasonable, we draw links between Internet censorship and related domains, such as traffic obfuscation.
    \item We summarize trends and challenges of Internet censorship and its measurement, including both, technical, societal and human aspects.
\end{enumerate}

The rest of the paper is structured as follows. Sect.~\ref{sect:relwrk} covers the related work. {Sect.~\ref{Sect:PaperSelectionProcedure} describes our literature selection procedure while} Sect.~\ref{sect:basics} introduces fundamentals and a taxonomy of censor capabilities. Next, Sect.~\ref{sect:censortechniques} surveys censorship techniques, whereas Sect.~\ref{sect:measurementmeth} reviews mechanisms to detect censorship. {We cover the censorship of circumvention tools and the measurement for these tools' censorship in Sect.~\ref{Sect:CircumventionTools}.} Measurement platforms and datasets are summarized in Sect.~\ref{sect:measurementplatforms}. Sect.~\ref{sect:trends} discusses trends and challenges and Sect.~\ref{sect:concl} draws conclusions.

\section{Related Work}\label{sect:relwrk}
\begin{table*}[!ht]
    \centering
    \caption{Comparison of Existing Surveys' Major Focus}
    \label{tab:surveycmp:overview}
\resizebox{1.0\textwidth}{!}{
    \begin{tabular}{|r|c|c|c|p{4.75cm}|c|c|c|}
        \hline 
        \textbf{Author(s)} & \textbf{Ref.} & \textbf{Year} & \textbf{Published} & \textbf{Main Focus} & \textbf{Int.~Cens.} & \textbf{Inf.~Hid.} & \textbf{Traf.~Obfs.} \\ \hline
        Subramanian & \cite{subramanian2011growth} & 2011 & CIIMA & General Internet censorship & \checkmark & - & - \\ \hline
        Aceto and Pescapé & \cite{Aceto2015} & 2015 & Comp.~Netw. & Censorship detection & \checkmark & - & - \\ \hline
        Tschantz \ea & \cite{tschantz2016sok} & 2016 & IEEE S\&P & Evaluation of circumvention methods & 
        \checkmark & \checkmark & \checkmark \\ \hline
        Mazurczyk \ea & \cite{NIHbook} & 2016 & textbook & Network information hiding & - & \checkmark & \checkmark \\ \hline
        Khattak \ea & \cite{Khattak:2016} & 2016 & PoPETS & Censorship circumvention systems & \checkmark & \checkmark & \checkmark \\ \hline
        Zittrain \ea & \cite{Zittrain2017shifting} & 2017 & research report & Country-based comparison; HTTPS & \checkmark & - & - \\ \hline
        Niaki \ea & \cite{ICLab:SP20} & 2020 & IEEE S\&P & Censorship measurement & \checkmark & - & - \\ \hline
        Ververis \ea & \cite{ververis2020cross} & 2020 & Policy \& Internet & Country-based comparison of censorship & \checkmark & - & - \\ \hline
        Karunanayake \ea & \cite{karunanayake2021anonymisation} & 2021 & COMST & Tor de-anonymization attacks & \checkmark & - & - \\ \hline
        Master and Garman & \cite{master2023worldwide} & 2023 & FOCI & Current censorship in 70 countries & \checkmark & - & - \\ \hline
        {Hall \ea} & \cite{rfc9505} & {2023} & {IRTF RFC} & {Internet censorship methods} & {\checkmark} & {-} & {-} \\ \hline
        {Wrana \ea} & \cite{wrana2025sok} & {2025} & {PETS} & {Future Internet architectures} & {\checkmark} & {(\checkmark)} & {-} \\ \hline
        \textbf{This Paper} & - & 2025 & - & Censorship and its measurement & 
        \checkmark & \checkmark & \checkmark \\ \hline
    \end{tabular}
}
\end{table*}

Surveys on Internet censorship have been published by several authors. We will first summarize the key characteristics of these surveys and then show how our paper improves over these existing surveys.
A fraction of the available surveys discusses (mostly) political aspects of Internet censorship (e.g., \cite{Zuchora2010,bitso2013trends}) but we decided to exclude such works due to the technical focus of our paper.

Tab.~\ref{tab:surveycmp:overview} provides an overview of the existing surveys in the field, highlighting their major focus. We compare the Internet protocols covered by the various reviews and censorship detection techniques in Tab.~\ref{tab:surveycmp:techniques}. 


\begin{table*}[!ht]
    \caption{Comparison of Network Protocol-based Techniques Covered by Existing Surveys. (Checkmarks in brackets indicate that the topic is partially addressed but not fully surveyed. This case appeared mostly when papers had a different focus, such as performing a cross-country analysis.)}
    \label{tab:surveycmp:techniques}
\resizebox{1.0\textwidth}{!}{
    \begin{tabular}{|r|c|c||c|c|c|c|c|c|c|c|c|c|c||c|c|c|c|c|c|c|c|c|c|c|}
        \hline 
        \multicolumn{3}{|c|}{\textbf{General Aspects}} & \multicolumn{11}{|c|}{\textbf{Censorship Meth./Countermeasures}} & \multicolumn{11}{|c|}{\textbf{Measurement Meth.}} \\ \hline
        \textbf{Author(s)} & \textbf{Ref.} & \TRot{\shortstack[l]{\textbf{Coverage of}\\\textbf{last five years}}} & \TRot{\textbf{Taxonomy}} & \TRot{\textbf{IPv4}} & \TRot{\textbf{IPv6}} & \TRot{\textbf{BGP}} & \TRot{\shortstack[l]{\textbf{TCP, UDP,}\\\textbf{QUIC}}} & \TRot{\textbf{DNS}} & \TRot{\textbf{HTTP(S)}} & \TRot{\textbf{TLS (or: SSL)}} & \TRot{\shortstack[l]{\textbf{Circumvention}\\\textbf{Tools / VPNs}}} & \TRot{\shortstack[l]{\textbf{Other Appl.}\\\textbf{Protocols}}} & \TRot{\textbf{Active Probing}} & \TRot{\textbf{Taxonomy}} & \TRot{\textbf{IPv4}} & \TRot{\textbf{IPv6}} & \TRot{\textbf{BGP}} & \TRot{\shortstack[l]{\textbf{TCP, UDP,}\\\textbf{QUIC}}} & \TRot{\textbf{DNS}} & \TRot{\textbf{HTTP(S)}} & \TRot{\textbf{TLS (or: SSL)}} & \TRot{\shortstack[l]{\textbf{Circumvention}\\\textbf{Tools / VPNs}}} & \TRot{\shortstack[l]{\textbf{Other Appl.}\\\textbf{Protocols}}} & \TRot{\textbf{Active Probing}}\\ \hline
        Subramanian & \cite{subramanian2011growth} & - & - & - & - & - & - & - & - & - & - & - & - & - & - & - & - & - & - & - & - & - & - & -\\ \hline
        Aceto and Pescapé & \cite{Aceto2015} & - & \checkmark & \checkmark & - & \checkmark & \checkmark & \checkmark & \checkmark & \checkmark & - & - & (\checkmark) & \checkmark & \checkmark & - & - & \checkmark & \checkmark & \checkmark & \checkmark & (\checkmark) & (\checkmark) & (\checkmark) \\ \hline
        Tschantz \ea & \cite{tschantz2016sok} & - & (\checkmark) & \checkmark & - & - & (\checkmark) & \checkmark & \checkmark & \checkmark & \checkmark & \checkmark & \checkmark & - & - & - & - & - & - & - & - & - & - & - \\ \hline
        Mazurczyk \ea & \cite{NIHbook} & - & \checkmark & \checkmark & \checkmark & - & \checkmark & - & (\checkmark) & - & \checkmark & \checkmark & - & - & - & - & - & - & - & - & - & - & - & - \\ \hline
        Khattak \ea & \cite{Khattak:2016} & - & - & (\checkmark) & - & (\checkmark) & (\checkmark) & (\checkmark) & (\checkmark) & - & \checkmark & (\checkmark) & (\checkmark) & - & - & - & - & - & - & - & - & - & - & - \\ \hline
        Zittrain \ea & \cite{Zittrain2017shifting} & - & - & - & - & - & - & - & (\checkmark) & - & - & - & - & - & - & - & - & - & - & (\checkmark) & - & - & - & -\\ \hline
        Niaki \ea & \cite{ICLab:SP20} & - & - & (\checkmark) & - & - & \checkmark & (\checkmark) & (\checkmark) & - & - & - & - & - & \checkmark - & - & & \checkmark & \checkmark & \checkmark & - & - & - & - \\ \hline
        Ververis \ea & \cite{ververis2020cross} & - & - & - & - & - & - & - & - & - & - & - & - & - & - & - & - & - & - & - & - & - & - & - \\ \hline
        Karunanayake \ea & \cite{karunanayake2021anonymisation} & 1y & (\checkmark) & - & - & (\checkmark) & - & (\checkmark) & (\checkmark) & - & \checkmark & - & - & - & - & - & - & - & - & - & - & - & - & - \\ \hline
        Master and Garman & \cite{master2023worldwide} & 3y & (\checkmark) & (\checkmark) & (\checkmark) & (\checkmark) & (\checkmark) & - & (\checkmark) & (\checkmark) & (\checkmark) & (\checkmark) & - & - & - & - & - & - & - & - & - & - & - & - \\ \hline
        {Hall \ea} & \cite{rfc9505} & {3y} & {(\checkmark)} & {\checkmark} & {(\checkmark)} & {\checkmark} & {\checkmark} & {\checkmark} & {\checkmark} & {\checkmark} & {(\checkmark)} & {(\checkmark)} & {(\checkmark)} & - & - & - & - & - & - & - & - & - & - & - \\ \hline
        {Wrana \ea} & \cite{wrana2025sok} & {\checkmark} & - & {(\checkmark)} & - & {(\checkmark)} & - & {(\checkmark)} & - & - & {(\checkmark)} & {(\checkmark)} & - & - & - & - & - & - & - & - & - & - & - & -\\ \hline
        \textbf{This Paper} & - & \checkmark & \checkmark & \checkmark & \checkmark & \checkmark & \checkmark & \checkmark & \checkmark & \checkmark & \checkmark & \checkmark & \checkmark & \checkmark & \checkmark & \checkmark & \checkmark & \checkmark & \checkmark & \checkmark & \checkmark & \checkmark & \checkmark & \checkmark \\ \hline
    \end{tabular}
}
\end{table*}

In a 2011-survey, Subramanian \cite{subramanian2011growth} summarizes heterogeneous aspects of Internet censorship for selected countries, including circumvention methods. 
Aceto and Pescapé \cite{Aceto2015} provide a comprehensive survey on Internet censorship detection but also cover censorship methodologies. Dated 2015, the survey covers a plethora of fundamental methods for censorship detection but lacks developments of the last ten years. 
{Moreover, the authors neither discuss information hiding and traffic obfuscation methods nor do they cover newer protocols (e.g., IPv6 and QUIC) or human aspects of Internet censorship}. 
Tschantz \ea \cite{tschantz2016sok} survey censorship \emph{circumvention} approaches, especially their related evaluation methodology. They further discuss circumvention approaches within the context of real-world censor's known capabilities of 2016 and conclude that there is a disconnect between the goals of sophisticated circumvention methods and the censor's methods. They present three key reasons for this mismatch, i.e., (i)~that censors focus on the discovery and setup procedure of circumvention channels while circumvention research mostly focuses on the channel itself; (ii) that censors apply cheap passive censorship approaches instead of complex ones; and (iii) that censors try not to risk false blocking while research covers several attacks with collateral damage \cite{tschantz2016sok}. 
In their textbook, Mazurczyk \ea~\cite{NIHbook} discuss network information hiding methods, including network steganography and traffic obfuscation. Regarding the countermeasures, the work mostly focuses on the detection, elimination, and prevention of covert channels on different levels of the TCP/IP stack, including specific covert channel/circumvention tools. 
Khattak \ea~\cite{Khattak:2016} survey censorship-resistant communication tools both from practical and theoretical viewpoints. 
%
A 2017-survey by Zittrain \ea \cite{Zittrain2017shifting} conducts a comparison of censorship on different nations and also investigates the impact of HTTPS on Internet censorship. 
While not a survey \emph{per se}, Niaki \ea provide an overview of several censorship measurement methods employed by \emph{ICLAB} \cite{ICLab:SP20}, featuring approaches for prominent communication protocols. 
Ververis \ea conduct a cross-country comparison of Internet censorship \cite{ververis2020cross}. Their survey focuses on France, Turkey, and Iran. 
%
%
Karunanayake \ea \cite{karunanayake2021anonymisation} provide a comprehensive survey of de-anonymization attacks on Tor. Their work features a well-grounded taxonomy of Tor attacks. 
Master and Garman \cite{master2023worldwide} conduct a cross-sectional study on Internet censorship of $70$ countries. They also summarize the applied censorship methods and found that censors still apply filtering methods that are easy to bypass and perform total Internet shutdowns. {Hall \ea \cite{rfc9505} provide a survey-styled RFC covering several network-level aspects of Internet censorship methods. While obfuscation methods are briefly mentioned, information hiding techniques, measurement techniques, and human aspects are largely neglected.} 
Finally, Wrana \ea study the potential for censorship in future Internet architectures, including NDN, SCION, XIA, and NEBULA.
%
%

In comparison to the above-mentioned surveys, we do not only provide the most up-to-date coverage of censorship {and measurement methodology (including several 2025-publications and novel trends, such as regional censorship or detection of vulnerabilities in censorship systems)} that features the broadest set of network protocols. We also draw links to related disciplines, such as information hiding. {Finally, to the best of the authors knowledge, our survey is the only one that explicitly covers human aspects linked to Internet censorship.}

\section{Literature Selection}\label{Sect:PaperSelectionProcedure}

{Our literature selection process is sketched in Fig.~\ref{fig:methodology}. First, we conducted a query on the dblp computer science literature repository for the keywords \texttt{censor*} AND (\texttt{network} OR \texttt{internet} OR \texttt{block*} OR \texttt{firewall} OR \texttt{circumvent*} OR \texttt{measure*}). We considered papers that had been published in the 2010-2024 range. Additionally, we queried Google Scholar and Web of Science (WoS) with the term \texttt{Internet censorship} (in both cases, we sorted the listing of results ``by relevance'', excluded papers dated earlier than 2010, and considered the first 250 results). All queries were conducted in October 2024 and redundancies were removed. Note that we included pre-prints/technical reports. Next, we manually monitored high-rank conferences and journals as well as specialized events, such as FOCI, for freshly appearing papers, which led to the inclusion of several late-2024- and 2025-papers. All found papers were filtered, i.e., we excluded papers on political aspects of censorship, papers for which other attributes (such as author names) led to their inclusion, papers on censorship circumvention approaches (without discussing relevant measurement aspects) and papers on entirely different topics that matched our keywords. In addition, we performed a snowballing procedure with all survey papers that we found as well as with papers cited at least 50 times according to WoS to detect additional relevant works that could be included. Moreover, we manually scanned Google Scholar for relevant yet dated papers that appeared before 2010 so that we were able to put current techniques into historic context. %
All papers were read, and a team discussion was held to decide on the exclusion of border-case papers. %
Thankfully, the Reviewers pointed us towards a small number of papers that had not been found by our methodology, such as RFC 9505 \cite{rfc9505}. These papers had been included in the present version of the work.}

\usetikzlibrary{positioning, shapes.geometric, shapes.symbols, arrows.meta, calc, backgrounds, fit, calc, decorations.pathreplacing,shadows}

\begin{figure}[h!t] 
    \centering
    \begin{tikzpicture}[scale=0.7, transform shape] 
        \begingroup 
        \hypersetup{hidelinks} 
        \tikzstyle{stage_frame}=[
            draw=black!60, thick, dashed, inner sep=1pt 
        ]
        \tikzstyle{stage_title}=[ 
            font=\large\bfseries, text=black!80, anchor=west, fill=white
        ]
        \tikzstyle{process_box}=[ 
            rectangle, rounded corners=2pt, draw=black!70, thick, fill=white, 
            minimum height=2em, text centered, font=\small, inner sep=1pt, top color=black!7, bottom color=black!2
        ]
        \tikzstyle{artifact_box}=[ 
            process_box, top color=orange!40!orange!60, bottom color=orange!40!orange!20, font=\small\itshape, align=center, rounded corners=0pt
        ]
        \tikzstyle{core_concept_box}=[ 
            process_box, top color=orange!100!cyan!40, bottom color=orange!100!cyan!20, font=\small\bfseries
        ]
        \tikzstyle{output_box}=[ 
            process_box, top color=black!20, bottom color=black!10, align=center, minimum height=3em
        ]
        \tikzstyle{flow_arrow}=[ 
            draw, thick, -Stealth, color=black!50, tension=100
        ]
        \tikzstyle{arrow_label}=[ 
            font=\tiny\itshape, fill=white, inner ysep=0.25pt
        ]

        \begin{scope}[yshift=0cm]
            \node[stage_frame, minimum width=\columnwidth-2pt, minimum height=7.75cm] (frame1) at (0,0) {};
            \node[stage_title] at ([xshift=5pt]frame1.north west) {1. Paper Identification};

            \node[process_box, text width=8em, below=0.5cm of frame1.north, xshift=-2.15cm] (gs) {\faGoogle \ Google Scholar};
            \node[process_box, text width=8em, below=0.3cm of gs] (ieee) {\faDatabase \ Web of Science};
            \node[process_box, text width=8em, below=0.3cm of ieee] (dblp) {\faDatabase \ DBLP};
            \node[process_box, text width=8em, below=0.5cm of frame1.north, xshift=2.15cm] (monitoring) {\faFile \ Venue Monitoring};
            \node[process_box, text width=8em, below=0.3cm of monitoring] (sok) {\faBookOpen \ Related SoKs};

            \node[process_box, text width=9em, below=3.0cm] (merge) at ($(gs)!0.5!(monitoring)$) {\faSearch Search \& Merge};
            \node[process_box,text width=15em,  below=0.8cm of merge] (filter) {\faFilter Inclusion/Exclusion Criteria};
            \node[artifact_box, text width=\columnwidth-4pt, above=0pt of frame1.south] (corpus) {\faFileArchive \ Corpus of Relevant Papers};
            
            \draw[flow_arrow] (gs.east) to[out=-30, in=90] (merge);
            \draw[flow_arrow] (ieee.east) to[out=-30, in=90] (merge);
            \draw[flow_arrow] (dblp.east) to[out=-30, in=90] (merge);
            \draw[flow_arrow] (monitoring.west) to[out=210, in=90] (merge);
            \draw[flow_arrow] (sok.west) to[out=210, in=90] (merge);
            
            \draw[flow_arrow] (merge) -- node[arrow_label, pos=0.3] {\large 500+} (filter);
            \draw[flow_arrow] (filter) -- node[arrow_label, pos=0.3] {\large 250+} (corpus);
        \end{scope}

        \begin{scope}[yshift=-6.5cm] 
            \node[stage_frame, minimum width=\columnwidth-2pt, minimum height=3.5cm] (frame2) at (0,0) {};
            \node[stage_title] at ([xshift=5pt]frame2.north west) {2. Analysis};

            \node[process_box, text width=10em, below=1.4cm of frame2.north, minimum height=3.0em] (team) at (0, 2.5) {\faUsers \ Manual Reading and Team Discussion};
            
            \node[artifact_box, text width=\columnwidth-4pt, above=0pt of frame2.south] (csv_data) {\faTable Completed List of Papers};

            \draw[flow_arrow] (team) -- node[arrow_label, pos=0.3] {\large 200+} (csv_data);
        \end{scope}

        \draw[flow_arrow, color=black!90, line width=1.5pt] (corpus.south) -- (team.north);

    \endgroup 
    \end{tikzpicture}
    \caption{Paper selection methodology}
    \label{fig:methodology}
\end{figure}

\section{Fundamentals}\label{sect:basics}

During the years, Internet censorship has developed its own jargon and also started a vivid research area overlapping across several disciplines (e.g., privacy, network security and traffic engineering). In the following, we present the basic terminology as well as the capabilities of a censor organized within a suitable taxonomy.  

\subsection{Basic Terminology}
Aceto and Pescapé refer to \textbf{Internet censorship} as the \emph{intentional impairing or blocking of access to online resources and services} \cite{Aceto2015}. They further define \textbf{censorship circumvention} as the \emph{process of nullifying the censoring action, i.e., accessing the unmodified target -- or an equivalent copy -- despite the presence of a censoring system}. In this context, the \textbf{target} is an \emph{online resource or service}. The target can refer to nodes/endpoints, sites/URLs, protocols and other information that is required for a user to access it, while the altered experience of end-users when accessing a target is called a \textbf{symptom} \cite{Aceto2015}.

{Note that while Internet censorship, in a broader sense, also covers governmental/authority actions that moderate or remove content, we focus on pure network-level censorship and thus exclude political aspects of Internet censorship. However, we highlight relevant human aspects at the end of the paper.}

{Censorship can also be described by Simmons' \emph{Prisoners' Problem} \cite{DBLP:conf/crypto/Simmons83}. In this setting, Alice and Bob are prisoners, located in isolated cells. The only way they can exchange messages is through the \textbf{warden}. The warden can read and manipulate/drop these messages. The goal of Alice and Bob is to communicate by handing slightly manipulated messages to the warden. The manipulations represent secret information. The warden delivers these benign-looking messages while not recognizing their embedded secret message. In a censorship setting, Alice would aim to access a target (Bob) while the censor (the warden) influences that access. The term \emph{warden} is widely used for censoring actions but also for adversaries and countermeasures that target censorship circumvention.}

{Hall \ea split the censorship process into three stages: \emph{prescription}, \emph{identification}, and \emph{interference} \cite{rfc9505} that correspond to the phases in which censors determine the targets they want to block, how they detect if users try accessing such targets, and how they handle access attempts to these targets, respectively.}
Following Aceto \ea, censorship can {also} be split
into client-, server-, network-side and self-censorship \cite{Aceto2015}. 
\textbf{Client-side} censorship refers to censorship actions performed on the network user's client, e.g., by means of a keyword filter or a personal firewall that blocks access to targets. Instead of such \emph{additional} software components, censors can also offer \textbf{replacements for popular tools}, such as an own instant messaging or video conferencing tool with censorship functionalities already built-in.
In contrast, \textbf{server-side} censorship is integrated into a (target's) server or applied to it. {A technical example is a server-internal source address filter} that prohibit access to selected resources {for national users}.
\textbf{Network-side} censorship operates as part of the routing infrastructure. For instance, a censor could drop network packets that aim to reach undesired destinations. Similarly, undesired flows can be slowed down or redirected to monitoring networks. 
Finally, one form of censorship is \textbf{self-censorship}. In such a case, a user \emph{self-restrict[s his/her] possibilities of the expression due to fear of punishment, retaliation, or other negative consequences} \cite{Aceto2015}. {It is important to note, however, that self-censorship should be understood as a behavioral reaction to existing (perceived) censorship and surveillance, not as an intentional strategy imposed by censors. At the same time, it reinforces and extends the reach of censorship by amplifying its effects.}

\subsection{General Capabilities of a Censor}
\label{Sec:Foundamentals:GeneralCapabilities}
\begin{figure}
    \centering
    \includegraphics[width=1.00\linewidth]{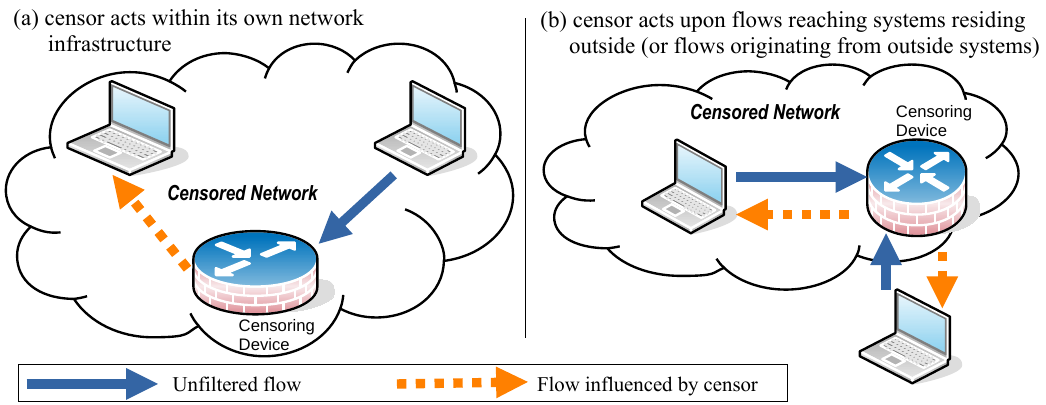}
    \caption{A censor's fundamental influence on communication between nodes}
    \label{fig:censorgeneral}
\end{figure}

The capabilities of censors reportedly vary \cite{Khattak:2016}.
In general, a censor can both influence the communication within its own network infrastructure (Fig.~\ref{fig:censorgeneral}(a)) or the communication to and from external nodes with internal ones (Fig.~\ref{fig:censorgeneral}(b)). Note that the device that separates the censored from the public network is called the \textbf{network perimeter}, and its filtering actions \textbf{perimeter filtering} \cite{weinberg2012stegotorus}.

Khattak \ea describe a censor in an abstract model that (i)~classifies traffic to detect flows that are relevant for blocking attempts; (ii) contains a cost function that considers collateral damage to a potential censorship action; and (iii) a decision function for determining the censorship action to be taken under consideration of outputs from the classifier and the cost function \cite{Khattak:2016}.

In general, a censor can perform four types of fundamental censoring actions  \cite{Aceto2015,NIHbook,rfc9505}:

\begin{enumerate}
    \item \textbf{eliminate} the target's content (e.g., taking down a particular website){, which is usually caused by an administrative order.}
    \item \textbf{prevent access} to a non-eliminated target (e.g., blocking access to a host or the resolution of related DNS records, blocking VPN protocols and circumvention tools used to bypass filters, deactivating Internet access in a whole region etc.).
    \item \textbf{limit} the communication with a target (e.g., slowing down a live stream of a protest), this is sometimes called \textbf{soft blocking}.
    \item \textbf{monitor (fingerprint) behavior} {and identify} end-users by passive or active means (e.g., plain eavesdropping, redirecting traffic through analytical networks, or injecting a Trojan that leaks monitored data).
\end{enumerate}

\begin{figure}
    \centering    
    \includegraphics[width=1.0\linewidth,trim={0cm 1.3cm 0cm 1.3cm},clip]{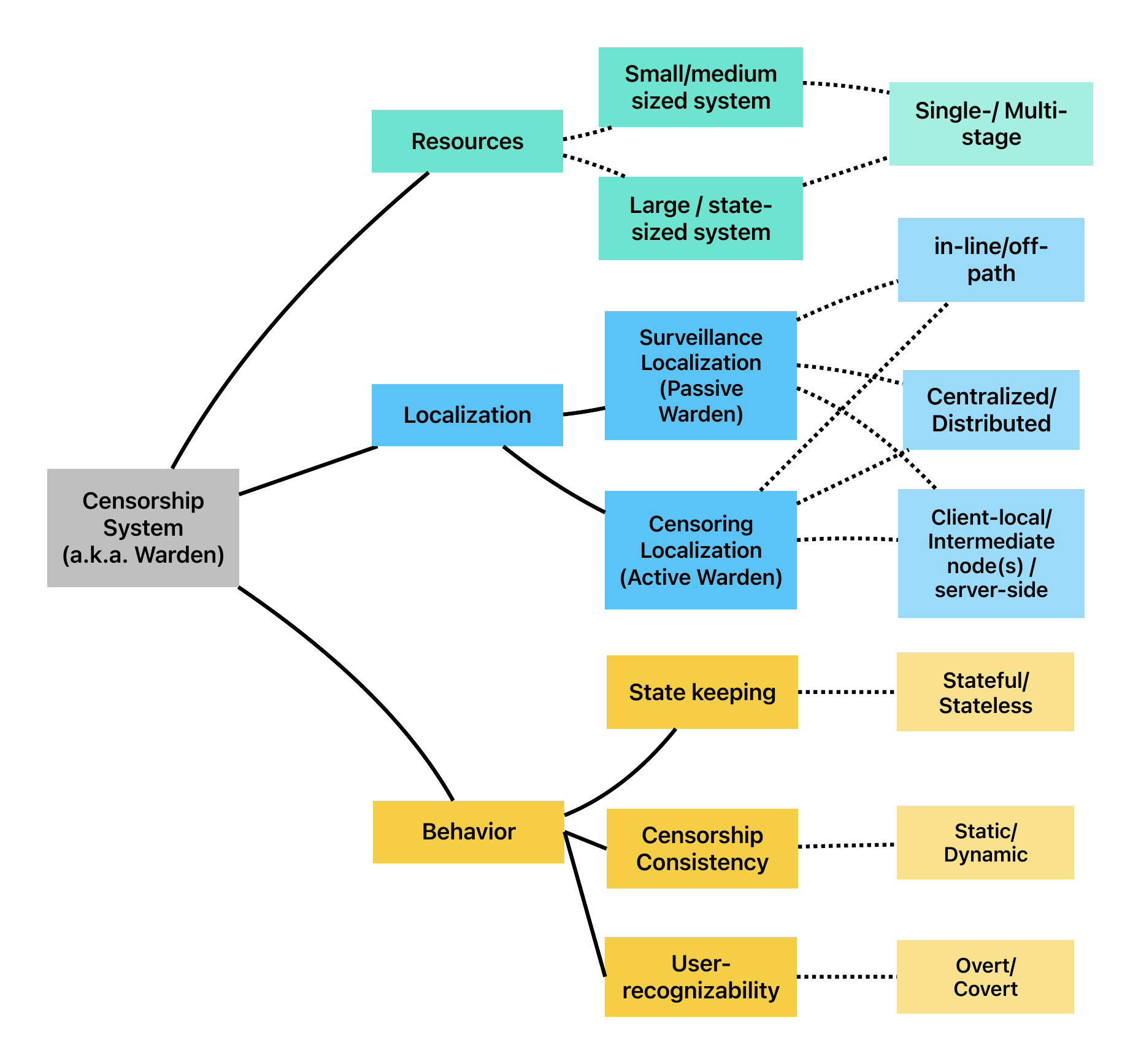}
    \caption{High-level censorship system taxonomy}
    \label{fig:tax:censorgeneric}
\end{figure}

\paragraph*{Taxonomy of censorship methods}
Censorship actions can be categorized in a more fine-grained manner. For this purpose, we reviewed and merged both, literature on censorship methodology \cite{Aceto2015,houmansadr2013parrot} and on network information hiding \cite{FGCS:DynWarden,NIHbook}. Such a merge turned out to be effective, especially due to the overlap between the two research areas. The resulting taxonomy is shown in Fig.~\ref{fig:tax:censorgeneric}.

First of all, we categorize censorship systems based on their (known) \textbf{censoring resources}. {The fact that resources are a key issue for censors (and other complex filtering systems) was mentioned by several papers, e.g., \cite{Aceto2015,FGCS:DynWarden,Wu:23:GFW}. Censorship systems} can either be small (e.g., a SME, a single NGO, a local government) or large/state-level (including regional/national governments). This aspect is driven by the proposal of Houmansadr \ea to distinguish between local, state-level oblivious and state-level omniscient censors~\cite{houmansadr2013parrot}, which was also adapted to the case of network information hiding in~\cite{FGCS:DynWarden}. Note that state-level oblivious and omniscient censors differ in terms of their resources (the latter can \emph{aggregate data collected at different network locations and store all intercepted traffic for offline, computationally expensive analysis} \cite{houmansadr2013parrot}). However, for our high-level view, it is enough to distinguish between the two cases mentioned above.
Further, these two types of censor resource categories can employ \textbf{single-stage or a multi-stage censoring} (where a first-stage performs quick and superficial flow analysis and follow-up stages perform analyses in depth) {\cite{rfc9505,weinberg2012stegotorus,houmansadr2013parrot}}. 

While Aceto and Pescapé already include \textbf{localization} aspects, such as symptoms and triggers, the relevant generic aspects are the surveillance location {(in the sense of technical monitoring)} and the censoring location of the censorship system. Both can be either \textbf{in-line/in-path/on-path} (some authors differentiate between in-path and on-path, depending on the possibility of the censor to \emph{modify} packets or passively monitoring them {\cite{ZhongjieEtAl:2017:YourStateIsNotMine}}) or \textbf{off-line/off-path} {\cite{rfc9505},}\cite{ICLab:SP20,Jin:2021:CensMeasurement}. However, we extended the categorization with ideas of Mazurczyk \ea \cite{FGCS:DynWarden}, where they differentiate between \textbf{centralized and distributed} systems. They further differentiate between localhost/intermediate node and network systems, which we merged with Aceto and Pescapés differentiation of client-side, server-side and network-side censorship, so that we finally have the options \textbf{client-local}, \textbf{intermediate node(s)} and \textbf{server-side}.

Regarding the \textbf{behavior} of the censorship system, most works {(e.g., \cite{Aceto2015,rfc9505,ZhongjieEtAl:2017:YourStateIsNotMine})} differentiate the \textbf{state-keeping} functionality (either \textbf{stateless} or \textbf{stateful}). However, only {few works explicitly considered different ``\textbf{consistencies}'' (or: ``dynamics'') \cite{FGCS:DynWarden,ZhongjieEtAl:2017:YourStateIsNotMine}. In this category,} censorship can either be \textbf{static} (i.e., the behavior of the censorship system remains (mostly) unchanged over time) or \textbf{dynamic} (the behavior of the censorship system changes over time, e.g., to increase uncertainty of circumventing tool users or to rapidly adjust to new circumvention trends).

Finally, following Niaki \ea \cite{ICLab:SP20}, censorship systems can operate in an \textbf{overt} way, i.e., they inform the user that censorship is happening, for instance through a blocking web page, or \textbf{covertly}, i.e., causing misleading errors so that the user does not recognize that censorship happens. Accordingly, we call this category \textbf{user-recognizability}.

\paragraph*{Censorship Costs} Censorship is linked to certain (potential) costs, of which we identified three \emph{major} categories:

\begin{enumerate}
    \item Blocking and limitation techniques are often based on heuristics. Such heuristics are never ``perfect'' and might cause undesired side effects, e.g., due to imperfect regular expressions used in a filter. Moreover, there is an imperfect target collection procedure (e.g., incomplete or outdated list of IP address /websites to block).
    The two costs (side effects) are (i) preventing Internet access to targets not meant to be blocked (\textbf{overblocking} due to \emph{false-positives}), and (ii) not covering all desired targets (\textbf{under\-block\-ing} due to \emph{false-negatives}) \cite{master2023worldwide,mohajeri2012skypemorph,tschantz2016sok,weinberg2012stegotorus}. {A special form of overblocking called \emph{censorship leakage} occurs when international traffic transits through a  country but is unintentionally influenced by that country's censorship \cite{lipphardt:25:1800cens}.}
    \item Monitoring/blocking flows and limiting throughput results in \textbf{resource costs} (equipment, staff, investments and operational costs) \cite{ScrambleSuit,weinberg2012stegotorus,tschantz2016sok}. For this reason, a censor might not be able to run heuristics and blocking attempts on every packet or flow that passes through its network infrastructure \cite{houmansadr2013parrot,MassBrowser,figueira2022stegozoa,mohajeri2012skypemorph}. Thus, there are often at least two-stages of blocking: a non-detailed level with quick heuristics that are able to check many flows/packets. More interesting flows are then analyzed in-depth by more costly means, e.g., caching flows and conduct expensive post-analysis \cite{houmansadr2013parrot,weinberg2012stegotorus}. Blocking \emph{all} traffic (or a large fraction), e.g., during the \emph{Arab Spring}, is often not considered attractive for the censor although was practically applied \cite{mohajeri2012skypemorph,weinberg2012stegotorus}. A study performed in 2023 has shown that total Internet shutdown is still applied in several countries \cite{master2023worldwide}.
    \item Censorship measures \textbf{decrease the reputation} of an entity acting as a censor, as well as the trust in it.
\end{enumerate}

\section{{Network-level} Censorship Techniques}\label{sect:censortechniques}
A censor can select censorship techniques that target the users, hardware, software, or the network. For instance, country-level ``real name'' policies \cite{Aceto2015} can foster self-censorship. Hardware delivered to end-users could be manipulated or replaced before it reaches the customer. {Moreover}, a software manipulation can be realized by deploying a backdoor in its code.

{In contrast to these methods, we focus on} \emph{network}-level censorship {targeting the Internet, transport and application layers}. In the network context, a censor could interrupt{, redirect, alter, or record a node's transmission to specific destinations, e.g., through the physical manipulation of cables} \cite{master2023worldwide}. Most methods focus on the network level protocols above the physical and data link layers. This is why we survey censorship techniques residing on the Internet layer or above. {After introducing the various censorship techniques, we also explain their limitations.}

\subsection{IPv4/IPv6-based Methodology}\label{sect:censmeth:IP}
One of the most fundamental methods for Internet censorship is \textbf{discarding IPv4/IPv6 packets} destined to an undesired target \cite{Aceto2015,Khattak:2016}. The target can reside outside the censor's network (such attempts have been made early on, see, e.g.,~\cite{ang1996censorship}). Moreover, the censor can prevent communication of selected nodes \emph{within} its network (e.g., nodes of a group of journalists and their chat service node). The censor can also \textbf{re-route} such packets through an analysis network for further \textbf{inspection}. This can be achieved through altered routing tables on a router or by sending ICMP redirect messages (ICMPv4 type 5 / ICMPv6 type 137). However, ICMP redirects are often ignored by hosts who receive these messages due to security reasons.

A censor can also \textbf{introduce a disruptive ICMPv4 or ICMPv6 error response}, e.g., ICMPv4 type 3 or ICMPv6 type 1 (\emph{destination unreachable}) with desired code options (\emph{network / host / port unreachable}, \emph{destination network unknown}, or \emph{communication administratively prohibited}), or ICMPv4 type 11 / ICMPv6 type 3 (\emph{time exceeded}), even if the packet would actually reach its destination \cite{raman2022network,Aceto2015}. Finally, the censor can \textbf{decrease the Quality of Service (QoS) for
selected flows} between sources and targets, which can be done by increasing packet loss, delays, or jitter as well as by reducing the bandwidth  \cite{Aceto2015,raman2022network,Khattak:2016,rfc9505}.

\subsection{Routing Protocol-based Methodology}\label{sect:censmeth:BGP}
Routing-based censorship techniques focus almost exclusively on the \emph{Border Gateway Protocol} (BGP), which is the major routing protocol in the modern Internet. BGP routers exchange information about how they reach destination networks, taking care of the routing at large, i.e., between \emph{Autonomous Systems} (AS) \cite{Cho:BGPhijclassif}.
Therefore, BGP routers propagate \emph{prefixes} (network identifier with a network mask) jointly with distance information.

Censorship methods for BGP have been summarized already by Aceto \ea\cite{Aceto2015}. The first method is to let a country's ASes \textbf{disappear} from the Internet by preventing the propagation of BGP updates, with users experiencing ICMP \emph{network unreachable} errors. 
A second method is \emph{prefix hijacking} \cite{ballani2007study,Khattak:2016,Cho:BGPhijclassif,Aceto2015}. Traffic that should be directed towards an AS $B$ is redirected to an AS $C$ instead---through false propagation of a prefix not owned by the censor \cite{ballani2007study,Cho:BGPhijclassif}. A censor would need to propagate a short path for the hijacked prefix, as this increases the chances for neighbor routers to consider the route to $C$ \cite{Cho:BGPhijclassif}. This allows for \textbf{blackhole attacks} so that a censor renders specific destinations unreachable for end-users~\cite{Aceto2015}. Alternatively, a censor might transparently \textbf{redirect} traffic through its own AS $C$ for \textbf{inspection}, and finally forwards the traffic to $B$ so that users barely notice any alterations or delay \cite{ballani2007study,Cho:BGPhijclassif}. Users could experience \emph{time exceeded} errors if redirected routing paths are too long \cite{Aceto2015}. 
While such blackholing and interception attacks can appear between ASes, a regional censor might also apply censoring methods on regional or even local routing environments, potentially even those running the \emph{Open Shortest Path First} (OSPF) protocol.

\subsection{UDP-, TCP- and QUIC-based Methodology}
UDP- and TCP-based censorship techniques are often applied in the form of \textbf{port filtering}, and usually combined with specific IP addresses, forming tuples, such as (client IP, client port, server IP, server port) that are (temporarily) blocked~\cite{master2023worldwide}. As in the case of other protocols, \textbf{eavesdropping and consecutive traffic analysis} can also be conducted on the transport layer, e.g., to fingerprint the type of end-user device or the involved application-layer tool through packet sizes or other meta-data. Alternatively, undesired flows can also be \textbf{throttled}, e.g., by dropping selected UDP datagrams or delaying TCP acknowledgment segments on a router that is located on the path between end-user and target. Further, \textbf{hijacking} of transport-layer flows can happen at the start of a flow/connection or later, e.g., during or right after a TCP handshake was completed or in the middle of a download when some keywords were detected in the payload. Undesired datagrams and segments can be discarded and connections terminated (e.g., injecting spoofed TCP segments with active \texttt{RST} or \texttt{FIN} flags to terminate/close connections) \cite{Aceto2015,raman2022network,Khattak:2016}. Elmenhorst \ea report \textbf{blackholing} against QUIC connections that interrupt the handshake \cite{Elmenhorst21:HTTP3:QUIC}. The resulting symptom for an end-user depends on the time of tampering. For instance, dropping packets during the TCP connection establishment/QUIC handshake phase lets the service appear unavailable (e.g., causing a timeout) \cite{Elmenhorst21:HTTP3:QUIC} but dropping packets in the middle of a file download would let the connection appear unstable.

\subsection{DNS-based Methodology}
Censors can modify DNS servers as well as DNS entries (called \emph{resource records}) in multiple ways. The most relevant types of resource records for censors are typically \texttt{A} (provides the IPv4 address of a hostname) and \texttt{AAAA} (provides the IPv6 address of a hostname). According to Aceto \ea and Khattak \ea, resource records of a censor-controlled DNS server can be influenced in several ways: (i) \textbf{resource record removal} ({technical deletion from the DNS zone database}) or \textbf{pretended in-existence} (sending an \texttt{NXDOMAIN} response), even if the resource record could be fetched from an external authoritative DNS server (that is explicitly not contacted by the censor's DNS server), (ii) \textbf{response alteration:} resource record's resolution can be modified so that it refers to an IP address that provides a ``block page'' or an ``error page'' (from an end-user perspective, only the content of the provided website differs), (iii) \textbf{returning a ``failing IP''}, e.g., a non-routable in-house IP address, and (iv) \textbf{returning a {network-level} surveillance system's IP address} to further investigate traffic of clients \cite{Aceto2015,Khattak:2016}. Such a surveillance system can also act as a transparent proxy that forwards all traffic to a client's desired destination (and vice versa) but records and/or modifies inspected traffic.

If the censor has no direct control of the DNS server but has enough visibility over DNS requests, it can also attempt a \textbf{DNS injection attacks on the side} \cite{Aceto2015,Jin:2021:CensMeasurement}. In this case, when the censor notices a DNS request sent from a client's resolver to a DNS server, it might reply faster than the uncensored DNS server, especially if that server needs to forward the query to DNS root or TLD servers \cite{Aceto2015}. The client would accept the quickest response, resulting in the adoption of the fake record. Such an attack has been reported as being part of China's UDP-based DNS filtering, where forged responses for censored domains are injected before the legitimate ones arrive \cite{274650}.
Finally, in case a DNS transfer is conducted through TCP (e.g., because it exceeds the maximum allowed size of UDP-based DNS packets), an on-path censor could also \textbf{inject TCP \texttt{RST} segments} \cite{Jin:2021:CensMeasurement}.

\subsection{HTTP-based Methodology and Web-content Filtering}\label{Sect:CensorMeth:HTTP}
HTTP-based filtering was reported to be the most detected censorship mechanism in an analysis by Master and Garman published in 2023~\cite{master2023worldwide}.

When the censor controls the target, it can process {\textbf{header and payload filtering}, i.e., block packets of connections that contain undesired keywords/content, such as \texttt{Host:} parameters \cite{rfc9505}}. In case of unencrypted HTTP connections, the censor has multiple options: if the censor \emph{has control over the server}, it can \textbf{influence the {transfer of} content hosted at the server} (e.g., modify the HTML content during transmission~\cite{Khattak:2016}), it can also \textbf{alter the behavior of the server}, and it can \textbf{record the exchanged traffic} between client and server for further analysis. In case of behavioral alternations, the censor can return a {\tt 30x} HTTP response code that represents a \textbf{redirect}~\cite{Aceto2015}, i.e., the client is redirected to some error page, blocking page or a fake copy of a website where the client is brought in direct contact with the analysis backend. The censor can also \textbf{pretend that a website does not exist or that access to the website is forbidden} (response codes \texttt{404} and \texttt{403}, respectively) \cite{Aceto2015,Khattak:2016}.
Further, the censor could \textbf{pretend that there is an internal server error} (response code \texttt{500}).\\
Instead, if the censor has \emph{no} control of the target server but is located on a gateway between the client and the server, it can \textbf{eavesdrop/monitor} the connection and can influence the content of HTTP connections by \textbf{injecting fake response codes} as described before. This also works if the end user uses \emph{circumvention sites} (i.e., non-blocked websites that display bridge targets as proxies \cite{IntZensurChina}). However, the injection needs to be conducted in a just-in-time fashion so that the faked server response arrives before the server's original response.
To this end, the censor's gateway could handle the whole handshake with the client and pretend to be the server while the original packets are not forwarded to the server, at all.
An alternative scenario mentioned by Aceto and Pescapé \cite{Aceto2015} is one in which the censor installs a \textbf{transparent proxy} on the path between the client and the server. This provides the advantage that a server's responses never directly reach the client.

\begin{figure}[!tp] 
    \centering
    \includegraphics[width=\columnwidth,trim={0cm 0.4cm 0cm 0.4cm},clip]{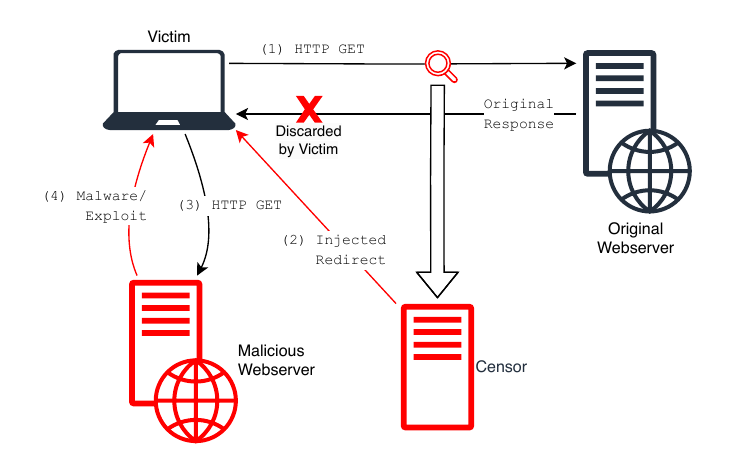}
    \caption{Man-on-the-Side (MotS) attack. The client (i) initiates an HTTP request to the legitimate server. The censor intercepts the traffic and (ii) injects a TCP segment containing a redirect to a malicious server. The client's original request may reach the server, and the server will respond. However, the censor's response arrives earlier and is thus considered a duplicate by the client and discarded. The client then connects (iii) to the malicious server, which (iv) provides censored content and/or delivers malware.}
    \label{fig:mots-attack}
\end{figure}

Finally, a censor could redirect a user through a \textbf{Man-on-the-Side (MotS) attack} that was revealed to be used by the NSA \cite{Wuebbeling:MotS}. The attack itself has been known since 1985 and is carried out on the HTTP and TCP levels jointly. The goal is to redirect a user who aims to visit a target website to a censor-controlled website where the user's browser is tricked to download a drive-by-exploit that infects the user with malware. The core idea is that the censor replies quicker to the user's HTTP packets than the actual target, which requires the censor to inject a TCP segment. The original reply of the target arrives later than the injected one, and is thus considered a duplicate and discarded by the client. The injected segment contains the source address of the server and is destined to the IP of the client; it contains a redirect to the website of the attacking censor. The hosted replica is optically indistinguishable from the original target. Fig. \ref{fig:mots-attack} illustrates this process.

In addition to HTTP, HTTPS censorship has been discovered more often in recent years, probably rooted in the fact that the majority of HTTP traffic is now TLS encrypted \cite{master2023worldwide}. When HTTPS is used, a censor can still perform a set of actions: (i) it can render HTTPS connections to undesired targets unattractive by \textbf{dropping randomly selected packets or disrupting connection establishment} through TCP \texttt{RST} packets \cite{TsaiEtAl:NDSS24:ModelingDetectingIntCens}; (ii) it can \textbf{try to gain access to encrypted communication} \cite{TsaiEtAl:NDSS24:ModelingDetectingIntCens} as discussed in Sect.~\ref{Sect:Censorship:TLS}. 
The above-mentioned \textbf{MotS attack} could also be conducted in case of HTTPS if the censor injects its TCP packet early during the connection establishment phase (the initial TCP SYN packet is still unencrypted, and the censor could spoof the whole TCP handshake and then the first packet that contains the redirect). 

Censorship can also be conducted in a user-tailored manner, where users are first tracked and then handled individually. An important scenario refers to the adoption of the \emph{HTTP Strict Transport Security} (HSTS) protocol. So-called \textbf{HSTS supercookies} can be used for censor-triggered tracking attempts \cite{davitt2024costrictor,syverson2018hsts}. In HSTS, websites let a browser store a unique HSTS bit that the browser transmits to a website during future visits. For creating a supercookie, websites can further include (invisible) subcomponents of other websites that also store such HSTS bits, combining the unique value stored in the browser that can later on be used by the censor to identify the client while browsing.

\subsection{Other Application-layer Protocols and Platforms}
{A censor's actions on this level are related to traffic, user accounts, and content. To allow for a better understanding, we decided to include selected non-purely technical aspects that provide a slightly broader view in this subsection.}

\paragraph{{Keyword Filtering}}
Like in case of HTTP and DNS, all unencrypted communication (including e-mail message bodies transferred via SMTP, IMAP or POP3, chat content or FTP filenames and payload \cite{master2023worldwide}) can be \textbf{analyzed by content filters} (keywords or other forms of content and hash values {\cite{master2023worldwide,rfc9505}}). %
For instance, unencrypted communications have been censored early on \cite{ang1996censorship}. Traffic matching such filters can be \textbf{blocked or redirected to monitoring networks}.

AI-driven {payload} analysis is increasingly being deployed to enhance {simple filter heuristics}, enabling sentiment analysis and image detection within communications \cite{chenAccuracyBiasesAIBased2025}. For example, AI can now analyze chat logs for nuanced critiques or automatically detect images associated with protest movements. This allows censors to move beyond explicit keywords and suppress a broader range of dissenting communications.

\paragraph{{Blocking and Throttling}}
Some popular services are also blocked with crude heuristics. For instance, Telegram was censored by \textbf{blocking large sets of IP addresses} associated with nodes responsible for providing the required computing, network, and storage resources. As a result, Telegram users applied several methods, including obfuscation as well as \emph{IP hopping} so that IP addresses change quickly, which requires blocking IP addresses with massive collateral damage \cite{ermoshinaTelegramBanHow2021}. Another communication service (Twitter, now X) was reportedly subject to \textbf{throttling}, and a censor identified Twitter-usage through the Server Name Indication of TLS (see Sect.~\ref{Sect:Censorship:TLS}) when related domains were identified, e.g., \texttt{twimg.com}, \texttt{twitter.com} and \texttt{t.co}, \cite{xueThrottlingTwitterEmerging2021}.

\paragraph{{Access Revocation and Content Deletion by Authorities}}
{Another early attempt that can still be found in today's Internet censorship is} deleting anonymous posts, performing or revoking access to Usenet, e-mail or FTP \cite{ang1996censorship} { or other online services, which are technical actions ordered by authorities}.
A {related} strategy of censors is to \textbf{delete or modify content} (videos, textual content{, e-mails}) of online platforms {or servers} if under their own management or request such removals from operating companies.
With the rise of online gaming platforms such as \emph{Steam} or \emph{PSN}, they have become targets for censorship, too. Some of these censorship systems are in place to prevent addiction to online gaming platforms, while other censorship systems aim to limit the exchange of political statements. {In such settings, censorship is} conducted based on age entries and IP address blocking \cite{fengStudyChinasCensorship}.
{If} undesired messages are {detected, the censor could also \textbf{pretend that a message was delivered} but actually delete it.}

\paragraph{{Replacement of Tools}}
{Despite not being directly targeting the network layer, this censorship approach allows to perform a direct control over the resulting traffic. Specifically, a censor can interfere with both the development process or the distribution stage of a specific tool, in order to produce ``censored'' versions.}
In case of popular video and text chat services, a censor could block access to the particular original tools (e.g., their download sites) and \textbf{offer replacement tools} with censorship functionality already built in. This was done for \emph{TOM-Skype} (Chinese clone {of the discontinued Skype audio-video chat tool}) and the instant messenger \emph{SinaUC} \cite{aase2012whiskey,Aceto2015}. The Chinese e-mail and chat platforms \emph{QQMail} and \emph{WeChat} conduct censorship based on keywords and keyword combinations \cite{QQMailCens}. In addition, the LINE chat tool, the microblogging platform \emph{Sina Weibo} and live streaming platforms \emph{YY}, \emph{9158}, \emph{SinaShow} and \emph{GuaGua} as well as Chinese online games face censorship \cite{Knockel:ChineseToolsCensGames,knockel2015ChineseToolsCensVideo}.

\subsection{TLS-based Methodology}\label{Sect:Censorship:TLS}
The majority of modern operating systems already provide a list of certificates signed by a trusted Certificate Authority (CA), rendering attacks on TLS-based communications challenging, especially unnoticed ones as browsers and other tools inform users about certificates that have not been signed by a trusted CA. For this reason, a censor would need to \textbf{deploy its own certificate}, e.g., for a websites' replica on the client. This can be done through malware, but is not easy to accomplish. Through DNS cache poisoning, the client could then be redirected to the replica site hosted on a server with an IP address that is owned by the censor, and the faked certificate for the website must be used by the client at this point. Without DNS cache poisoning, the censor could still try to trick the user to visit a replica site using a domain containing homoglyphs, i.e., a domain name with characters that \emph{look} like the ones of the original website (e.g. \texttt{target-website.xyz}) but contain letters that are drawn from another alphabet, e.g., replacing a Latin $e$ with a Greek $\epsilon$: \texttt{target-w\textepsilon{}bsite.xyz}.

In case of a man-in-the-middle (MitM) scenario, the censor could also conduct a \textbf{STRIPTLS attack} (also known as \emph{STARTTLS stripping attack} and \emph{STARTTLS downgrade attack}) in which the censor removes the STARTTLS command from packets so that the TLS handshake is not initiated \cite{Margolis2018:MTASTS}. Alternatively, the censor could apply a \textbf{downgrade attack} during the TLS handshake so that the client believes that the server only supports an outdated TLS/SSL version and/or less secure ciphers. {However, the options for cipher suite downgrades are limited since TLS/1.3 only allows the selection of five up-to-date ciphers: four based on AES 128/256 using GCM/CCM and SHA256/SHA384 as well as one using CHACHA20\_POLY\-1305\_SHA256) \cite{rfc8446}.}

Further, a censor can determine the virtual hostname of the server if the \textbf{server name indication} (SNI) extension is used by the client \cite{master2023worldwide,Chai:ESNI,Jin:WWW21:EncrDNSCensorship,xueThrottlingTwitterEmerging2021,Jin:2021:CensMeasurement,rfc9505}. While the encrypted virtual hostname is sent to the HTTPS server using a \texttt{Host} string during the HTTP request, the SNI is transmitted earlier during the TLS \emph{ClientHello} message and tells the server the desired virtual hostname in plaintext. The purpose of SNI is that the server can select the correct certificate of that virtual host \cite{rfc6066} which is then used for encrypting the following HTTP request.

There is also an \textbf{Encrypted SNI} (ESNI). ESNI is available since TLS/1.3. As the established TLS channel cannot be used to encrypt the SNI (this would require to first exchange keys but they will be exchanged with succeeding packets, resulting in a chicken or the egg dilemma), one needs to employ a \emph{separate} channel for exchanging a secret used to encrypt the SNI. ESNI works as follows \cite{ietf-tls-esni-22,Chai:ESNI}: The administrator of the web server must publish a public key for the particular domain on a DNS server (typically in the form of a \texttt{TXT} record). The client fetches that key before connecting to the web server. The client uses the ESNI (instead of SNI) extension in the \emph{ClientHello} message to encrypt the SNI with a symmetric key derived from the server’s public key and a key selected by the client through a hybrid public key encryption (HPKE) as defined in RFC 9180 \cite{rfc9180}. 
However, a censor can still block connections that contain the ESNI extension in the \emph{ClientHello} message (although the censor would act blindly, i.e., not knowing the particular domain that was requested). It could also \textbf{reduce the QoS} of the connection. Moreover, the censor could also first  \textbf{spoof the target's DNS key entry} and act as a MitM attacker between client and server. Further security considerations on ESNI can be found in \cite{ietf-tls-esni-22}.

{Finally, for TLS versions before 1.3, censors can determine the target through the certificate transferred as part of the server's response to a ClientHello message \cite{rfc9505}.}

\subsection{Multi-stage Censorship with Active Probing}
\begin{figure}[!tp] 
    \centering
    \includegraphics[width=0.9\columnwidth,trim={0cm 0.4cm 0cm 0.6cm},clip]{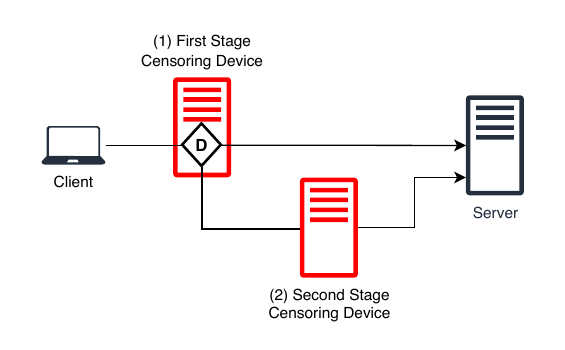}
    \caption{Multi-stage censorship: The first stage censoring device (i) conducts a high-level censorship decision (D) to either allow, block or forward the flow to a second stage censoring device (ii) for in-depth analysis.}
    \label{fig:multi-stage}
\end{figure}

As mentioned in Sect.~\ref{Sec:Foundamentals:GeneralCapabilities}, censorship is sometimes realized in a multi-stage fashion, in which the first stage is a performance-optimized stage that allows for scanning a massive amount of traffic in real-time. Flows demanding further inspection are then redirected into a second (or further) stages. Fig.~\ref{fig:multi-stage} visualizes the concept. 
For instance, Aceto and Pescapé, reported that HTTP can be censored/monitored in a second stage dedicated to filtering \cite{Aceto2015}. In such a setup, the first stage redirects the client to a fake HTTP proxy using either DNS hijacking (client requests a blacklisted IP of a webserver) or BGP hijacking (client sends a packet to a blacklisted destination, port $80$ or $8080$).

However, by using only passive observations, a censor cannot necessarily state if a client is talking to a ``benign'' node or to a circumvention proxy (e.g., a Tor bridge or a VPN endpoint). To this aim, \textbf{active probing} refers to attempts that involve an active interaction with such nodes to infer information about the services they run \cite{Fifield:PhD}. 
Active probing also provides a link to related research disciplines: a similarity of active probing and network information hiding methods is that an adversary (censor) tries to infer (and optionally influence) the covert communication protocol (e.g., a proxy protocol or a covert channel protocol) \cite{NIHbook}, which optimally leads to gained information about involved parties and the content or type of communication. 

In general, censors utilize a second-stage infrastructure for such probing approaches. Fifield highlights that active probing is linked to certain benefits for a censor \cite{Fifield:PhD}: (i) probing can be conducted asynchronously (time-delayed) to trigger observations, and (ii) the risk of false-positive classification (and consequential blocking) must be considered a collateral damage, but the overall blocking precision is improved by active probing.
During active probing, a censor sends different types of probe requests, depending on the protocol. For instance, a censor could send tailored HTTP requests to a node that is suspected to be a HTTP proxy. 

\paragraph*{Case Study}
China developed one of the most sophisticated monitoring and censorship systems in the world \cite{zuboff:age}. Reportedly, the \emph{Great Firewall} (GFW) utilized active probing mechanisms already in 2010 by initiating own connections to remote nodes to infer if they provide proxy-related functionalities \cite{Fifield:PhD}. 
%
In \cite{Wu:23:GFW}, Wu \ea characterize the GFW when used to block websites and filter contents~\cite{185061}. As fully encrypted protocols are critical for censorship circumvention, the GFW aims to detect and block this type of traffic.

The GFW operates using crude heuristics to exempt traffic that is unlikely to be fully encrypted, and then blocks the remaining non-exempted traffic in real time \cite{Wu:23:GFW,GFWReport}. The censorship system includes both passive and active components that function independently. For passive censorship, the GFW is capable of performing purely passive detection, traffic analysis, and real-time blocking of fully encrypted traffic. In parallel, the GFW employs active probing as part of its active censorship. Both active and passive censorship are based on efficient and largely identical heuristics, with active censorship having additional packet length rules. Specifically, the GFW only probed experimenter's connections with 200-byte random data, but not with 2- or 50-byte random data \cite{Wu:23:GFW}.

There are five specific rules that the GFW employs to exempt non-encrypted traffic from being blocked \cite{Wu:23:GFW}: The first exemption rule (Ex 1) measures the entropy of traffic packets by counting the set bits (\emph{popcount}). If a bit string fall between 3.4 and 4.6 bits per byte, the traffic will be blocked because it appears too similar to randomized data, which is characteristic of encrypted traffic. Ex 2 exempts traffic from blocking if the first 6 bytes of the connection fall within the printable ASCII range. Ex 3 exempts traffic if more than half of all bytes in the first packet fall within the printable ASCII range. 
Ex 4 exempts traffic if there are more than 20 contiguous bytes that are printable. Ex 5 exempts two popular protocols, TLS (as long as the first three bytes of the connection match the TLS \emph{ClientHello} message) and HTTP (at least for the HTTP methods \texttt{GET}, \texttt{PUT}, \texttt{POST}, and \texttt{HEAD}), as the GFW is able to infer these protocols from the first 3-6 bytes of the client's packet. Lange \ea figured out that China does not block any unencrypted HTTP/2 traffic \cite{censors-ignore-unencrypted-http-2-traffic}.

The GFW can block traffic on all ports. It conducts \textbf{residual censorship} \cite{BockEtAl:2021:YourCensorIsMyCensor,Chai:ESNI,ZhongjieEtAl:2017:YourStateIsNotMine}, meaning that the same 3-tuple (client IP, server IP, server port) will be blocked for 120 or 180 seconds, and the timer does not get reset during this period.\footnote{Another form of residual censorship is \emph{long-lived} residual censorship that can be extended to 30\;min and beyond \cite{bhaskar2024understanding}.} The GFW limits the number of connections it blocks residually, resulting in shorter blocking times when several connections are blocked in parallel. 
However, residual censorship makes it difficult for a proxy user to successfully connect once detected, as all involved connections between client and proxy are affected~\cite{Wu:23:GFW}.

The goals of the blocking strategies deployed by the GFW are to mitigate false positives and reduce operational costs. This is mainly achieved by 1) only blocking specific IP ranges, such as those of popular data centers that offer proxy connections (affected IP ranges), and 2) using a probabilistic blocking strategy that blocks only $26\%$ of the affected IP ranges \cite{Wu:23:GFW}. This approach results in a lower true positive rate but also reduces false positives. 

\subsection{Limitations of Censorship Techniques}\label{Sect:CensMethLimits}

{The presented  censorship techniques are also characterized by a wide range of limitations. In the following, we review and discuss them. Tab.~\ref{tab:SummaryCensTechLimits} summarizes the main limitations, organized on a per-layer basis.}

\begin{table*}
\footnotesize
\centering
\caption{Summary of censorship techniques' limitations according to literature (\CIRCLE: relevant, \LEFTcircle: partially relevant, \Circle: not relevant)}
\label{tab:SummaryCensTechLimits}

\begin{tblr}{
  width = \linewidth,
  colspec = {Q[70]||Q[90]Q[80]Q[45]Q[110]Q[40]Q[130]},
  column{1} = {r},
  column{even} = {c},
  column{3} = {c},
  column{5} = {c},
  column{7} = {c},
  vline{1} = {2-9}{},
  vline{2} = {1-9}{},
  vline{3-8} = {1}{},
  vline{3-8} = {2-9}{},
  hline{1} = {2-7}{},
  hline{2-10} = {-}{},
}
                       & \textbf{Censor Device Location-Dependence} & \textbf{Blocklist Incompleteness} & \textbf{Over\-blocking} & \textbf{Crypto\-graphic Hurdles} & \textbf{State\-holding} & \textbf{Notes}                                      \\
IPv4/v6                & \CIRCLE                                 & \CIRCLE                        & \CIRCLE              & -                             & \CIRCLE              & -                                                             \\
Routing                & \CIRCLE                                 &  \Circle                                 & \CIRCLE              & RPKI                           &  \Circle                       &  - \\
TCP/UDP/ QUIC          & \LEFTcircle                               & \CIRCLE                        & \CIRCLE              & \Circle                              & \CIRCLE              & QUIC challenging to block                                     \\
DNS                    & \LEFTcircle                               & \CIRCLE                        & \LEFTcircle            &  DNSSec, DNSCurve, DoH, DoT    & \Circle                       & - \\
HTTP(S)                & \LEFTcircle                               & \CIRCLE                        & \LEFTcircle            & Encrypted HTTPS traffic        & \Circle                       & HTTP smuggling, content trans\-formations (website screenshots) \\
Other Appl. Protocols  & \LEFTcircle                               & \CIRCLE                        & \CIRCLE              & TLS-based protocols            & \LEFTcircle            & Replacement tools costly to develop                           \\
TLS                    & \LEFTcircle                               & \CIRCLE                        & \CIRCLE              & ECH, ESNI                      & \Circle                       & -  \\
Multistage Censorship  & \LEFTcircle                               & \Circle                                 & \Circle                       & \Circle                              & \CIRCLE              & Obfuscation is an issue                                          
\end{tblr}
\end{table*}

\paragraph*{IP-level}
Aceto and Pescapé point out that an IP-level filter must be located on the path between client and target, which is not necessarily the case for alternative attempts~\cite{Aceto2015}, such as BGP-based hijacks (cf.\ Sect.~\ref{sect:censmeth:BGP}). They further mention the fact that IP filtering can result in overblocking, e.g., when an IP of a virtual host that also serves legitimate websites is blocked \cite{Aceto2015,Elmenhorst21:HTTP3:QUIC}.\footnote{Cf.~Sect.\ref{Sect:Censorship:TLS} on censoring virtual hosts while TLS is used.} This is confirmed by Master and Garman in a 2023-study that has shown a decline in IP blocking \cite{master2023worldwide}. Master and Garman mention three reasons for the decline: (i) blocklist maintenance of ephemeral IP addresses, (ii) collateral damage when blocking a CDN's IP range\footnote{For instance, Wu \ea were able to determine collateral damage of $0.6\%$ of all connections for China's filter rules \cite{Wu:23:GFW}.}, and (iii) deployment of IPv6 with the resulting growth in addresses. One additional issue refers to resource limits when a high number of fragmented packets arrive: when a censorship device tries to reassemble a high number of packets of which only fragments (IP packets with the \texttt{MF} flag set to `1') have arrived, a stateholding problem \cite{NormalisierungsPaper} arises.

\paragraph*{Routing}
BGP-based traffic interception became more challenging in recent years, especially due to the implementation of Resource PKI (RPKI) which prevents {some} hijacking methods \cite{Cho:BGPhijclassif}. {While RPKI prevents attackers from announcing a targeted IP prefix as an origin, it is still possible for an attacker to maliciously announce a short AS-path through themselves towards the actual origin AS.} Further, detection of BGP hijacks has been studied extensively \cite{ballani2007study,Cho:BGPhijclassif}.
Master and Garman found in a 2023 study \cite{master2023worldwide} that among 70 analyzed countries, BGP-based censorship was the least utilized among eight selected censoring methods. They believe that this might be linked to the fact that BGP announcements impact the routing of several more BGP routers outside the censor's AS.
{As highlighted by Al-Musawi \ea \cite{al-musawiNovelDataDrivenGeolocation2025}, even sophisticated BGP analysis may not reveal the complete picture of censorship evasion, necessitating the integration of additional data sources such as IP geolocation and shutdown records.}

\paragraph*{UDP/TCP/QUIC}
Maintaining a port blocking list is difficult as targets can change their assigned ports over time and can be coupled to ephemeral IP addresses (see Sect.~\ref{sect:censmeth:IP}). Further, due to resource limits, there is a state-holding problem (like with any other stateful filter devices, such as traffic normalizers \cite{NormalisierungsPaper}), i.e., censor devices cannot keep track of an unlimited number of TCP connections' states. Resource limits in terms of computing power and memory also limit how the censor is capable of handling inconsistent TCP retransmissions~\cite{NormalisierungsPaper} (i.e., when a filtering device observes retransmissions of segments with different payloads which can be challenging when fragments are to be re-assembled for further analysis). A minor issue is the cold-start problem, i.e., the evaluation of already existing connections after bootstrapping the normalizer~\cite{NormalisierungsPaper}. While UDP- and TCP-based blocking attempts are pretty common, blocking methods for QUIC are partially in their infancy. Elmenhorst \ea mention in a 2021-study that QUIC is not necessarily considered by some censors due to its novelty \cite{Elmenhorst21:HTTP3:QUIC}. However, the authors expect a growing trend in QUIC blocking.

\paragraph*{DNS} As with IP blocklists, maintaining DNS blocklists faces the problem of incompleteness and outdated entries. Moreover, manipulating DNS is more challenging when DNSSec or DNSCurve are used, as responses are signed by an authoritative server and thus can be verified for correctness by a client.
As a result, injecting a faked response is not directly feasible for a censor~\cite{Aceto2015} as the censor usually does not own the private key of that particular DNS server. However, a censor might still block packets directed to a particular DNS server \cite{Aceto2015} (DNSCurve can encrypt DNS requests, so the censor can only drop such packets blindly).
When \emph{DNS over HTTP} (DoH) and \emph{DNS over TLS} (DoT) are used, the packets between the client and the resolver are secured, but not necessarily the packets between the resolver and the queued nameserver(s), allowing DNS manipulations between these systems at least for on-path scenarios
\cite{Jin:WWW21:EncrDNSCensorship}.

\paragraph*{HTTP/HTTPS}
While censoring attacks on plain HTTP are easy to accomplish and thus attractive to the censor, several of the HTTPS-based attacks are more challenging to realize (see Sect.~\ref{Sect:Censorship:TLS}) as deploying faked certificates requires additional steps, and already established encrypted connections cannot be decrypted by the censor. 
Further, if circumvention sites (that provide proxy front-ends for circumvention \cite{IntZensurChina}) transfer blocked target websites in image form (screenshots), their blockage becomes non-trivial as keyword filtering cannot be applied anymore. 
Recent work has also shown that censorship can be circumvented through \emph{HTTP smuggling} \cite{muller2024turning}, where an HTTP request is nested inside the payload of a surrounding HTTP request (the \texttt{content-length} refers to a larger request, but the transfer encoding splits the nested from the surrounding request, and the censor is assumed to investigate only the surrounding request).

\paragraph*{Other Application Layer Protocols}
Blocking attempts like in the case of Telegram can easily bring larger collateral damage with them. Blocking and limiting is also challenging as there is a plethora of censorship circumvention methods available and since new online platforms appear quickly, requiring adjustments of censorship methodology.
TLS-secured application-layer protocols need more effort and face the TLS-related challenges (see Sect.~\ref{Sect:Censorship:TLS}). The development of own (replacement) tools with built-in censorship is costly for the censor and only attractive to end-users if the originals are either not accessible (blocked) or the replacement tools are of similar quality.

\paragraph*{TLS}
{Master and Garman mentioned that if an Encrypted Client Hello (ECH) is used, SNI-based filtering can be expected to be eliminated \cite{master2023worldwide}. However, ECH is still an IETF draft \cite{ietf-tls-esni-22} and may also introduce new censorship challenges. For example, the censor can determine that ECH is being used and, as such, they may choose to simply block all connections using ECH. 
Beyond ECH, an attempt was made to at least encrypt the SNI itself with \emph{Encrypted SNI} (ESNI). ESNI censorship was analyzed by Chai \ea in 2019, revealing indeed less blocked sites as in case of using SNI. ESNI attacks are either blind or more challenging (DNS poisoning combined with a MitM transparent proxy).

{Another limitation is \emph{domain fronting},\label{sect:domainfronting} where CDN-based proxies are used to access the target but also several legitimate websites, so that blocking such a proxy results in \emph{unacceptable collateral damage} \cite{fifield2015:DomainFronting,Fifield:PhD}. This is essentially realized by constructing a TLS handshake that contains an allowed SNI value, but the TLS-encrypted HTTP request contains a \texttt{Host} value that indicates a forbidden website to which traffic is finally routed \cite{fifield2015:DomainFronting,rfc9505}. Xie \ea consider this approach a network covert channel \cite{Xie:2024:CCinDomainFronting}.
However, it would be imaginable that censors aim to enumerate the number of devices behind CDNs (attacks like \emph{structure analysis} \cite{Quick:LoadBalancing} already allow counting QUIC devices behind load balancers in real-world CDN settings) to quantify the collateral damage and aid their decision-making process.
Finally, users might omit SNI values so that censors cannot easily determine targets, while blocking such connections could lead to overblocking \cite{rfc9505}.}

\paragraph*{Multi-stage Censorship}
Several methods have been implemented to render active probing and other sophisticated detection methods more challenging to a censor. {One of these methods is to apply domain fronting (Sect.~\ref{sect:domainfronting}).} Domain fronting has been used for several circumvention systems, including Tor, Psiphon, and Lantern \cite{Fifield:PhD,fifield2015:DomainFronting}.

Another approach to challenge active probing is to minimize indicators of circumvention traffic, e.g., by employing \emph{sophisticated covert channels} (steganography methods) that employ multiple hiding patterns sequentially or in parallel, e.g., through temporal or spatial modulation \cite{CSURpaper,TowardsPatternInsights}, as well as \emph{obfuscation extensions}, such as the Tor pluggable transport \texttt{obfs4} \cite{Xie:2024:CCinDomainFronting}. For instance, one temporal modulation is to frequently change proxy addresses, as realized by \emph{Snowflake} \cite{bocovich2024snowflake,Fifield:PhD}, which can be considered a form of \emph{host-based scattering}{\cite{TowardsPatternInsights}}. Another form is to employ multiple flows, protocols, and circumvention methods simultaneously, like done by \emph{StegoTorus} \cite{weinberg2012stegotorus}, hiding pattern hopping \cite{CSURpaper}, \emph{Turbo Tunnel} \cite{Fifield:TurboTunnel} and \emph{Raceboat} \cite{vines2024communication}. However, censors also adjusted their active probing methods to identify obfuscation modules such as \texttt{obfs2} and \texttt{obfs3}~\cite{Fifield:PhD}.


\section{{Network-level} Censorship Measurement}\label{sect:measurementmeth}
The previous section described censorship techniques. We will now investigate how the presence of these censorship techniques can be inferred, that is, \emph{detected} or \emph{measured}. 
Note that we neither discuss generic issues of Internet measurement as they have been, for instance, discussed by Paxson \cite{PaxsonIMC04:InternetMeasurement,Paxson:SIGMETRICS98}, nor do we cover data analysis aspects, which are covered by standard textbooks, e.g., \cite{Collins:NetSecDataAnalysis}. Instead, we will focus on the measurement of censorship-specific artifacts. 
%

\subsection{Measuring IPv4/IPv6-based Censorship}\label{sect:measure:IPlevel}
Measurement of IP-based censorship in a device-agnostic manner presents a significant challenge due to the inherent diversity and dynamic nature of the censorship infrastructure. Censors employ a wide array of devices with unique configurations and filtering mechanisms. In addition, the path between a client and a target server often traverses multiple ISPs, transit providers, and {Content Delivery Networks (CDNs)}, making it difficult to pinpoint the exact location and nature of IP-based interference. For this reason, IP-based censorship measurements often focus on specific devices or installations and their location within a network \cite{raman2022network}. As such, this section will discuss several key approaches to measuring IP-based censorship, including \emph{(i)} reachability tests, \emph{(ii)} QoS degradation measurements, and \emph{(iii)} techniques to identify the location of censor devices within the network.

\textbf{\emph{(i)} Measuring Reachability}
The most fundamental approach to determine the reachability of hosts is to conduct \textbf{scans of IP ranges} from within an AS that faces censorship. For instance, a volunteer residing in such an AS could send ICMPv4/v6 echo request packets either to the hosts of a whole subnet or to a list of selected hosts on the public Internet that are suspected to be blocked. The sheer fact whether a valid ICMP echo reply was received is an indicator of the reachability of an IP. However, such trivial probes are often error-prone. First, sending a high number of ICMP requests can trigger ICMP rate limiting \cite{ICMPratelimiting,guo2017detecting}, which is often used to mitigate the impact of DoS attacks by suppressing ICMP response packets. ICMP rate limiting requires an intermediate router to drop a fraction of probe packets and requires the probe device to send requests with a slow rate \cite{guo2017detecting} and with a repetition of packets, thus slowing down such scans. Second, the appearance of a high number of ICMP probe packets within a short fraction of time can lead to a probe device's IP being blocked residually. Detection of ICMP rate limitation during measurements can be done by determining (randomized) probe loss by comparing slow-rate test probes with faster-rate test probes \cite{guo2017detecting}.

Liang \ea developed \emph{Pathfinder} \cite{liang2024pathfinder}, which measures the reachability of websites but with a focus on routing paths (and is thus discussed in this section). Pathfinder is a system that utilizes different routing paths within a country to determine whether a target is blocked on all of these routing paths. {They discovered an inconsistency in several policies deployed by countries, which is commonly referred to as the \textbf{censorship inconsistency problem}~\cite{liang2024pathfinder}}. This issue is rooted in the fact that different ISPs do not necessarily implement exactly the same blocking policies. Bhaskar and Pearce report that censorship inconsistencies have been recognized for several protocols \cite{bhaskar2024understanding}. Liang \ea argue that previous research neglected the fact that even \emph{different} vantage points used to emit probe packets still take overlapping routes if the packets are destined to the same target (vantage point location was early on reported to potentially skew the interpretation of measurement results \cite{PaxsonIMC04:InternetMeasurement}). Their approach is to utilize vantage points in the form of residential SOCKS proxy IP servers so that they are located around the world. In addition, they deployed a set of control servers that represent targets and reside in uncensored environments. So-called clients feed a list of destination domains to the vantage points. Next, these vantage points try to reach the target servers and conduct HTTP requests, in which the \texttt{Host} header field contains an entry of the domain list provided by the client. This enables the censor to spot the domain name in the header and can trigger a filtering reaction of the censor device. As vantage points are distributed and control servers are distributed, too, and reachable with different IPs, Pathfinder can measure several paths between vantage points and target systems and infer blocking policies on different censor-operated gateways. {Moreover, the authors deployed measurement in a way that it covers different hosting providers so that the influence of peering-policies on the censoring behavior can be estimated~\cite{liang2024pathfinder}.} 
Elmenhorst \ea used a mix of Virtual Private Server (VPS), Virtual Private Network (VPN), and personal devices of volunteers \cite{Elmenhorst21:HTTP3:QUIC}. They did find that VPN and VPS used for testing often were less censored compared to the devices of the volunteers.

\textbf{\emph{(ii)} Measuring QoS Degradations}
QoS degradation can take different forms, e.g., packet loss or bandwidth throttling. It is feasible to measure the QoS decrease by comparing current with historical transmissions. For instance, a current flow's unacknowledged TCP packets that were retransmitted through the TCP reliability measures can be compared to historic flows. A higher number of retransmissions indicates more packet drops (loss). 
Throttling of flows can be measured similarly. Either in a historic manner (comparing throughput to the same destination for flows of the past) or for long-lasting flows so that the change of throughput of a flow can be monitored. 
However, an increase in packet loss and a reduction in throughput can also be the result of unintended or legitimate network changes in the network, such as legitimate alterations of routing paths.

Among other factors, Anderson examines latency, packet loss, and out-of-order delivery, and provides an example strategy to identify throttling mechanisms in Iran \cite{andersonDimmingInternetDetecting2013}.

\textbf{\emph{(iii)} Identifying Censor Device Location}
To identify the location of a censor, a common approach is to apply \textbf{\texttt{traceroute}-like approaches} \cite{Aceto2015}. These approaches make use of the original datagrams returned (``quoted packet'' in the ICMP error response messages) to detect the presence and potential actions of middleboxes within a network path. This quoted packet is sent due to many routers following the RFCs 792 and 1812. A discrepancy between the original sent packet and the quoted packet can indicate the presence of a censor. This works because when a probe packet's Time-To-Live (TTL) expires, a router along the path sends back an ICMP Time Exceeded message. 
The quoted packet feature includes a portion of the original probe packet within this ICMP message. 
If a censorship device modifies the packet before its TTL expires, those modifications will be reflected in the quoted packet received by the measurement tool.
For example, if a censor injects a reset (RST) flag into a TCP packet to disrupt a connection, the quoted packet in the subsequent ICMP Time Exceeded message will contain the injected RST packet.
By comparing the sent packet with the quoted packet at each hop, researchers can pinpoint the location where the modification occurred, thus identifying the potential location of the censorship device. Detal \ea implement this technique with their tool named \texttt{tracebox}~\cite{detalRevealingMiddleboxInterference2013}. However, it is crucial to note that while TCP RST injection is a censorship technique, it would not typically trigger an ICMP Destination Unreachable message (type 3, code 3) from the censor itself. ICMP Destination Unreachable is usually sent in response to issues like a closed port or a non-existent host. Instead, the RST packet sent by the censor directly terminates the connection on the client side. The quoted packet revealing the RST would be contained in the ICMP Time Exceeded message from the router downstream from the censor, where the probe packet's TTL finally expires.

Raman \ea builds on this concept with a more sophisticated detection technique \cite{raman2022network}. 
CenTrace actively probes for censorship by \textbf{sending HTTP(S) requests with varying TTL values and content}. It aims to pinpoint the location of blocking by identifying the network hop where requests for censored domains fail, while requests for uncensored domains succeed. This active probing allows CenTrace to specifically target HTTP(S) censorship, whereas \texttt{tracebox} is more general-purpose and might detect various types of middlebox interference, not just censorship. The authors do however state, that their approach would be easily extended to other protocols such as DNS or SSH.

A simple scenario of this method is shown in Fig.~\ref{fig:traceroute}, which is split into two experiments using two domains \emph{(a)} and \emph{(b)}. The part of the control domain \emph{(a)} represents the unchanged \emph{baseline} where no censoring occurs. The part of the test domain \emph{(b)} shows that a censor intercepts requests to the server with an RST response \emph{in the path}. This behavior indicates that R3 acts as a censor. For the test domain, the TTL to be tested can be incremented stepwise until a RST is received.

\begin{figure}[!tp] 
    \centering
    \input{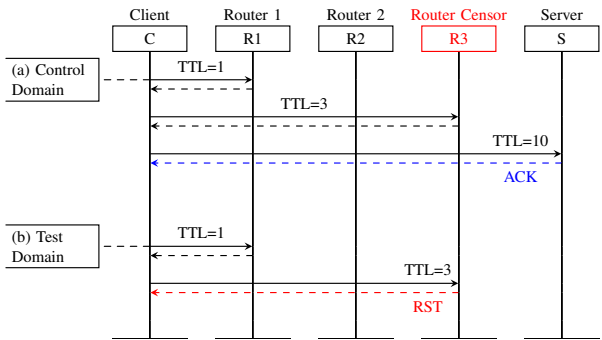}
    \caption{Example case for \texttt{traceroute}-like censor location detection like \cite{detalRevealingMiddleboxInterference2013}. Visualization inspired by Raman \ea~\cite{raman2022network}}
    \label{fig:traceroute}
\end{figure}

Recently, censors started to actively detect and block IP-based location probes. The possibilities for them to do so are manyfold, ranging from detecting abnormal traffic patterns associated with probing to actively disrupting the tools and techniques used for fingerprinting. Censors can control request rates, examine network traffic for abnormal patterns, or discard certain packets to conceal their actions. They can alter ICMP responses and quoted packets to deceive \texttt{traceroute}-like tools, complicating the task of tracing network routes back to the censorship origins. Furthermore, censors use deep packet inspection, block specific signatures, and apply protocol-specific filters to detect and eliminate known fingerprinting tools. After discussing the techniques mentioned in this paragraph, Amich \ea present \emph{DeResistor}, a tool that employs machine learning techniques for traffic and pattern generation that minimizes the risk of being detected \cite{amichDeResistorDetectionResistantProbing2023}. They leverage an open source tool called \emph{Geneva} \cite{Bock:Geneva} as an engine to find evasion strategies and in this context protects Geneva from being detected in turn.

{Wampler \ea~\cite{wampler2025protoscan} investigated how vantage points can be suitable to evaluate whether the \textbf{version of the IP protocol} plays a major role. In more detail, bidirectional measurements performed over DNS, HTTP, and TLS traffic when served via IPv4/IPv6 exhibit substantial differences in censorship. In fact, many state-level censors fully block and filter IPv4 traffic but seem not to consider IPv6. On one hand, this could be due to the limited volume of users relying upon IPv6 (e.g., Morocco, Tanzania, Kuwait, and Turkey). On the other hand, such a behavior suggests that some countries implement blockages of IPv6 traffic within a specific ISP, thus making it harder to have a precise estimate (see, e.g., the case of Russia). Apart from providing a prime attempt at understanding the IPv6 ``readiness'' of the main blocking mechanisms, this investigation highlights an important aspect concerning censorship circumvention. Specifically, in states where only IPv4 is blocked, v4/v6 transitional mechanisms, e.g., $6$-to-$4$ tunneling, may offer a way to stay undetected or unfiltered.}

Apart from trying to \emph{directly} detecting a censor, a valid strategy could be the continuous monitoring of possibly affected infrastructures. This implies the deployment of a distributed measurement infrastructure to monitor Internet censorship through active probing techniques. These probes emulate the behavior of regular users, attempting to access websites and online services while meticulously documenting their experiences.
Advanced Internet censorship measurement systems, like ICLab (cf.\ Sect.~\ref{sect:measurementplatforms}), enhance this idea by simplifying both the setup and execution processes.


\subsection{Measuring Routing-based Censorship}

Routing-based censorship can be subtle and hard to detect, as they manipulate the fabric of Internet traffic flows. 

To \textbf{detect cut-out or cut-off ASes or routes}, researchers can rely on the combination of various datasets with applied anomaly detection \cite{ballaniStudyPrefixHijacking2007}. Using these, they can try to derive the application of censorship. Combining data from various sources, such as BGP routing tables, \texttt{traceroute} measurements, DNS responses, and even publicly available information such as IP geolocation and AS relationships, allows researchers to corroborate findings and uncover subtle patterns that could be missed when analyzing individual datasets in isolation. For example, anomalies in BGP updates can be cross-referenced with \texttt{traceroute} data to pinpoint the location of suspicious route manipulation.

Dainotti \ea use BGP control plane data, data plane recordings, and a geolocation database to detect changes over time in Lybia \cite{dainottiAnalysisCountrywideInternet2011}. They mapped Libyan IP addresses (including those geolocated in Libya via MaxMind~\cite{MaxmindWebsite}) to ASes and prefixes, tracking prefix visibility in BGP tables, and correlating this with changes in darknet traffic from Libya. The combined analysis revealed a sequence of BGP withdrawals coinciding with traffic drops, confirming BGP-based blocking. However, other outages, outside of the BGP-blocking induced ones, suggested the additional use of packet filtering. Although traceroute data provided supporting evidence, its limitations hindered the fine-grained analysis of shorter outages. This multifaceted approach enabled them to expose details about the blackout's timing, methods (BGP disruption and packet filtering), and attribution to actions within Libya.

Al-Musawi \ea further present a comprehensive survey on BGP anomaly detection approaches using these types of datasets~\cite{al-musawiBGPAnomalyDetection2017}. They categorized the techniques by approach, employed BGP features and effectiveness in detecting anomalies and pinpointing their origins. An examination of 21 major techniques led to their classification into five types: time series analysis, machine learning, statistical pattern recognition, validation of historical BGP data, and reachability checks.
Their findings revealed that no single technique effectively combined real-time detection, differentiation between anomaly types, and source identification. Although some techniques excelled in one or two of these areas, none achieved all three. This highlighted the need for next-generation detection systems to incorporate a combination of approaches and leverage multiple data sources (updates to BGP, Routing Information Base (RIB) tables, Internet Routing Registry (IRR) databases, etc.) and features (AS-PATH length, prefix origin changes, volume anomalies, etc.) for more comprehensive and accurate BGP anomaly detection. Their classification framework and analysis provided a valuable contribution to understanding the strengths and weaknesses of current techniques and identifying key requirements for future research.

\textbf{Historical and time-focused data} can also be employed to detect BGP hijacking. Shapira \ea uses such datasets 
to develop a novel BGP hijacking detection methodology. Their approach utilizes an unsupervised deep learning model trained on large-scale datasets of BGP routing information. Specifically, they employ AS-level paths and prefix origin data derived from BGP update messages collected over extended periods \cite{shapiraAP2VecUnsupervisedApproach2022}. Another approach was taken by Moriano \ea with their algorithm that detects the burstiness of BGP updates. This enabled them to detect large-scale updates of malicious intent \cite{morianoUsingBurstyAnnouncements2021}.
Their method leverages the observation that BGP announcements associated with disruptive events, like prefix hijacking, tend to occur in concentrated bursts of activity, followed by periods of relative quiet. Instead of simply looking at the volume of announcements, which can be misleading due to benign events like session resets, their algorithm analyzes the timing of these announcements. This burstiness metric, combined with the analysis of the affected prefixes, allows them to detect hijacking events with higher precision than volume-based methods.

Schlamp \ea \cite{schlampHEAPReliableAssessment2016} proposed HEAP (Hijacking Event Analysis Program), a framework for \textbf{assessing the validity of suspected BGP hijacking incidents}. Introducing a formal model for Internet routing, they classify various attack types and highlight the often-overlooked threat of subprefix hijacking {(announcing more specific prefixes)}. HEAP utilizes Internet Routing Registries, topological analysis, and SSL/TLS certificate validation to distinguish legitimate routing events from malicious hijack attempts, thus reducing false positives in existing detection systems. In addition, the detection of BGP hijacks has been studied extensively \cite{ballani2007study,Cho:BGPhijclassif,safaeipourComprehensiveSurveyRecent2023}.

While not having a direct impact on the measurement of an actual route or BGP censorship, it is worth noting that routing can be inconsistent. Bhaskar and Pearce highlight the crucial role of Equal-Cost Multi-Path routing in influencing censorship measurements. Their work demonstrates that varying source IP and port, which influence a packet's flow identifier (Flow-ID), can significantly alter the paths taken through a network and, consequently, the observed censorship results. To address this, they developed \emph{Monocle}, a route-stable measurement and traceroute platform \cite{bhaskar2024understanding}. By controlling the Flow-ID parameters and therefore the route taken by packets, Monocle enables consistent measurement of DNS, HTTP, and HTTPS censorship, including detection of DNS manipulation, RST injection, packet drops, and block pages. This approach allows for more accurate assessment of censorship by isolating routing-induced variations from actual blocking behavior.

\subsection{Measuring UDP-/TCP-/QUIC-based Censorship}\label{sect:measure:TCPUDPQUIC}

The transport layer of the OSI model features censorship methods based on UDP, TCP and QUIC. These protocols can be used for censorship based on triggers on the same layer or on other/higher layers (due to interference). Therefore, these protocols can also be used to \emph{measure} censorship on their own or other layers of the TCP/IP model.

Several approaches focus on \textbf{TCP packets with RST or FIN flags being set}. 
A straightforward approach is trying to establish a TCP connection to a target using a tool like \texttt{netcat} and observing the resulting traffic (whether the TCP handshake is completed, a TCP RST is received \cite{ICLab:SP20,bhaskar2024understanding} or a TCP FIN is received at some point in time \cite{ICLab:SP20}). A typical indicator for censorship also comes with the observation of duplicated TCP response packets, i.e., a censor making sure that the injected TCP RST packet arrives before the original TCP SYN-ACK sent by the target \cite{ICLab:SP20}. However, Niaki \ea point out that it is feasible to determine if a target is down instead of censored if control nodes are also receiving TCP RST or ICMP destination unreachable responses at the same timeframe \cite{ICLab:SP20}. Further, Bhaskar and Pearce motivate an approach where probing and comparing control and vantage point traffic over multiple routes is essential for minimizing the chance for false conclusions on censorship \cite{bhaskar2024understanding}. For instance, RST packets can occur or be omitted due to normal network errors/packet loss. Bhaskar and Pearce suggest repeating measurements multiple times while enforcing larger (30\;min) time gaps between successive probes.

Clayton \ea used this approach in \cite{clayton2006gfc} to test the GFW. They sent forged TCP-nested HTTP \texttt{GET} requests (through \texttt{netcat}) to web servers behind the GFW with and without ``bad words'' and observed the resulting TCP-RST packets.

Ensafi \ea propose a slightly more involved method in~\cite{Ensafi2014}. They show that it is possible to detect the injection of TCP-RST packets by observing the increase in IPv4 Identifier (IP-ID) values after sending TCP-SYN-ACK packets with spoofed source IPs. This method can detect if no blocking, client to server or server to client blocking is applied by analyzing the amount of which the IP-ID increased. Fig.~\ref{fig:rst_scan} depicts a visual representation of the three possibilities.
In our example, an increase of 2 indicates no blocking, an increase of 1 indicates server to client blocking and an increase of 4 indicates client to server blocking.
It is to note that server-to-client blocking can overshadow client to server blocking. Thus, in case of bidirectional blocking, this measurement method will only indicate server to client blocking.

\begin{figure*}[!t]
\centering
    \input{figures/rst_scan}
    \input{figures/rst_scan2}
    \input{figures/rst_scan3}
    \caption{TCP censorship measurement {based on} Ensafi \ea\cite{Ensafi2014}. {Red} indicates blocked packets.}
    \label{fig:rst_scan}
\end{figure*}

In 2023 Nourin \ea \cite{nourin2023measuring} used multiple approaches to measure the censorship methods of Turkmenistan with their \emph{TMC}\label{sect:TMC} system. A simple approach similar to Clayton \ea works by sending out HTTP \texttt{GET} requests containing blocked words and examining the resulting TCP-RST packets. 
They also used an approach that omits the actual TCP three-way handshake and directly sends out two consecutive PSH+ACK packets. The first PSH+ACK packet containing a forbidden \texttt{Host} header, the second packet follows with 5 to 29 seconds of delay. The second packet will then be answered by a TCP RST packet. This behavior stems from the residual blocking functionality of the Turkmen censor.



Another indicator of TCP connection tampering are \textbf{TCP sequence number collisions} after a handshake has been completed successfully while RST and FIN flags are absent~\cite{ICLab:SP20}. This is because if a censor's blocking page response for HTTP is sent to the client, the target's original response would not match the blocking page, resulting in larger/smaller TCP response packets and thus sequence numbers.

Elmenhorst \ea \cite{Elmenhorst21:HTTP3:QUIC} initiated a connection with servers and recorded when a \textbf{timeout, reset or routing error} would occur. They categorized their artifacts into \textbf{TCP handshake timeouts, TLS handshake timeouts, QUIC handshake timeouts, connection resets and routing errors}. 
Depending on when a connection processes an error or when a timeout occurred and whether it occurred for TLS and/or QUIC, they could gain insights on the type of censorship method. If both TLS and QUIC were blocked, it would point to IP-based censorship. If only TLS connections were blocked, the timing of the error can point towards TLS-SNI based or TCP-port-based blocking.

Even if the censorship does not occur based on TCP or UDP, these protocols can still be used to measure the censor. 
Polverini and Pottenger noted in \cite{Polverini2011} that the DNS responses of requests to DNS servers outside of China had \textbf{corrupted UDP checksums}. This corruption was a result of tampering with the DNS packets and can therefore be used to measure DNS based censorship.

Finally, multiple forms of \textbf{port scanning} using traditional techniques can be used for detecting blocked ports (e.g., by scanning a list of destinations for ports of interest)~\cite{Aceto2015}.

\subsection{Measuring DNS-based Censorship}\label{sub:DNS_measure}

A straightforward approach for measuring DNS-based censorship is to deploy measurement units in particular regions of the world and let these try to \textbf{resolve certain DNS hostnames on a regular basis}. The responses and the changes in responses are analyzed over time, e.g., detecting if \texttt{NXDOMAIN} is returned or whether the resolved IP address refers to a blocking page or known censorship page. Some authors, e.g., Nabi~\cite{Nabi2013}, go one step further and \textbf{compare regional DNS responses with results from international DNS resolvers} from less censored (optimally: uncensored) networks. Nabi also found that the results from international resolvers were partially tampered \cite{Nabi2013}. Similarly, Kührer \ea~\cite{kuhrer2015going} compared the DNS responses gained from open DNS resolvers with those of a ground truth gained from trusted resolvers (see also \cite{Pearce2017}). 
In~\cite{Pearce2017}, Pearce \ea crafted DNS queries that they obtained from a publicly available list from \emph{Citizen Lab}, cf.~\cite{CitizenLabTestList}, enriched with domain names from the Alexa Top 10,000 list. These queries were sent to geographically distributed DNS resolvers. The authors consider a DNS response as manipulated when it fulfills two criteria: (i) it must be inconsistent with control data, and (ii) the returned information must be clearly flawed (e.g., non-matching TLS certificates). This approach minimizes false positives. Another study by Berger and Shavitt \cite{berger2024measuringdnscensorshipgenerative} measures specific censorship of generative AI platforms by performing DNS queries, and found that especially Russia and China block some of these platforms.

A different approach is to \textbf{send DNS queries to any IPs within the censored network}. These IP addresses do not need to be those of DNS servers, they do not even need to be active at the time of probing. The goal here is to see if a false reply is received, one to a non-routable address, such as \texttt{127.0.0.1} or any other reserved IP address. Nourin \ea used such an approach in \cite{nourin2023measuring} to measure the DNS part of Turkmenistan's censorship system, which injects a blocking DNS response to all forbidden queries, no matter their destination IP.

Hoang \ea used a similar idea in~\cite{USESEC21:GFWatch}. They sent DNS queries from the outside into China's censored network using two target nodes controlled by the researchers. These hosts had \emph{no} DNS server software installed, so any DNS replies must come from the GFW. These domains are then tested again from inside China by using other researcher controlled machines.

Similarly, Bhaskar and Pearce sent requests for DNS \texttt{A} resource records to IP destinations that also do \emph{not} run DNS resolvers \cite{Bhaskar2022}. They inferred the presence of a censorship system if there was no response for a non-filtered target but one for a ``sensitive'' target as this must have been injected by the censor device, although there was no actual DNS resolver that would have sent a response.

There are also measurements of DNS tampering that aim to minimize false-positives by enhancing the observation period. Niaki \ea \cite{ICLab:SP20} monitored \texttt{NXDOMAIN} as well as inhouse IP address responses and compared responses received by vantage points with those received by control nodes. To avoid false-positives, only those responses were labeled as ``tampering'' if they have been observed for a duration of seven days while only the control nodes received globally routable addresses.

Another method for inferring DNS censorship was also introduced by Niaki \ea \cite{ICLab:SP20}: they detect if a vantage point receives \textbf{two responses for the same DNS request that belong to different ASes}. The assumption of Niaki \ea is that an on-path censor injects the DNS response on the fly while residing in another AS than the DNS server that was actually indicated by the destination IP. A special case reported by the same authors is when the \textbf{vantage point and the control nodes both receive addresses from different AS} while both of these addresses are globally routable. The reason for such an observation can be caused by CDN providers that direct traffic to data centers that are closer to the client~\cite{ICLab:SP20}. However, Niaki \ea observed that censors utilize a limited number of IP addresses for redirecting clients to block pages.
To distinguish between false positives (such as CDNs) and cencorship, they record a set of websites that resolve to a single IP address when requested from a vantage point. If more than $\theta$ of those websites resolve to IPs from different ASes when accesed from a control node, Niaki \ea consider these websites as being tampered. $\theta$ was optimized empirically to achieve a low false-positive rate, with $\theta=11$ providing satisfying results \cite{ICLab:SP20}. 

In~\cite{Bhaskar2022}, Bhaskar and Pearce focus on a different aspect of DNS censorship measurement by investigating the \textbf{impact of source port and source IP address on the censorship behavior}. 
They discovered that varying source parameters influence the path taken by probe queries. As a next step, they evaluated the censorship behavior based on these changing routes. Their results indicate that the changes in source parameters also influence the censorship behavior. Bhaskar and Pearce found that most changes are ``all or nothing'', meaning some combinations of parameters will experience full censorship while others will experience none.

Fan \ea introduced Wallbleed~\cite{WallBleed2025}, a buffer over-read vulnerability in some GFWs middle boxes performing DNS injections. This bug allowed the researchers to extract up to 125 bytes from the middle boxes' memory per request, allowing a more in-depth understanding of the GFWs inner workings. The exploit was triggered by sending malformed DNS requests with mismatched length fields for domain labels. During their two-year study, they found data that resembled x86\_64 pointers and x86\_64 instructions. Although the authors do not believe that these intrusions were code of the GFW. They also found out that the GFW would leak previously analyzed network traffic, proving an additional risk to user privacy. 
The bug that was exploited in Wallbleed has been fixed in 2024 and can no longer be used, but it shows a general idea how censorship can also be measured.

CDNs, load balancing and other methods can influence DNS resolution.
For this reason, Tsai \ea came up with an approach for \textbf{measuring DNS manipulation based on TLS certificates} \cite{tsai:PETS23:CERTainty}. 
The general idea here is to check the TLS certificates presented by the target servers because a valid TLS certificate is considered a strong indicator for a legitimate connection to the actual server. In their measurements, the authors did not encounter any valid certificates for blocked pages.
Untrusted certificates with matching hostnames were a sign of blocked pages, which the authors confirmed with known block page fingerprints.
Trusted certificates with mismatched hostnames were considered a sign of DNS manipulation.
Untrusted certificates with mismatched hostnames were considered a strong sign of DNS manipulation.
Lastly, the authors also considered invalid ``control'' certificates. I.e., certificates that are also invalid when resolving from an uncensored device. These certificates mostly stem from misconfigured web servers and therefore cannot be considered a sign of DNS censorship.

A recent method to enhance the detection of DNS tampering was introduced by Calle \ea \cite{calle2024toward}. The authors utilize OONI~\cite{filasto2012ooni} probe data (cf.\ Sect.~\ref{sec:OONI}) in conjunction with supervised and unsupervised machine learning classification. Based on features of the OONI dataset, the models successfully learned expert-defined heuristics for known tampering patterns and even identified potential new censorship signatures. For unsupervised learning, they adopted Isolation Random Forest and the One-Class Support Vector Machine which complement each other. The authors argue that these models are beneficial for unlabeled data, as DNS manipulation ground truth data is hard to obtain. They further employ XGBoost as a supervised learning model that uses decision trees for accurate predictions.

\subsection{Measuring HTTP(S) and Web-content Filtering}\label{sect:measure:HTTP}

The most common approach to measuring HTTP-based censorship involves using tools that \textbf{mimic a standard web browsing behavior}, especially sending HTTP \texttt{GET} requests to target websites, and analyzing the responses.

Censorship measurement can also implement the so-called \textbf{HTTP fuzzing}. This is a technique that alternates/randomizes different types of inputs to a system. For HTTP in particular, this can yield insight into the behavior of the censor. Listing~\ref{lst:fuzzing} sketches a toy example of the approach.

\begin{lstlisting}[
  caption=Fuzzing HTTP request example (inspired by \cite{jermynAutosondaDiscoveringRules2017}.),
  label=lst:fuzzing,
]
GET / HTTP/1.1\r\nHost: www.target.com\r\n\r\n
GEt / HTTP/1.1\r\nHost: www.target.com\r\n\r\n
get/ HTTP/1.1\r\nHost: www.target.com\r\n\r\n
\end{lstlisting}

The fuzzing requests sent to the server do not necessarily have to conform to the HTTP protocol standard, but rather are used to inspect the response behavior of a censor. HTTP fuzzers commonly utilize accessible tools such as Python scripting, \texttt{wget}, or \texttt{curl} to create the required HTTP requests. Fuzzing involves examining the entire content of the HTTP response to identify potential indicators of censorship, such as the presence of block pages, error messages, or content substitution. An example of censorship-oriented HTTP fuzzing is provided by Jermy \ea with their software named \emph{Autosonda}. Autosonda can reverse-engineer HTTP filtering rules and fingerprint censorship middleboxes by carefully crafted network traffic probing \cite{jermynAutosondaDiscoveringRules2017}. They discovered limitations of filter rules in multiple ways. Examples include introducing (unnecessarily) long target hostnames, testing multiple variants of \texttt{\textbackslash{}r\textbackslash{}n} character combinations and placements, and introducing keywords in the \texttt{Host} parameter.

There are additional approaches available for analyzing HTTP-served content for censorship. One method is to determine the \textbf{structure of HTML tags} inside websites, as censors tend to use the same blocking page structure on different systems and with different detailed content \cite{ICLab:SP20}. Niaki \ea created vectors that represent HTML tag structures, such as \texttt{<body><ul>...}, and could identify the typical vectors for several block pages \cite{ICLab:SP20}. The same authors identified \textbf{textual similarity clusters} in websites by processing the HTML structure with locality sensitive hashing (LSH) which was used to center candidate pages around clusters of block pages. This approach worked as censors tend to apply the same (or similar) phrases in their blocking pages but required manual inspection~\cite{ICLab:SP20}. A third method by Niaki \ea is to analyze the \textbf{URL-to-country ratio} to identify unknown block pages. To calculate this ratio, the authors took clusters not centered around a known block page and counted the URLs that point to a page in that cluster and divided that number by the number of countries with a page in that cluster. The list of URLs was then sorted and inspected manually again. Newly discovered block pages had ratios between 1.0 and 286 \cite{ICLab:SP20}.

Finally, a 2024-study by Lange \ea measured whether HTTP/2 is censored if it is unencrypted. Although unencrypted HTTP/2 is not used by any popular browser, more than 6\% of the tested websites supported unencrypted HTTP/2 communication \cite{censors-ignore-unencrypted-http-2-traffic}. Indeed, unencrypted HTTP/2 communication can be observed and tampered with easily by a censor, but the authors found that it is not affected by blocking in both China and Iran, rendering it an option to, e.g., download circumvention software.

Beyond content analysis, researchers employ a controlled measurement technique sometimes referred to as \textbf{censorship tomography} to pinpoint censorship triggers \cite{Aceto2015}. Although this term is not formally defined in the literature and can be confused with network tomography \cite{castroNetworkTomographyRecent2004}, we use it here to describe the following method: This involves setting up a client within the censored network and a helper server outside, both under the control of the researcher. The client sends probes, such as HTTP requests, to the helper server, embedding potential censorship triggers, such as specific keywords, URLs, or modified headers. The helper server responds in a controlled manner, sending normal responses when assumed triggers are absent and simulated censorship responses (block pages, errors, or altered content) when triggers are present. By analyzing the behavior of the network between the client, the helper server, and optionally a target website, researchers can pinpoint the triggers. This analysis focuses on variations in routing paths, packet loss or modification, and timing differences between probes with and without triggers. For example, consistent packet loss for probes containing a specific keyword would suggest that the keyword is a censorship trigger. This controlled approach using a helper server isolates triggers more effectively than directly probing potentially censored websites.

Measurement of \textbf{HTTPS-based censorship} is significantly more challenging than that of HTTP due to the inherent encryption provided by TLS. As Aceto and Pescaé point out in their survey, the very nature of HTTPS obscures the content of communication, making traditional keyword-based detection or content analysis ineffective \cite{Aceto2015}. 
However, examining the TLS handshake process itself can reveal potential triggers for censorship \cite{nourin2023measuring}. Censors might block specific cipher suites, target the Server Name Indication (SNI) field, or inject invalid certificates. Please refer to Sect.~\ref{Sect:Measure:TLS} for TLS-based censorship measurement.

It is also feasible to conduct \textbf{protocol-overlapping comparisons} (e.g., HTTPS over TCP vs.\ HTTPS over QUIC). The introduction and adoption of HTTP3 or QUIC promises many benefits regarding Internet censorship, in particular due to its integrated encryption mechanisms. However, Elmenhorst \ea found that, despite its novelty, this protocol is already filtered and censored in some networks \cite{Elmenhorst21:HTTP3:QUIC}. They detected these censors by conducting HTTPS-based requests and subsequently redid them using QUIC. Distinguishing the response behavior made it possible to conclude whether or not a QUIC censor is involved and if one of the protocols is censored more strictly than the other.

Finally, \textbf{cache proxies} might be deployed by Internet Service Providers (ISPs) to reduce network latency for clients. Such cache proxies can infer with censorship measurements, as vantage systems might receive a reply from such a cache proxy instead of the original target. Jin \ea describe a method to detect such cache proxies \cite{Jin:2021:CensMeasurement}: The authors set up an authoritative DNS server for a reference domain as well as an HTTP service for that domain. The idea is that this web service provides a different website than another web service controlled by the authors (called \emph{disguiser} server). Vantage points are then used to retrieve the website using the authoritative DNS server for resolving the IP. If a cache proxy is present, it will cache the DNS and/or web content of that connection. Afterward, the vantage point conducts a cache probe by trying to retrieve the webpage from the disguiser server (that would reply with a different response than the actual webserver of the domain). If the response is the one of the original webserver (instead of the disguiser-provided response), a cache proxy is considered to be involved. Note that this method might fail to detect cache proxies if a cache proxy is configured to only cache specific (popular) websites \cite{Jin:2021:CensMeasurement}.

\paragraph*{Notes}
To omit redundant coverage, we also refer the reader to Sect.~\ref{sect:measure:IPlevel} where we explain an approach developed by Liang \ea called \emph{Pathfinder} \cite{liang2024pathfinder}, which measures the reachability of HTTP sites based on routing paths. Further, we already discussed how HTTP website censoring can be identified by sending HTTP \texttt{GET} requests containing \texttt{Host} header fields or potentially blocked keywords \cite{nourin2023measuring} while monitoring the resulting RST and FIN flags in TCP response packets as well as TCP timeouts in Sect.~\ref{sect:measure:TCPUDPQUIC}. Similarly, we discussed the detection of residual HTTP-based censorship using TCP flags in that section. For this reason, we will not cover these strategies redundantly.

\subsection{Measuring Censorship of Other Application-layer Protocols and Platforms}

We can broadly distribute the remaining application layer protocols into four distinct categories: \emph{(a)} communications, \emph{(b)}~multimedia, \emph{(c)} file sharing, \emph{(d)} gaming and \emph{(e)} miscellaneous. In the following, we will briefly elaborate on detections of censorship on these protocols via short examples.

In the domain of \emph{(a)} communications, we include instant messaging (e.g., \emph{WhatsApp}, \emph{Telegram}, \emph{Signal}, VoIP as well as textual social media platforms such as \emph{X} formerly known as Twitter). A typical attempt to detect censorship of these services is based on sending \textbf{IP probes} to the affected servers. For example, Ermoshina \ea discusses {the blocking} of Telegram, a widely used chat application, in Russia \cite{ermoshinaTelegramBanHow2021}. As the attempt to block this application was mostly based on IP addresses that used the chat protocol (and resulted in massive collateral damage), the detection of censorship was trivial \cite{ermoshinaTelegramBanHow2021}. {A recent work by Lipphardt \ea~\cite{lipphardt2025can} investigates censorship techniques deployed in Saudi Arabia and United Arab Emirates against VoIP communications. In more detail, the work analyzes conferencing tools like Zoom and Google Meet, as well as A/V services offered by IM platforms like Telegram and WhatsApp. The investigation found that detection can be done by evaluating whether Session Traversal Utilities for Network Address Translators (STUN) traffic has been selectively blocked. In fact, inspecting \textbf{dropped STUN} requests revealed the presence of middleboxes capable of filtering traversal traffic needed for point-to-point conversations, without impacting other companion features like text and media sharing. }

Another approach is to \textbf{monitor the throughput} of a service over time to detect a decline that could be caused by censor-induced throttling.
Such a case is presented by Xue \ea, which set up a measurement platform that continuously compared the available bandwidth of Twitter domains (e.g, \texttt{t.co} or \texttt{twimg.com}) with non-Twitter domains \cite{xueThrottlingTwitterEmerging2021}. The authors operated a measurement node in the US and eight vantage points in Russia. To measure throttling, they applied a ``record and reply'' method that was proposed by Kakhki \ea \cite{Kakhki15}: first, unthrottled traffic is recorded using control nodes; second, the traffic is replied by vantage points. Each vantage point first sends the reply traffic with sensitive TLS SNI values to the measurement node of the researchers. This is done to cause throttling. Afterwards, the vantage points send the same traffic (but with bit-flipped payload) so that throttling is \emph{not} caused. When the vantage points send traffic to the measurement server, the server injects TLS ClientHello messages with sensitive \textbf{SNI values} and modulates the TTL values of the injected packets. This is a \texttt{traceroute}-like approach, where packets have been injected using NFQueue. By decrementing TTL values for succeeding injections, authors were able to narrow down throttling agents' locations in Russia~\cite{xueThrottlingTwitterEmerging2021}. They discovered that throttling limited the throughput to around 130-150 kbps and that seven of their eight vantage points faced throttling. They also discovered some collateral damage in the filter policies, e.g., the censor would also throttle domains such as \texttt{throttle\textbf{twitter.com}} as it would match \texttt{*twitter.com}.

Video streaming and multimedia \emph{(b)} platforms such as \emph{YouTube}, \emph{NetFlix}, \emph{Spotify} and \emph{Twitch} as well as VC tools such as \emph{Zoom} or \emph{Facetime} are also subject to excessive censorship in certain regions. {Note that censorship of such platforms often comes in the form of authority-driven content takedowns rather than a pure technical filtering.} Aceto \ea highlight a procedure to detect video content that disappears over time: an MIT project called \emph{YouTomb} \textbf{documented previously existing videos} and determine their current availability \cite{Aceto2015}. {In other words, while a government order to remove content is not directly observable, disappearing online content can be identified by low-frequent polling for the content's existence.} However, such attempts can result in false-positives since videos can also disappear when users behind YouTube channels decide to delete their videos, e.g., because they become outdated. Moreover, the measurement strategies mentioned above \emph{(a)} can be used (sending probe packets to servers and monitoring throughput).

Filesharing \emph{(c)} protocols are often used to distribute and trade illegal and legal files of all kinds. Many countries therefore have a high incentive to limit the usage of these kinds of application. However, since these services often utilize a peer-to-peer architecture, censorship measurement of the involved protocols is particularly challenging and usually involves sending probe packets. One better way of measuring its censorship is to \textbf{participate as a peer}. Dischinger \ea~\cite{dischingerDetectingBittorrentBlocking2008} describe one such approach: Their method involves a purpose-built tool named \emph{BTTest} that synthetically generates BitTorrent flows and measures whether those flows are aborted for no apparent reason. One might also measure throughput.

Regarding \emph{(d)} (online gaming platforms), {almost no dedicated network-level approaches have been developed. A measurement strategy could be to \textbf{send potentially censored keywords to gaming platform's chat services} while \textbf{monitoring the appearance of connection resets and throttling}. In contrast to such approaches,} Feng \ea conducted a \textbf{user study} based on surveys in China asking participants about online gaming censorship \cite{fengStudyChinasCensorship}. They discovered that most participants are capable of identifying occurring censorship, but also that most are able to circumvent those in various ways, e.g., by using VPNs and by faking their ID number or the ``age'' entry used in their accounts \cite{fengStudyChinasCensorship}.

Some miscellaneous protocols \emph{(e)} have also been used to measure censorship. Among these are \textbf{UDP-based Echo and Discard services}, as they are used by the Censored Planet project (cf.\ Sect.\ \ref{sect:measurementplatforms}) checks for keyword-based blocking using the Echo service. The Discard service is used similarly, but, as payload is only transferred into one direction, it can be used to measure whether the direction of a flow influences the blocking.

\subsection{Measuring TLS-based Censorship}\label{Sect:Measure:TLS}

One example where \textbf{TLS is used to measure censorship of another protocol} was already mentioned above in Sect.~\ref{sub:DNS_measure} (Tsai \ea \cite{tsai:PETS23:CERTainty} evaluated the validity of TLS certificates as part of their DNS censorship measurement). However, in this section, we will focus on methods that use TLS to measure censorship.

While the usage of HTTPS limits censorship, triggers based on the \textbf{SNI of the TLS ClientHello message} are still analyzable by censors (see Sect.~\ref{Sect:Censorship:TLS}) and are thus considered for measurement.
Chai \ea \cite{Chai:ESNI} measured SNI-based filtering capabilities of the GFW. To this end, they exploited the bidirectional nature of the GFW by initiating TLS handshakes from outside of China to a server inside of China under their control to trigger censorship responses. They discovered that certain SNI values would result in TCP \texttt{RST} packets from the GFW, indicating censorship.
Additionally, they were able to verify that the GFW is stateful, as the SNI-based filtering would only trigger when a 3-way-handshake was performed before the offending TLS ClientHello message was sent. The method could be applied to measure the statefulness of other censors as well.

Chai \ea also investigated \textbf{Encrypted SNI (ESNI)} censorship. They used Firefox 64.0 and controlled it with the Selenium Python library to check if pages would support ESNI when accessed from the US and then when accessed from a potentially censored vantage point. In their tests, they did not encounter any ESNI based censorship.

While researching new circumvention methods for SNI-based censorship, Niere \ea discovered previously unseen behavior of the GFW, which they attribute to three different censorship middleboxes \cite{niereTransportLayerObscurity2025}. Two of these middleboxes have been described in literature before, the last one was not described yet. In their research the authors applied different manipulations to TLS packets like fragmentation, header field length manipulation, domain fronting or SNI removal with the goal of stopping the GFW from analyzing the SNI and therefore circumventing the censorship. They found that domains would be censored differently, depending on the circumvention method they employed. The requests were censored with either three TCP \texttt{RST} packets with residual blocking, a single \texttt{RST} without residual blocking or a new approach with a single \texttt{RST} and residual blocking. By combining various TLS manipulations, the authors were able to individually target each of the middleboxes by exploiting differences in their TLS parsing behavior. This allowed a deeper understanding of the inner workings of the GFW by highlighting the existence of a third middlebox and also by targeting specific parts of the GFW for measurements.

Another method to measure censorship is to \textbf{check whether STRIPTLS attacks or downgrade attacks are conducted}, as those are often connected to censorship approaches. While not directly focusing on censorship detection, Nikiforakis \ea proposed \emph{HProxy} \cite{Niki2010HProxy}, a client-side tool to detect STRIPTLS attacks. The tool works by creating profiles of the TLS behavior of previously visited sites and comparing them to the current request and responses. This works when such an attack is newly engaged, as the tool must learn the correct (attack-free) behavior first.
In general, STRIPTLS as well as downgrade attacks, can be measured by comparing handshakes with targets conducted by control nodes against those conducted by vantage points.

In \cite{Holz2012TLS}, Holz \ea proposed \emph{Crossbear}, a tool for \hbox{SSL/TLS} MitM detection and localization. The general idea here is to have a multitude of so called ``hunters'' in the form of Firefox add-ons or standalone tools that \textbf{connect to webservers from multiple vantage points and collect TLS certificates}. These certificates are then compared to a stored version of the certificate. If a discrepancy is detected, the position of the attacker can be approximated by intersecting recorded routes of different hunters with and without detected tampering. Filasto and Appelbaum mention using Crossbear in \cite{filasto2012ooni} to detect content blocking.

As mentioned in Sect.~\ref{sect:measure:TCPUDPQUIC} and Sect.~\ref{sect:measure:HTTP}, Elmenhorst \ea \cite{Elmenhorst21:HTTP3:QUIC} compared HTTP/TLS based censorship with QUIC-based censorship and were able to infer if a communication was blocked on the IP, transport or TLS-level. For the TLS part, they focused on the most common ways a connection could be interrupted: a timeout during the TLS handshake or a connection reset during the handshake.

\subsection{Measuring Censor-side Active Probing Infrastructure}
\label{sect:measure:actprob}
Fifield reported that measurements of active probing infrastructure have been conducted by analyzing source IP addresses of active probing systems and analyzing the received probing packets and their overlaps in terms of payload \cite{Fifield:PhD}. It turned out that, despite a large number of probing IPs were discovered, these were \emph{managed by only a few underlying processes,} which was concluded due to shared patterns in metadata (TCP ISNs and timestamps). Connecting to the identified source IP addresses led to no reaction (except a few systems which were responsive but appeared benign) \cite{Fifield:PhD}.

\subsection{Limitations of Measurement Methodology}
As presented in Section~\ref{sect:measure:IPlevel} to Section~\ref{sect:measure:actprob}, each layer of the TCP/IP protocol architecture can be targeted by a censor. However, not all censorship approaches exhibit the same effectiveness, especially due to the different complexity of the involved protocols and use cases. This also reflects in the correctness and quality of the related measurements methods. To provide an overall view, Tab.~\ref{tab:SummaryMeasurementLimits} summarizes the main cons of observed for each major protocols. As it can be seen, probing device location and genuine network malfunctions are an issue common for all the considered protocol/layers. For the sake of brevity, in the following, we solely address the most relevant limitations.

\begin{table}[!th]
    \centering
    \footnotesize
    \begin{tabular}{|r||c|c|c|c|}
\hline
{\shortstack[r]{\textbf{Measurement}\\ \textbf{Level}}} & \TRot{\shortstack[l]{\textbf{Vantage Point}\\\textbf{Location-}\\\textbf{dependence}}} & \TRot{\shortstack[l]{\textbf{Network} \\\textbf{Malfunctions/}\\\textbf{Routing}}} & \TRot{\shortstack[l]{\textbf{CDNs,}\\ \textbf{load balancing,}\\ \textbf{proxies, NAT} }} & \TRot{\shortstack[l]{\textbf{Dataset}\\\textbf{quality}}} \\ \hline
IP & \CIRCLE & \CIRCLE & \CIRCLE & \CIRCLE \\ \hline
Routing & \CIRCLE & \CIRCLE & \CIRCLE & \CIRCLE \\ \hline
TCP/UDP/QUIC & \CIRCLE & \CIRCLE &  \Circle &  \Circle \\ \hline
DNS & \CIRCLE & \CIRCLE & \CIRCLE & \LEFTcircle \\ \hline
HTTP(S) & \CIRCLE & \CIRCLE & \CIRCLE & \LEFTcircle \\ \hline
Other Appl. & \CIRCLE & \CIRCLE & \LEFTcircle & \LEFTcircle \\ \hline
TLS & \CIRCLE & \CIRCLE &  \Circle &  \Circle \\ \hline
Multi-stage Cens. & \CIRCLE & \CIRCLE & \CIRCLE & \LEFTcircle \\ \hline
    \end{tabular}
    \caption{{Summary of measurement methods' limitations according to literature (\CIRCLE: relevant, \LEFTcircle: partially relevant, \Circle: not relevant)}.}
    \label{tab:SummaryMeasurementLimits}
\end{table}

\paragraph*{IP/IPv6}
Although IP-based censorship measurement provides valuable information, its effectiveness is inherently limited by the dynamic and often opaque nature of today’s Internet architecture. 
The path traversed by sent packets cannot be ensured by the sender, leaving room for changes in routing paths for packets even of the same flow. This challenges the detection of blocking attempts on particular routes \cite{liang2024pathfinder}. 
The increasing reliance on CDNs makes it difficult to pinpoint the exact location of blocking, as a single IP address might serve content for multiple domains across geographically diverse servers. Similarly, the widespread use of Network Address Translation (NAT) obscures the true origin of connections, hindering efforts to attribute censorship to specific actors or networks.
NAT is commonly used to allow multiple devices within a private network to share a few public IP addresses. This is particularly relevant in scenarios where the client performing the measurement is behind a NAT device, as it makes it appear as if the connection originates from one of the NAT gateway's public IPs, but the client cannot necessarily control the use of a particular IP and its consistent use over multiple probing sessions. 
Load balancing techniques further complicate the measurement by distributing traffic across multiple servers, making it challenging to discern whether inconsistencies are due to censorship or legitimate network management. Moreover, the accuracy of IP geolocation databases, often used to map IP addresses to geographic locations, can be unreliable, especially for shared or dynamically assigned IP addresses, leading to {the potential} misattribution of blocking events {\cite{al-musawiNovelDataDrivenGeolocation2025}}. These limitations underscore the need for more sophisticated measurement strategies that combine IP-level analysis with deeper packet inspection, side-channel observations, and data-driven approaches to obtain a more complete understanding of censorship practices. As highlighted by Al-Musawi \ea \cite{al-musawiNovelDataDrivenGeolocation2025} even when applied to BGP manipulation, the need for constant vigilance in cross-referencing data is important.


\paragraph*{Routing}
The distinction of malicious activity from an error can be difficult. As Cho \ea report that simple human error can and has been the cause of routing anomalies \cite{Cho:BGPhijclassif}. They give an example of an operator mistyping an 8 instead of a 9 in his own prefix. Moreover, pinpointing the actual censor can be hard, as he might be several hops away from the actually affected network. Further, as many detections make use of datasets, the detection can only be as good as the quality of these datasets. Acquiring up-to-date, sensible and extensive datasets can be challenging. 
As in case of detecting IP-based blocking, routing paths for probe traffic usually cannot be ensured \cite{liang2024pathfinder}, i.e., a test can barely fix a set of particular routers to be traversed in a given order.\footnote{Even if this could be changed by using IP source routing, a censor could still ignore the source route setting. Moreover, such packets might be considered suspicious and would potentially be dropped by intermediate routers.} 
While Monocle \cite{bhaskar2024understanding} offers a significant advancement in route-stable measurement, challenges remain in fully understanding and quantifying routing-based censorship. The dynamic nature of routing, where paths can shift due to network conditions or traffic engineering, makes it difficult to capture a complete and static view of censorship infrastructure.

\paragraph*{TCP/UDP/QUIC}
Depending on what kind of measurement techniques are used, different limitations apply.
Timeout failures can be an indicator of censorship, but it can also be the result of genuine network malfunctions, and thus, measurements are repeated regularly to observe timeouts over a longer time \cite{Elmenhorst21:HTTP3:QUIC}. 
Moreover, duplicated TCP response packets can also be caused by legitimate reasons, such as due to HTTP load balancers \cite{ICLab:SP20}, and even if vantage points and control nodes do not gain matching responses after sending initial SYN packets to a target, this could still be caused by network errors and can be considered as ``uncertain'' \cite{ICLab:SP20}. 
Methods like proposed by Ensafi \ea \cite{Ensafi2014} that require specific setups of client and server behavior (global IP-ID on the client, an open port on the server) obviously depend on such circumstances. 
Another, more general, limitation is the directionality of the censorship method. For instance, Nourin \ea \cite{nourin2023measuring} report that the Turkmen firewall is bidirectional.

\paragraph*{DNS}
As pointed out in \cite{ACSAC24:WorldOfCensorship,ICLab:SP20}, detecting DNS manipulation faces issues due to CDNs, content localization and load balancing as control measurements might become outdated quickly or are routed to regional data centers.
As the same domain might be resolved to different IP addresses based on time of day or probe location, it might be difficult to decide if a mismatch in probe results stems from censorship or legitimate reasons. This is why we recommend the additional evaluation of TLS certificates as done by Tsai \ea \cite{tsai:PETS23:CERTainty} and described above.

A different limitation that was described by Nourin \ea in~\cite{nourin2023measuring} is that a censor might learn that a certain host is used as a measurement probe and suspend the censoring for this host. The authors noticed this with their DNS measurements, but the same idea can be applied to other measurement techniques as well. This is similar to malware detecting the presence of a debugger or virtual machine and suspending any malicious behavior.

Further, Tang \ea \cite{tang2024automatic} show that the generation of domain probe lists, which was often crowed-sourced and manual, can be improved by automation. High-quality domain probe lists are not only essential to gain useful measurement results but also require continuous adjustments.

\paragraph*{HTTP(S)}
Websites frequently use dynamic content, making it difficult to establish a stable baseline for comparison. Variations in timestamps, personalized elements, and server-side updates can trigger false positives, mistaking legitimate changes for censorship.
For HTTPS, the encryption poses the biggest challenge for measurement. The inspection of content is infeasible without sophisticated techniques, requiring a reliance on metadata, handshake analysis, and side-channel observations (also cf.\ Sect.~\ref{Sect:Measure:TLS}).

\paragraph*{Other Application Layer Protocols {and Platforms}}
Measuring online platforms' and application layer protocols' censorship is often driven by user-agents that implement the particular protocols and utilize APIs of the platforms. {Closed platforms cannot be inspected with automated tools or -- alternatively -- would require massive reverse engineering effort to develop measurement tools.} For this reason, measuring every single service must be considered a {costly} effort, especially since online platforms' popularity changes quickly and new platforms emerge regularly. Further, geo-blocking must be considered, e.g., through geo-diverse vantage points. Finally, the participation as a \emph{peer}, like in case of \emph{BTTest}, must mimic typical user behavior to gain high-quality measurement results and often requires manual account registration.

\paragraph*{TLS}
A general issue, that is mentioned in works discussing TLS-based measurements, is the small amount of vantage points used for a measurement \cite{Chai:ESNI,tsai:PETS23:CERTainty}. Additional geolocation-diverse vantage points could lead to a more complete view of the censorship methods, as some censors differentiate based on the source or destination of packets and in some cases other routes to the target can result in different censorship behavior.

Measurements based on the validity of TLS certificates could be falsified by state-level actors employing rogue CAs that find their way into the trust stores of mainstream browsers like Chrome or Firefox. Researchers have to rely on the browser developers to keep those trust stores clean \cite{tsai:PETS23:CERTainty}, as such rogue CAs could falsify the reference data.

Given the above approaches, each provides an own vector of insight (e.g., faked TLS certificate or presence of STRIPTLS attacks). It would be beneficial to \emph{combine} the ideas of existing TLS-based measurements to overcome this limitation. For instance, a large set of geo-diverse vantage points and control nodes could be used to compare as many SNI values as possible while at the same time TLS handshake results are recorded and compared to detect downgrade attacks and further try to detect STRIPTLS attacks.

\section{{Censorship and Censorship Measurement of Circumvention Tools}}\label{Sect:CircumventionTools}

{In this section, we will first introduce censorship techniques for different types of circumvention tools. Afterward, we explain the methodology that can be used to measure censorship of circumvention tools. Following that, we discuss limitations of these censorship and measurement techniques.}

\subsection{{Censorship of Circumvention Tools}}
{In this section, we discuss the censorship of VPNs and other circumvention tools, such as Tor, I2P and Psiphone. Since the techniques for different tools do overlap, we decided to cover them together.} 
We concentrate on the most important censorship methods and refer the reader to specific surveys and papers when it comes to related topics, e.g., {(Tor-)deanonymization} attacks \cite{karunanayake2021anonymisation}, {surveillance aspects, analyses of circumvention-related} court case \cite{Tippe:PETS25}, or surveys on censorship \emph{circumvention} itself {\cite{Khattak:2016,tschantz2016sok}}.


\textbf{Covered circumvention approaches:}
In addition to classic VPNs, there is a set of additional circumvention tools that are not pure VPN solutions. Such tools either (i) provide a cryptographic transmission that (like VPNs) provides an encrypted tunnel, optionally with sender/receiver anonymity, (ii) obfuscate the traffic, (iii) hide the traffic steganographically through covert channels, or (iv) combine two or three of these ideas. Over multiple decades, \emph{The Onion Router} (Tor) \cite{dingledine2004tor,robertson2017darkweb} has emerged as a widely used tool to enhance online privacy and bypass censorship. Its decentralized network architecture, which routes traffic through multiple layers of encryption (hence the ``onion'' analogy), makes it difficult to trace user activity and block access to websites \cite{dingledine2004tor,tschantz2016sok,robertson2017darkweb,Bishop:CSAS}. Tor can be enhanced with modules that provide traffic obfuscation or steganographic enhancements. In addition to the main Tor implementation, other variants of Tor exist, e.g., \emph{Arti}\footnote{\url{https://tpo.pages.torproject.net/core/arti/}}, which is a Rust-based implementation. However, in this article, we will not distinguish between Tor implementations. We further subsume Tor-related enhancements in this section, such as proxy services for non-onion services, e.g., \emph{Onionspray}\footnote{\url{https://onionservices.torproject.org/apps/web/onionspray/}} and \emph{Oniongroove}\footnote{\url{https://onionservices.torproject.org/apps/web/oniongroove/}}. Hyphanet (also called \emph{Freenet}) as well as its fork I2P (\emph{Invisible Internet Project}) operate on a mix-network structure similar to onion routing. Both tools provide several enhancements and features. Another popular censorship circumvention tool is \emph{Psiphone}\footnote{\url{https://psiphon.ca/th/index.html}}, which supports several different features to bypass censorship, including VPN functionality with tunnels, utilization of proxy servers, and traffic obfuscation.

{
\paragraph{Blocking Access to well-known Services}
One general idea for censoring circumvention tools is to block the access to well-known entry points of those tools.}

{
As VPNs utilize a centralized setup, it is easy to see that} VPN endpoints, the servers to which VPN clients connect, are frequently targeted by censors and network operators looking to restrict VPN usage \cite{winterMeasuringCircumventingInternet2014}. \textbf{Blocking VPN endpoints} {already stops the initial creation of VPN} tunnels that {are used} to bypass censorship or access restricted content. Common blocking methods include \textbf{IP address blacklisting and port filtering}, where the censor maintains a list of known IPs and ports of the VPN server and blocks connections to them \cite{gaoVPNTrafficClassification2020}. This {blocking} approach {can be countered by} VPN providers simply shifting to new IP addresses, leading to a constant cat-and-mouse game. {This aspect also affects other circumvention tools with well-known, central servers.}
{Censors might also use active traffic analysis techniques to identify VPN entry servers. Identifying VPN endpoints can be done through \textbf{active probing} \cite{Fifield:PhD}, where censors send connection requests to suspected VPN servers and analyze the responses (or lack thereof).}

A specific detail of Tor are \textbf{directory servers}. They are crucial for bootstrapping {a connection} into the Tor network, and thus a frequent target of censorship. Censors can block access to these servers by blacklisting their IP addresses or domain names, preventing clients from retrieving directory information necessary to build Tor circuits. Winter \cite{winterMeasuringCircumventingInternet2014} observed that the GFW blocks access to most Tor directory authorities by filtering their IP addresses, thereby preventing direct access to the Tor network from within China. This necessitates the use of Tor bridges {(i.e., Tor relays that are not listed publicly)} or alternative methods to bootstrap into the network, and highlights the importance of diverse entry points for circumventing censorship. However, a censor could also try to publicly announce malicious Tor bridges through the very same channels. Additional potential attacks against directory servers have been described early on \cite{dingledine2004tor}, including not only protocol-level attacks but also causing distrust to cause directory server dissent.

{
\paragraph{Traffic-based Blocking}
Instead of blocking the access to a circumvention service based on lists of known entry points, it is also possible for censors to detect circumvention tools by analyzing and fingerprinting their traffic. This can be used to block the newfound entry points or to drop/disrupt active connections.}

{Such tactics can involve} \textbf{Deep Packet Inspection (DPI) and VPN fingerprinting}. By analyzing packet headers and payloads, and aggregating packets into flows, DPI systems can identify VPN protocols (like OpenVPN or WireGuard) and block or throttle the corresponding traffic, even if it is encrypted {\cite{rfc9505}}.
For example, a DPI system might identify OpenVPN traffic by looking for specific opcodes (used to indicate the message type) or packet sizes within the encrypted TLS control channel.
As Xue \ea demonstrate, even ``obfuscated'' VPN services, designed to evade detection, can be fingerprinted and blocked through these methods, highlighting the ongoing challenge of maintaining unrestricted access to VPN technology \cite{xueOpenVPNOpenVPN2024}.
They hereby use a two-phase framework. First, the censor passively fingerprints flows by analyzing OpenVPN opcode sequences and OpenVPN ACK packet patterns within the initial handshake, exploiting OpenVPN's predictable structure even with encrypted payloads. Suspect flows are then actively probed, triggering unique timeout behaviors in OpenVPN servers due to their packet processing and handling of invalid lengths. These timeouts act as a side channel, confirming OpenVPN's presence even with probe resistance enabled. This combined approach achieves high accuracy with low false positives, effectively identifying encrypted VPN traffic even in the presence of common obfuscation techniques.

Gao \ea \cite{gaoVPNTrafficClassification2020} propose a statistical method for VPN traffic classification based on Payload Length Sequence (PLS). First, they analyze the payload length distribution of various VPN protocols and web browsing traffic. They observe that different VPNs exhibit distinct payload length patterns due to variations in protocol design and obfuscation techniques. Based on this, they construct a PLS feature vector by extracting the lengths of the first $n$ packets in a flow, excluding retransmitted packets and zero-length payloads. The value of $n$ is determined experimentally. This PLS vector captures the sequential order of payload lengths, which is a distinctive characteristic of different VPN protocols. They then use machine learning algorithms to train classifiers on a labeled dataset of VPN and web browsing traffic. The classifiers learn to associate specific PLS patterns with different VPN protocols, enabling them to classify new, unseen traffic flows. Their results show that Random Forest achieves the highest accuracy, reaching 98.68\% when using a PLS vector of length 6, indicating that this statistical approach based on payload length sequences is effective for VPN traffic classification, even for obfuscated VPNs.

{Similar} to the methods mentioned above, Saputra \ea demonstrate the application of \textbf{DPI to detect and block Tor traffic}, mirroring techniques used for VPN censorship \cite{saputraDetectingBlockingOnion2016}. Their method analyzes the TLS handshake within Tor connections, focusing on characteristics like cipher suites and server name indication (SNI) to distinguish Tor traffic from regular HTTPS. While effective against older Tor versions, this approach is less successful against newer iterations employing obfuscation techniques like ScrambleSuit (cf.\ \cite{ScrambleSuit}). 
Similar to VPN censorship, \textbf{active probing of suspected Tor bridges} could complement DPI-based detection, enabling censors to confirm bridge relay status and subsequently block them. Their proposed blocking mechanism combines DPI with IP blocking by extracting destination IP addresses from detected Tor traffic and adding them to proxy server blocklists. 
Ultimately, Saputra \ea's work \cite{saputraDetectingBlockingOnion2016} highlights the parallels between Tor and VPN censorship, showcasing how DPI and active probing are employed to target both, while also emphasizing the limitations of static detection methods in the ongoing arms race of censorship and circumvention. %

Moreover, attackers can leverage traffic analysis to observe communication patterns between nodes and the rendezvous point, potentially disrupting these interactions, as stated by Karunanayake \ea \cite{karunanayake2021anonymisation}. They further state that compromising the rendezvous point itself allows attackers to intercept, modify, or block communication, effectively disrupting network operations.

{Wails \ea \cite{wailsPreciselyDetectingCensorship2024} developed and evaluated a new approach for circumvention tool traffic identification. They developed a deep learning flow-based classifier that is combined with a host-based approach, collecting flows over time for each host. For their training, they used over 60 million real-world flows to more than 600,000 destinations. While the detection results for classical, solely flow-based detection remained low in a real-world scenario, the host-based approach had perfect recall and low false positives. The authors therefore raise concerns about host-based detections and argue for the development and research of circumvention systems that are resistant to such new detection approaches.}

{
\paragraph{Website Fingerprinting for Circumvention Tool Detection}
A special form of traffic-based blocking} attempts are \textbf{website fingerprinting (WFP)} attacks. WFP aims to detect which websites a circumvention user accesses and was discussed for several circumvention tools, such as Tor \cite{dingledine2004tor,panchenko2011website,SurveyTorWebsiteFingerprinting2023,Cherubin:Usenix22:TorWFPRealWorld} and Psiphone \cite{Ejeta17}. 

{
While WFP can mostly be considered a deanonymization attack that is used when a known circumvention tool is present, it can also be used to identify if a circumvention tool is used in the first place.} Ejeta and Kim \cite{Ejeta17} present a WFP attack on Psiphone: First, a censor (or a law-enforcement agency) visits the target website several times using a circumvention tool and records the generated traffic with a capturing tool such as \texttt{tcpdump}. Next, metadata is extracted from these flows to gain feature values: packet length uniqueness (i.e., assigning a packet length a value between ``1’’ (unique packet length) and ``0’’ non-unique packet length), total transmission (total incoming and outgoing packet count and total connection time), length of the {first 20} packets of a connection, packet order (total number of packets before the current packet of a sequence as well as the total number of incoming packets between an outgoing packet and its predecessor), concentration of outgoing packets (``number of outgoing packets in non-overlapping spans of 30 packets''), and bursts \cite{Ejeta17}. These metadata are used {to} generate a \emph{fingerprint} model of the particular website. {Afterward}, the censor needs to record the (assumed) traffic of an end-user visiting the same target and extract metadata from that flow as well to form another fingerprint used for comparison. Due to both traffic recordings never being an exact match (e.g., {because of} different encryption keys, network congestion etc.), Ejeta and Kim \cite{Ejeta17} used kNN and SVM classifiers to determine if both fingerprints match. The reported success rate was strongly {dependent on the circumvention tool used}. {They} gained good classification results for detecting \textit{Psiphon} traffic \emph{but} reached a low accuracy when they tried to identify visited website targets among a set of previously chosen ones.

\paragraph{{Detecting Steganography and Obfuscation Characteristics}}
Finally, there are many methods to detect, limit and prevent censorship circumvention based on steganography and traffic obfuscation. Such {circumvention} methods are applied in a way such that Tor or VPN traffic is encapsulated into a steganographic channel (serving as a tunnel) to let the traffic appear harder to detect. Similarly, traffic obfuscation {as well as the replacement of previously recorded benign traffic \cite{lorimer2021oustralopithecus}} can be used to let the circumvention traffic appear as legitimate traffic of another type (e.g., {circumvention tool traffic could appear as legitimate web traffic} if that is not blocked). To this end, many detection methods exist, both {based on} statistical methods and {based on} machine learning methods \cite{NIHbook,zander2007survey,CSURpaper,caviglione2021trends,GenericTaxonomy}. Further, attempts to block or limit such channels (decrease QoS or limit channel capacity) are often forms of traffic normalization \cite{NIHbook,CSURpaper,GenericTaxonomy}. For instance, traffic normalizers might remove or overwrite unused/reserved header fields to block simple forms of covert channels in these fields or slightly delay packets to limit the capacity of timing-based channels. 
To this end, the available methods consider a broad set of traffic features, such as packet lengths, inter-packet times, TCP and IPSec packet sequences, packet header field values, and the structure of dynamically composed packet headers \cite{CSURpaper,NIHbook,caviglione2021trends,GenericTaxonomy}---just to mention a few.

\subsection{Measuring Censorship of Circumvention Tools}
In the following, we discuss different methods to measure the censorship of circumvention tools. Again, as these methods overlap, we cover them together.

A particularly interesting target to censor are different types of \textbf{directory servers}, as these are essential for bootstrapping clients into a network. Examples include Tor directory servers and I2P reseed servers.\footnote{I2P reseed servers are used by newly joint relays during the bootstrap procedure and are a type of relay whose role is to provide information about other I2P relays \cite{hoangMeasuringI2PCensorship}.} 
Iszaevich, for example, presents a \textbf{user study} conducted in Mexico, where he analyzed whether the Tor \textbf{directory authorities} (DirAuths) are actively censored. To do so, he \textbf{distributed a questionnaire and a guide for \texttt{traceroute}-based measurements} \cite{iszaevich2019distributed} and found that a significant amount of DirAuths are censored on an IP level by censors.

In addition, Dunna \ea performed a study in which they \textbf{placed {multiple} previously unpublished Tor bridges} in the US, Canada, and Europe to connect to them through Virtual Private Servers (VPS) deployed in Chinese clouds \cite{dunnaAnalyzingChinasBlocking2018}. Within these VPS, they deployed Tor clients to connect to said Tor bridges. For probing connection initiation in a lightweight fashion, they additionally used the \emph{Tor Connection Initiation Simulator} (TCIS) to quickly derive the necessary metrics for their analysis (e.g., duration of blocking). They found that China blocks both unpublished and published relays. Moreover, the gained insights on the blocking allowed them to derive circumvention strategies \cite{dunnaAnalyzingChinasBlocking2018}.
Such active probing is a common and effective strategy to detect censorship for both, VPN tools (cf.~Xue \ea \cite{xueOpenVPNOpenVPN2024} on OpenVPN) and Tor-like tools, as it can reveal blocking through connection failures or timeouts. Network measurement platforms such as OONI \cite{filasto2012ooni} provide valuable data to detect Tor censorship by collecting measurements from geographically diverse vantage points.

Complementing Tor censorship measurements, Hoang \ea employed a novel approach using VPN Gate's distributed network of  \textbf{volunteer-run VPN servers to measure I2P blocking} globally \cite{hoangMeasuringI2PCensorship}. Their methodology involved four key techniques: 1) resolving I2P domain names and reseed server addresses using local and open DNS resolvers to detect DNS poisoning, 2) establishing HTTPS connections to the I2P homepage over VPN tunnels and monitoring the TLS handshake for SNI-based blocking, 3) capturing and analyzing network traffic for injected TCP RST/FIN packets during access attempts to I2P resources, including their own set of I2P relays, and 4) comparing fetched HTML content with legitimate I2P webpages to identify block page injection. This combination of DNS, TLS, TCP, and HTTP measurements, deployed through geographically diverse VPN Gate servers, allowed them to detect I2P censorship in several countries, including China, Iran, Oman, Qatar, and Kuwait, revealing various blocking techniques employed against the anonymity network.


Another aspect measurement studies focus on is the \textbf{delay between circumvention proxy appearance and a censor's blocking reaction}. For instance, {VPN entry points, I2P reseed servers and} Tor bridges appear at some time, but it would be valuable to gain insights into how long it takes for a certain censor to block {such new entry points}. {A Tor bridge-focused measurement} was designed by Fifield and Tsai \cite{FifieldTsai2016:DelayBlockingCircumvProxies,Fifield:PhD}. They implemented two scripts. The first script ran TCP connection attempts to Tor bridges every 20\;min from China, Iran, and Kazakhstan and reported gained error codes. In parallel, a control script made the same connection from the US. The second script realized a whole Tor-in-obfs4 connection from Kazakhstan. The authors considered only ``default bridges'', i.e., non-secret bridges known within the Tor community. Similar experiments could be conducted to reach proxies of other circumvention tools.


\subsection{Limitations}
\paragraph*{imitations of Circumvention Tool Censorship}
Blocking circumvention {tool} traffic, while a common censorship tactic, faces inherent limitations due to the distributed and adaptable nature of these technologies. Blocking VPN server IP addresses, as highlighted in Xue \ea \cite{xueOpenVPNOpenVPN2024}, is a continuous cat-and-mouse game, as VPN providers can easily shift to new IPs or deploy obfuscation techniques. Similarly, blocking Tor relays or bridges requires constant monitoring and updating of blocklists, as new relays and bridges appear frequently. Though effective for identifying VPN protocols or Tor traffic patterns (as shown by Saputra \ea \cite{saputraDetectingBlockingOnion2016}), DPI is computationally expensive and susceptible to evasion through obfuscation or traffic manipulation techniques. 
In addition, correlation attacks using metadata of traffic in form of volume or timing behaviors can be rendered difficult through obfuscation techniques, such as packet size modulation.

WFP faces several limitations that are practically relevant for censors. Generated website fingerprints can become outdated quickly, e.g., when new forum posts are added to a website hosting a discussion forum. 
Cherubin \ea highlight that several studies do not consider a real-world setting, e.g., relying on synthetic traffic and target websites that might not be visited by real Tor users (among other limitations) \cite{Cherubin:Usenix22:TorWFPRealWorld}. Similarly, Ejeta and Kim consider a scenario in which a user conducts no parallel activity while visiting a certain target website, and the censor is aware of the start and end time of a pageload~\cite{Ejeta17}. Further, WFP is usually conducted on flows that do not employ any obfuscation features, WFP does not perform well on such flows \cite{panchenko2011website}. Other issues, e.g., website caching behavior and lack of heterogeneous users, browsers, and systems, are covered in \cite{SurveyTorWebsiteFingerprinting2023} and \cite{Cherubin:Usenix22:TorWFPRealWorld}.

Furthermore, blocking access to VPN or Tor websites and documentation, while hindering discoverability, cannot prevent determined users from obtaining the software through alternative channels. Moreover, the decentralized nature of both VPNs {(various different providers)} and Tor, with multiple entry and exit points, makes it difficult for censors to achieve \emph{complete} blocking without resorting to drastic measures like nationwide Internet shutdowns, which often carry significant political and economic costs. 

\paragraph*{{Limitations of Circumvention Tool Censorship Measurement}}
These kinds of measurements have several implications and limitations. Iszaevich states that since measurements were conducted by local volunteers, misconfigured experiments or old clients are a source of error \cite{iszaevich2019distributed}. In relation, as Dunna \ea also state, the biggest issue is the locality of the experiments \cite{dunnaAnalyzingChinasBlocking2018}. For optimal measurements, people actually need to physically be in the country where censorship is applied. Depending on the laws of that country, this can be a significant risk.
Moreover, censors actively adapt their techniques, making detection a moving target. The use of obfuscation methods, such as those employed in some VPNs or Tor's ScrambleSuit \cite{winterMeasuringCircumventingInternet2014}, can make it difficult to distinguish VPN or Tor traffic from regular HTTPS, hindering DPI-based detection: obfuscation makes it more challenging to detect censorship itself, as the DPI systems may fail to recognize the censored traffic as such, leading to false negatives in censorship measurements. It also complicates the identification of specific blocking techniques, as the obfuscated traffic might not exhibit the clear signatures associated with methods such as RST injection or block page redirection.
As in case of previously covered methods, accurately measuring the impact of censorship, like throttling or blocking of specific servers, on circumvention tools can be challenging due to network fluctuations and the difficulty of establishing stable baselines. False positives and negatives are common, particularly when relying on automated measurement or limited vantage points. 

\section{Measurement Platforms \& Datasets}\label{sect:measurementplatforms}
Developing censorship circumvention methods requires an in-depth understanding of censorship. For this reason, the development of bypass tools must consider measurement datasets. These datasets can be used not only to tailor circumvention tools, but also to evaluate these tools. This section discusses the available censorship datasets and a comparison of these datasets. The above-mentioned datasets are current data or data from recent years.

\paragraph*{CensoredPlanet}
CensoredPlanet is a censorship measurement platform that captures and analyzes measurements from four different measurement techniques, namely Augur, Satellite/Iris \cite{scott2016satellite, Pearce2017}, Quack \cite{Ben2018Quack}, and Hyperquack \cite{2020HyperQuack}.
The goal is to complement existing datasets like OONI \cite{OONI} and ICLab \cite{ICLab:SP20} by providing a larger scale, coverage, and duration of measurements.
Until 2020 \cite{CensoredPlanet2020} they collected and published over 21 billion data points from over 20 months of measurement operations. However, they continuously publish new data on their platform that is available online \cite{CensoredPlanet}.

The platform captures baseline data from 2,000 domains using over 95,000 vantage points worldwide. This data is collected, aggregated and analyzed to provide a longitudinal view of censorship and allow distinguishing between localized and countrywide censors \cite{CensoredPlanet}.
In practice, they were able to reduce false positives, modeling the data as time series and using the Mann-Kendall test. With this, they were able to detect 15 censorship events over their measurement period, two thirds of which were not previously reported.

CensoredPlanet not only supports long-running broad measurements, it also has ``rapid focus capabilities'', allowing for timely measurements of specific networks. The authors used this to investigate changes in the blocking behavior of Cloudflare IPs from Turkmenistan.

CensoredPlanet measures censorship on six different protocols: IP, DNS, HTTP, HTTPS, Echo, and Discard. IP, DNS, HTTP and HTTPS were selected for their position as censorship targets. Echo is used to measure keyword-based censorship on the application layer by checking if the selected servers successfully reply to such requests containing potentially censored keywords. The Discard protocol is used to measure the directionality of the censor.

\paragraph*{CenDtect}
CenDtect utilizes the data of CensoredPlanet as input for their censorship detection model. Tsai \ea\cite{TsaiEtAl:NDSS24:ModelingDetectingIntCens} implemented a system that uses unsupervised machine learning.
They therefore implemented a clustering approach based on decision trees and DBScan. With this, they want to make the data easier to consume for other researchers, journalists or NGOs.
The output of the system consists of decision trees, which describe the censorship rules and a corresponding domain list showing which domains are affected by the censorship rule.
They also discuss the challenges of censorship datasets, like the lack of ground truth, the volatility of test lists and the sheer amount of data. 
While they do not provide a new dataset, they aim to augment existing data to make it more usable and are evaluating the possibility to directly integrate CenDtect into the CensordPlanet platform.
In consequence, we do not list CenDtect in Tab.~\ref{tab:datasetcmp}.

\paragraph*{Encore} Burnett \ea analyzed Internet censorship in \cite{10.1145/2785956.2787485}, specifically in the form of web filtering, particularly focusing on its extent. To achieve this, they introduced \emph{Encore}, which enables continuous and longitudinal measurement. A key advantage of the system is that users do not need to install custom software, emphasizing its ease of use.

Encore utilizes web browsers on Internet-connected devices as vantage points, allowing it to gather a large and diverse set of data. The major contribution of Burnett \ea is that meaningful conclusions about web filtering can be drawn from the side channels that exist in cross-origin requests.

An origin is defined as the protocol, port, and DNS domain (“host”) \cite{SameOriginPolicy}. Websites can send information to another origin using HTTP requests across origins. The cornerstone of Encore’s design is to use information leaked through these cross-origin embeddings to determine whether a client can successfully load resources from another origin. For example, an image embedded with the $<$img$>$ tag can trigger an onload event. Using this kind of information, Encore can make determinations about the extent of Internet censorship. The scope of web filtering can range from individual URLs to entire domains \cite{10.1145/2785956.2787485}.

\begin{figure}[!tp] 
    \centering
    \includegraphics[width=\columnwidth,trim={0cm 0.6cm 0cm 0.6cm},clip]{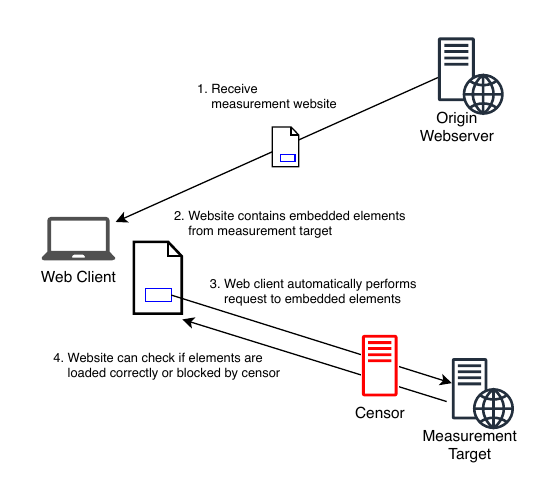}
    \caption{Encore measurement methodology (based on \cite{10.1145/2785956.2787485})}
    \label{fig:encore}
\end{figure}

Measuring web filtering with Encore involves three parties (Fig.~\ref{fig:encore}): (i) the web client, which acts as the measurement vantage point; (ii) the measurement target, which hosts a web resource suspected to be filtered; and (iii) the origin web server, which serves a web page to the client, instructing it on how to collect measurements with tasks, which are small programs (self-contained HTML and JavaScript snippets) that attempt to load a web resource.

As of 2015, Encore had recorded 141,626 measurements from 88,260 distinct IPs across 170 countries, including China, India, Egypt, South Korea, Iran, Pakistan, Turkey, and Saudi Arabia. However, Encore has a significant limitation: it can only observe the accessibility of individual web resources and cannot determine the exact method of filtering. This limitation positions Encore as a supplement to other censorship measurement systems, which are capable of more in-depth analyses but encounter greater challenges in deployment.

\paragraph*{GFWatch} 
GFWatch is specifically tailored to measure the behavior of the GFW by identifying censored domains and detecting forged IP addresses in fake DNS responses. The GFW employs DNS injection techniques, returning falsified responses. To summarize the function of GFWatch: Nodes inside China run DNS tests, whose results are then compared to DNS resolving results of nodes outside China to detect DNS-based filtering. This DNS poisoning not only impacts users within China but also pollutes global DNS resolvers like Google or Cloudflare. As a large-scale longitudinal measurement platform, GFWatch provides a continuously updated view of the GFW’s DNS-based blocking behavior \cite{USESEC21:GFWatch}.

Hoang \ea revealed that this manipulation of DNS records has far-reaching consequences beyond China’s borders \cite{USESEC21:GFWatch}. To mitigate its effects, they proposed strategies to detect poisoned responses, sanitize polluted DNS records, and assist in the development of circumvention tools to bypass the GFW’s censorship. To achieve its goal, GFWatch must continuously discover and test new domains as they emerge. For that, GFWatch uses top-level domain (TLD) zone files, which are updated daily. Additionally, GFWatch provides ongoing insights into the pool of forged IPs used by the GFW, helping researchers better understand and counteract China's DNS censorship tactics \cite{USESEC21:GFWatch}.

GFWatch probes the GFW from outside China to identify censored domains (step 1) and verifies these findings using controlled machines within China (step 2). It employs UDP-based DNS queries, as UDP is the default carrier for DNS and requires fewer network resources compared to TCP-based measurements. The procedure follows multiple steps: (i) The primary probing machine located in a US network sends DNS queries for test domains to two hosts controlled by Hoang \ea in China. These Chinese hosts lack DNS resolution capabilities, so any DNS responses received by the main prober originate from the GFW. Positioning the two Chinese hosts in different ASes helps address the GFW's centralized blocking policy and detect potential regional variations. (ii) GFWatch transfers detected censored domains to its Chinese hosts, which then probe these domains from within China towards the controlled machine in the US to confirm that the censorship occurs both inside and outside China. To account for potential inconsistencies, such as the GFW occasionally failing to block access under heavy load \cite{Niaki2020Triplet}, GFWatch tests each domain at least three times daily \cite{USESEC21:GFWatch}.

In \cite{GFWatch}, Hoang \ea tested a total of 534 million distinct domains. Over a nine-month period, they found that 311K individual domains were filtered. To analyze the evolution of blocking rules rather than just the number of censored domains, they identified the most general form of each censored domain (the so-called base domain) that triggers censorship. As a result, they discovered 138.7K base domains. Hoang \ea also found that 41K censored domains were overblocked due to textual similarity with a censored base domain, despite not being subdomains. The dominant censored categories were business, pornography, and information technology \cite{USESEC21:GFWatch}. 
They also found that forged IPs follow distinct patterns, rather than being injected entirely at random. They identified 11 groups: 10 static groups with consistent, predictable responses (e.g., \texttt{qcc.com.tw} always returned \texttt{89.31.55.106}) and one dynamic group with variable manipulations. Since the GFW frequently changes spoofed IPs within a certain range, DNS censorship can be effectively detected by comparing the returned IPs with the pool of spoofed addresses discovered by GFWatch \cite{USESEC21:GFWatch}.

Using this methodology, it is possible to sanitize polluted DNS records in the cache of public DNS resolvers. After identifying a censored domain, GFWatch can query the domain against popular DNS resolvers and verify whether their responses show any injection patterns. 
%

{
\paragraph*{GFWeb}
GFWeb \cite{GFWeb} is another system for performing measurements of HTTP and HTTPS censorship made by the GFW. Its goal is to augment existing measurements by including more domains and performing repeated trials over a longer time frame. Over a time span of $20$ months, GFWeb scanned more than $1$ billion domains and detected over $940$K and over $50$K censored domains for HTTP and HTTPS, respectively.
The measurement approach designed by the authors allowed them to trigger HTTP(S) censorship without performing an actual three-way handshake. To this aim, they performed tests by sending SYN and PSH/ACK packet pairs, with the latter containing a HTTP GET or a TLS client hello.
Additionally, GFWeb uses servers both inside and outside of China as vantage points, allowing it to scan both the ingress and egress behavior of the censor. This analysis demonstrated that the GFW acts in an asymmetrical manner, which was a previously unknown behavior. Other findings of this longitudinal study include changes in the filtering logic of the GFW. Previously there was considerable overblocking for some domains, which was found to be patched in later scans.
}

\paragraph*{ICLab} This Internet monitoring platform was introduced by Niaki et~al.~\cite{ICLab:SP20}. 
%
The project measures connectivity to sites. This includes a test of the DNS request-response behavior with local and public resolvers, TCP connection test, certificate chain tests (for HTTPS URLs), an evaluation of the HTTP response (headers and body), a traceroute log, and a capture of all packets exchanged during the measurement procedure \cite{ICLab:SP20}.
ICLab uses hosts located within the region that is being monitored as vantage points. These vantage points are split into 264 VPN endpoints and 17 endpoints operated by 17 volunteers. In both cases, these hosts run ICLab's measurement software (VPN points route traffic through a VPN while the volunteer-operated devices send the plain traffic probes) \cite{ICLab:SP20}. Since measurement result are compared by the authors, additional \emph{control nodes}, located in the US, Europe, and Asia are employed. 
The data collection stopped in 2020, with traceroute data being updated until 2021.

\paragraph*{IODA} The \emph{Internet Outage Detection and Analysis} project \cite{IODA} is a system designed for continuous monitoring of the Internet. It is operated by the Internet Intelligence Lab at Georgia Tech’s College of Computing. It detects large-scale outages at the network edge, including disruptions impacting ASes or entire regions within a country. IODA works in partnership with OONI \cite{OONIioda}. 
It uses three distinct and complementary sources of Internet measurement data (BGP, Active Probing, and Internet Pollution) to detect outages and has provided near-real-time visualizations of Internet connectivity on a public platform. 
Regarding BGP, IODA monitors the reachability of IP addresses by analyzing routing data from global sources such as RouteViews and RIPE RIS \cite{CAIDAioda}. It examines small network blocks to accurately detect whether parts of a network are offline or inaccessible. 
For active probing, IODA uses ICMP echo requests as part of the Trinocular method \cite{10.1145/2486001.2486017,IODA}. When initial probes receive no response, the system sends additional probes to verify the network's status. The Trinocular methodology can accurately distinguish between online, offline, and uncertain states. 
Considering Internet pollution, IODA analyzes Internet Background Radiation (IBR) \cite{CAIDAioda}, which is unsolicited traffic resulting from misconfigurations, malware propagation, and network scanning. It filters out spoofed and bursty traffic (e.g., botnet scans) to extract a liveness signal based on distinct source IP addresses, enabling the detection of whether networks or regions are online or offline. 

An example of the use of IODA is the investigation of complete Internet outages in Myanmar by Padmanabhan \ea in \cite{10.1145/3473604.3474562}. They analyzed the Coup Day in 2021. The outage was particularly evident in the BGP data sources, where the number of /24 address blocks reachable in Myanmar via BGP dropped from 695 to 376 - a 46 percent decrease. 
In \cite{MultistakeholderReportIran} IODA data highlights Internet outages in Iran between September 2022 and October 2022. The BGP data source reflects routing announcements or global routing information exchanged between ISP routers, which are typically highly stable. A sudden drop, which in this case occurred frequently, indicates a disruption in Internet connectivity.

\paragraph*{Kentik}
Kentik is a large network observability company and offers a cloud-based NetFlow analytics platform designed for network monitoring, capable of analyzing traffic data collected from sources such as company network routers \cite{MultistakeholderReportIran}. For example, Kentik provides a comprehensive overview of how data flows (traffic) within a network by leveraging aggregated NetFlow data from its customers. 

NetFlow is a network protocol used by routers and switches to collect information about network traffic. It records metadata about IP traffic flows that pass through a NetFlow-enabled device (such as a router, switch, or host). The data collected includes: source and destination IP addresses of the data packets, protocols used (e.g., TCP, UDP), packet count, data volume in bytes, and connection duration \cite{Kentik}. 
This information can be aggregated to identify trends and patterns \cite{CISCONetFlow, 10.1145/3473604.3474562}. The protocol and its logs provide detailed insights into network traffic, supporting monitoring, optimization, and security. When NetFlow data is sourced from multiple locations, it offers a composite view of the Internet from various perspectives. Using this aggregated data, Kentik enables detailed analyses to draw conclusions about outages and other network disruptions. Kentik’s customers, including tier-1 ISPs and global content providers, provide NetFlow data from Internet routers, making it possible to study censorship events and more \cite{Kentik}. 
As an example, Padmanabhan \ea used Kentik in \cite{10.1145/3473604.3474562} to analyze cellular data. To demonstrate that cellular connectivity had been heavily restricted, they compared traffic data from four major cellular providers with that of a leading non-cellular provider. Their analysis revealed a substantial reduction in traffic from cellular providers during both day and night. In contrast, the non-cellular provider experienced traffic drops only during the night. 
Further, in \cite{MultistakeholderReportIran} Kentik was used to show that traffic volume from Facebook’s network dropped substantially on September 21st, 2022, in Iran. On that day, the service WhatsApp, owned by Facebook, started showing signs of blocking.

The combination of IODA and Kentik offers significant advantages for detecting and analyzing network outages and censorship events. IODA specializes in macro-level analysis, monitoring unusual routing changes and identifying large-scale network outages across entire regions or ASes. It examines routing information to detect sudden changes in the reachability of network blocks. In contrast, Kentik focuses on micro-level analysis, providing granular insights into specific data patterns. This allows for detailed comparisons of data streams, complementing IODA's broader perspective. Together, these tools deliver a comprehensive approach to monitoring and analyzing network disruptions~\cite{MultistakeholderReportIran, 10.1145/3473604.3474562}.


\paragraph*{OONI}\label{sec:OONI}
The \emph{Open Observatory of Network Interference} 
\cite{filasto2012ooni,OONI} provides a \emph{Measurement Aggregation Toolkit} (MAT) \cite{OONI:MAT} that allows users to explore current and historic measurements of Internet censorship. MAT allows to investigate censorship specifically for selected ASes, countries, domains, dates and test methods. For instance, general Internet connectivity tests are available, but also tests for connecting to messenger services such as Facebook, Telegram, WhatsApp, or Signal. Further, certain middlebox-related tests, performance tests, tests for circumvention tools (Tor, including the Snowflake \cite{bocovich2024snowflake} pluggable transport, and Psiphon) as well as experimental tests (VPN connectivity and DNS check) can be analyzed regarding their results. OONI measurements are performed by volunteering users running OONI \emph{Probe} \cite{OONI:Probetool} in a mobile, command line or desktop version. Further, a browser-based probe tool and an API to access OONI data are provided by the project. In general, differences between current and control measurements are considered as \emph{anomalous} and can thus refer to a filtering attempt (invalid measurements can occur as well if a \emph{a measurement fails to execute or is unavailable for a particular query}) \cite{ACSAC24:WorldOfCensorship}. The OONI website further provides timely reports on identified censorship events.
{Notably, a recent project has integrated OONI data into the Internet Yellow Pages (IYP), which is a knowledge graph with a unified access methodology~\cite{lipphardt:25:1800cens}. The IYP allows an easier analysis of OONI data, as users do not need to autonomously perform any further data integration steps.}

\paragraph*{Psiphon}While Psiphon itself is an Internet censorship circumvention tool, the organization behind it also provides the \emph{Psiphon Data Engine} (PDE) \cite{PDE_online}. PDE is a platform that allows researchers to collectively analyze aggregated Internet censorship data. The platform collects censorship metrics, network efficiency measurements, connectivity tests and also collects records of social and political indicators regarding censorship events. So called ``partners'' can provide their measurement data to the platform.
PDE provides information about the usage statistics of Psiphon like the unique users per day per country or the tunnel median round trip time.
Currently, PDE lists OONI and M-Lab as partners.
In consequence, we do not separately list PDE in Tab.~\ref{tab:datasetcmp}.

{
\paragraph*{RIPE Atlas}
RIPE Atlas \cite{RIPEAtlas} is an Internet measurement platform allowing users to perform various connectivity measurements (e.g., ping, DNS, and traceroute) from over $13,000$ active community-hosted probes. Despite not being directly linked with Internet censorship goals, RIPE ATLAS offers a diverse set of probes, which turned out to be valuable for implementing censorship measurements. 
For example, Anderson \ea~\cite{AndersonAtlas} used RIPE Atlas to analyze blocking events in Russia and Turkey. Also, Gegenhuber \ea~\cite{GegenhuberAtlas} used Atlas to measure deanonymization risks of tor users in Russia.
}

\paragraph*{Tor Dataset} The Tor project publishes various usage data of the Tor network \cite{tormetrics}. These include statistics about user numbers connecting to the network. These numbers can be used to detect censorship events. The Tor project itself uses this data to present the ``Top-10 countries by possible censorship events''.

\paragraph*{Triplet Censors}
Niaki \ea \cite{Niaki2020Triplet} analyzed the DNS injection techniques of the GFW by observing the DNS results of the Alexa top 1M domains over a duration of nine months. They collected over 120 million forged DNS responses in their dataset. They published the dataset together with the code to allow for reproducibility studies and to foster the research in the domain.

\paragraph*{Comparison of Platforms/Datasets}
Tab.~\ref{tab:datasetcmp} compares the platforms and datasets discussed above. As can be seen, most of the datasets contain only DNS and HTTP data. Other probes are less frequent.
In contrast, OONI provides a broad dataset that is driven by community measurements.
Similarly, some datasets highlight only the censorship of a single country while others try to encompass as many countries as possible, many of those rely on volunteers which has a direct impact on the number of probes.
Most of the discussed datasets are still ongoing efforts, sometimes even providing real-time data to researchers.
Only the Tor dataset is considered ``passive'' as it only contains usage statistics and not active probe traffic.

\begin{table*}[!ht]
    \centering
    \caption{Comparison platforms/datasets used for censorship measurement. (Abbreviations: TP: traffic probe, WCA: web-content analysis; BLK: blocked page analysis; Notes: ($^1$) 13 countries with volunteer-operated devices)}
    \label{tab:datasetcmp}
\resizebox{1.0\textwidth}{!}{
\begin{tabular}{|r||c|c|c|c|c|c|c|c|c|c|c|} 
\hline
\textbf{Dataset}              & \rot{\textbf{CensoredPlanet}}               & \rot{\textbf{Encore}}          & \rot{\textbf{GFWatch}} & \rot{{\textbf{GFWeb}}}   & \rot{\textbf{ICLab}}            & \rot{\textbf{IODA}}   & \rot{\textbf{Kentik}} & \rot{\textbf{OONI}}                         & \rot{{\textbf{RIPE Atlas}}} & \rot{\textbf{Tor}} & \rot{\textbf{Triplet Censors}} \\ 
\hline
\hline
\textbf{Key Refs.}            & \cite{CensoredPlanet,CensoredPlanet2020}    & \cite{10.1145/2785956.2787485} & \cite{GFWatch,274650}  & \cite{GFWeb}                            & \cite{ICLab:SP20,ICLab:Website} & \cite{IODA}           & \cite{Kentik}         & \cite{filasto2012ooni}                      & \cite{RIPEAtlas}                      & \cite{tormetrics}  & \cite{Niaki2020Triplet}        \\
\textbf{Approach}             & active                                      & active                         & active                 & active                                  & active                          & active                & active                & active                                      & active                                & passive            & active                         \\
\textbf{Method}               & TP,WCA,BLK                                  & WCA,BLK                        & TP                     & TP                                      & TP,BLK                          & TP,BLK                & TP,BLK                & TP,WCA,BLK                                  & TP                                    & -                  & TP,BLK                         \\ 
\hline
\textbf{Latest Year}          & ongoing                                     & 2015                           & ongoing                & ongoing                                 & 2021                            & ongoing               & ongoing               & ongoing                                     & ongoing                              & ongoing            & 2020                           \\
\textbf{\# Countries}         & worldwide                                   & 170                            & 1                      & 1                                       & 62($^1$)                        & worldwide             & worldwide             & worldwide                                   & worldwide                             & worldwide          & 1                              \\
\textbf{Probe Nodes}          & 95,000+                                     & 88,260                         & 2+                     & 2+                                      & 281                             & 500+                  & n/a                   & community based                             & community based 13.000+               & n/a                & 1                              \\
\hline
\textbf{HTTP(S) Probes}       & \checkmark                                  & \checkmark                     & -                      & \checkmark                              & \checkmark                      & -                     & \checkmark            & \checkmark                                  & \checkmark                            & -                  & -                              \\
\textbf{QUIC Probes}          & -                                           & -                              & -                      & -                                       & -                               & -                     & -                     & \checkmark(\cite{Elmenhorst21:HTTP3:QUIC})  & -                                     & -                  & -                              \\
\textbf{DNS Probes}           & \checkmark                                  & \checkmark                     & \checkmark             & -                                       & \checkmark                      & -                     & \checkmark            & \checkmark                                  & \checkmark                            & -                  & \checkmark                     \\
\textbf{DoT/DoH Probes}       & -                                           & -                              & -                      & -                                       & -                               & -                     & \checkmark            & \checkmark                                  & -                                     & -                  & -                              \\
\textbf{Tor Probes}           & -                                           & -                              & -                      & -                                       & -                               & -                     & -                     & \checkmark                                  & -                                     & -                  & -                              \\
\textbf{Tor Snowflake/WebRTC} & -                                           & -                              & -                      & -                                       & -                               & -                     & -                     & \checkmark                                  & -                                     & -                  & -                              \\
\textbf{VPN Probes}           & -                                           & -                              & -                      & -                                       & \checkmark                      & -                     & -                     & \checkmark                                  & -                                     & -                  & -                              \\
\textbf{Traceroute}           & \checkmark                                  & -                              & -                      & -                                       & \checkmark                      & only routing problems & \checkmark            & \checkmark                                  & \checkmark                            & -                  & -                              \\
\hline
\end{tabular}
}
\end{table*}

\paragraph*{CensorLab}
While not strictly a measurement platform or dataset, CensorLab \cite{censorlab} is a censorship experiment testbed. The goal of CensorLab is to enable researchers to emulate various censorship scenarios to evaluate existing, but also simulate and test newly developed censorship techniques, corresponding circumvention and measurement methods. This allows a proactive approach to censorship research instead of a reactive one. Other benefits of a simulated environment are reduced costs and reduced risks to measurement volunteers.
CensorLab allows researchers to define the censors' behavior through block lists and so called ``censor programs'' which can emulate a wide range of different censorship approaches.

\paragraph*{Excluded Platforms and Datasets}
We excluded a set of censorship platforms and datasets from our analysis due to superficial or dated measurement methodology or content, i.e., data older than 2015, or because they simply are not operated / existent anymore.
For instance, in 2015, Aceto \ea\cite{Aceto2015} covered several measurement systems that we did not include in this paper, namely \emph{ConceptDoppler}, \emph{rTurtle}, \emph{Herdict}, \emph{Alkasir}, \emph{YouTomb}, \emph{Greatfire.org}, \emph{CensMon}, \emph{MOR}, \emph{Weiboscope}, \emph{Samizdat}, \emph{UBICA}, \emph{WCMT}, \emph{Spookyscan} and \emph{encore}.
We further excluded the \emph{OpenNet Initiative} (ONI) \cite{ONI} that monitors Internet censorship in the sense of traffic filtering. Only a small fraction of data in the form of a summary CSV file is available anymore \cite{ONIdataset}. The latest update of the ONI dataset is dated Sep-2013 and was thus excluded from our analysis. However, an investigation of the dated dataset can be found in \cite{gill2015characterizing}. 

\section{Trends and Future Challenges of Internet Censorship and its Measurement}\label{sect:trends}

In this section, we outline the major trends and future challenges of Internet censorship identified in the related literature. {We highlight both technical and human aspects, considering changes in user behavior alongside the rapidly evolving landscape of software architectures and network protocols.}

\subsection{Technical Aspects}

\paragraph{{Measurement Design and Sampling}}
{Quantifying} the impact of censorship requires {overcoming several} challenges. The main issues, discussed a decade ago by Aceto and Pescapé~\cite{Aceto2015}, {remain} valid today. For instance, measurements should be characterized by an adequate probing frequency and must include a set of hosts/endpoints representative of a geographical region.


Collecting network traffic and behavior patterns is {no longer primarily a problem} of volume. In fact, advancements in storage and processing, often offered through as-a-Service models, {have made it easier to} obtain effective traffic snapshots compared to ten years ago. %
\paragraph{{Infrastructure, Routing, and AI Challenges}}
{More pressing challenges stem from} the evolving nature of the Internet architecture. First, there is a potential censorship inconsistency problem, where different ASes, ISPs or regions of the same country differ in their censorship policy implementation. As reported by Tsai \ea \cite{TsaiEtAl:NDSS24:ModelingDetectingIntCens,ChinaRegionalCens25}, there is no global-scale ground truth available that can be used to validate a censorship dataset. 
{Hence,} certain authors and platforms (e.g., above-mentioned approach of CenDtect) explicitly address how they handle false-positives and false-negatives. %
Second, routing path {inconsistencies complicate} traffic probing, {often due to} the ubiquitous diffusion of CDNs and load balancing components.
As a partial workaround, advancements in AI to process large-scale complex information may support researchers in Internet censorship, e.g., to ``repair'' traffic plagued by mismatches. Yet, some state-level censors are expected to deploy mechanisms to influence the interaction with AI itself. 
%
For instance, censors may operate over the communication layer, e.g., by blocking certain IP ranges to prevent the availability of complete datasets for training Large Language Models (LLMs) or block access to these platforms, such as in China, Russia, Iran, and Israel \cite{berger2024measuringdnscensorshipgenerative,FOTN24}. Therefore, future research should consider the investigation of LLM-based censorship (as, e.g., initially done by Ahmed and Knockel~\cite{ahmed2024impact}). Considering LLMs (and AI in general), another expected trend is that censors work towards applying AI-based detection tools for nuanced critiques in textual and/or visual form \cite{chenAccuracyBiasesAIBased2025}, for which research is still in its infancy. Similar advancements can be anticipated for audio/video transmissions and content.

\paragraph{{Evolving Targets and Services}}
Another important challenge to face in the near future entails {the rapid evolution of} targets and services. Even if some works have shown how to promptly adapt to new communication protocols (e.g., for QUIC \cite{Elmenhorst21:HTTP3:QUIC}), the actual panorama of Internet services is changing at an unprecedented pace. Social media platforms, including gaming and video services, have tight popularity cycles, thus demanding for continuous modifications of the measurement techniques/algorithms.

{In addition}, measurement properties can {shift due to} alternating user-bases or methodological {adjustments} within datasets \cite{ACSAC24:WorldOfCensorship}.

\paragraph{{Integrating and Combining Datasets}}
{To address these challenges,} researchers conducted dataset-overlapping work. Crowder \ea \cite{ACSAC24:WorldOfCensorship} incorporated multiple of the above-mentioned datasets (Satellite, Hyperquack HTTP(S), Tor, OONI, and GFWatch) into a framework {for joint} statistical analysis to investigate censorship events. 
{Similarly, Lipphardt \ea \cite{lipphardt:25:1800cens} integrated OONI data into the IYP and mention that they plan to integrate data from additional measurement platforms in order to provide a unified accessibility through the query language \textit{Cypher}.}

\paragraph{{Adversarial Dynamics and Arms Race}}
The relevant corpus of works and the vivacity of censorship circumvention research is expected to fuel an ``arms race'' among states, regimes, researchers, and activists. In this vein, it is important to underline the suggestion of Aceto and Pescapé reported in~\cite{Aceto2015} stating that measurement approaches should consider that censors (could) identify probing devices and operate to actively hide the effects of blockages or filterings. {This adversarial dynamic is likely to intensify as measurement and circumvention techniques evolve.}

\paragraph{{Research Quality and Education}}
A final remark to improve the technical research is to plan for proper strategies to ensure quality and consistency. First, experiments devoted to gather Internet traffic should {follow clear} quality criteria~\cite{PaxsonIMC04:InternetMeasurement}. Gained results should then follow the \emph{FAIR} principle, i.e., research work (including, code, data, metadata, notebooks etc.) should be \emph{F}indable, \emph{A}ccessible, \emph{I}nteroperable, and \emph{R}eusable. A way to aid re-usability is to publish the artifacts on open repositories and provide companion software or traffic capture metadata.


{Finally}, we encourage starting to teach Internet censorship (measurement and circumvention) as a complement {in} network security and privacy {curricula}, especially to raise awareness and {reduce} inconsistent terminologies and scientific re-inventions~\cite{WCM:ResearchTribalWars}.


\subsection{Societal and Human Aspects}
\paragraph{{Beyond the Network Layer}}
{So far, we almost exclusively covered network-level techniques. However, to provide a deeper understanding of the topic, it is beneficial to also understand societal and human aspects of Internet censorship. While,} technical {measurements} map the prevalence of restrictions across different regions, the success, and accuracy also depends on human aspects involved. This includes how people adapt their online behavior in response to perceived threats, trust, fear, and resistance. 

Although digital technologies have transformed public discourse and how communities connect, the Internet is not a neutral communication tool. {More so, it is a} structure shaped by cultural differences, power imbalances and economic dynamics \cite{cairncross2002death,warf2011geographies,zook2007creative}. These factors influence how the Internet is constructed, controlled and experienced, with varying levels of censorship creating tensions between national regulations and its originally borderless nature \cite{warf2011geographies}. Additionally, geographical factors, such as historical legacy, legal frameworks, governmental structures, and societal norms, shape distinct patterns of restriction, surveillance, and control \cite{warf2011geographies}. In turn, the shift of power dynamics from nation states to global networks further emphasizes the {role of} human behavior in the success of censorship techniques, measurement, and circumvention {practices}.

\paragraph{{Motivations for Censorship Shape Measurement Challenges}} 
{Censorship is justified in diverse ways, shaping both enforcement mechanisms and user responses.} Motivations can be broadly categorized into four main categories \cite{Aceto2015,NIHbook,houmansadr2013parrot,ang1996censorship}{: (i) \emph{Political Control}, (ii) \emph{Social and Moral Regulation}, (iii) \emph{Parental and Company Controls}, and (iv) \emph{Economic Motivations}.} {Each motive influences what is censored, how it is applied, and how it is perceived. Political and social motives often fluctuate around events such as elections or crises, while economic and parental controls create subtler but still significant restrictions.}\\

(i) \emph{Political Control}: Governments, particularly repressive ones, strategically control Internet access to unwanted political opinions (e.g., NGOs, websites of foreign newspapers or social media). {The aim is to }limit its potential to empower citizens and prevent them from bypassing state-controlled media \cite{hobbs2018sudden,ryng2022internet,dal2022walking,alzaman2024rise,ashokkumar2020censoring}. Motivations for censorship range from suppressing political dissent and human rights activism to silencing criticism of the state or its officials, as seen in countries like China and Iran \cite{king2013censorship,he2024tactics}. Social media platforms, vital spaces for political mobilization, are often restricted or shut down entirely \cite{alzaman2024reasons,gillett2024investigating,poell2014social}. In such cases, measuring censorship requires more than merely detecting blocked news websites, but also involves understanding how fear influences open discourse and indirect forms of online manipulation \cite{gebhart2017internet,nisbet2017psychological}. This is especially crucial from a temporal perspective, as censorship practices strongly fluctuate around sensitive events like elections or protests \cite{FOTN24}.\\

(ii) \emph{Social and Moral Regulation}: {Another governmental justification for censorship is} expressing social and moral concerns. By claiming to protect societal values, public morality, or vulnerable groups \cite{king2013censorship,kubin2024political,lou2024negotiating}, mostly illegal content is targeted. This includes blocking websites linked to illicit trade (e.g., drugs, weapons), child exploitation, pornography, or platforms promoting hate speech. The definition of harmful content or what is deemed immoral varies across cultures \cite{kim2024digital,he2024tactics}, leading to uneven enforcement that disproportionately impacts marginalized groups \cite{warf2011geographies,clark2017shifting,busch2018internet}. Beyond safeguarding public morality, concerns about terrorism, national security, preventing crimes and maintaining social stability are increasingly cited \cite{wang2015internet,dal2022walking,meserve2020terrorism}. 
However, the relationship between national security laws and the Internet is complex, encompassing measures such as content takedowns and network shutdowns, which are responses to local conditions, such as armed conflicts or political unrest, reflecting historical tensions and contemporary power struggles. The use of centralized power structures to monitor and shape online content to very different extents, in some cases, is therefore rooted in fear of losing legitimacy \cite{dal2022walking}.\\

(iii) \emph{Parental and Company Controls}: Parental controls are often implemented at the household level or through private software filters, but some governments mandate or encourage Internet service providers or software vendors to offer content controls for minors \cite{rosenberg2001controlling,li2013parental,JMStV}. Although less intrusive than broader state censorship, these measures align with local regulatory environments.\\ 

(iv) \emph{Economic Motivations}: Other motivations for censorship can involve intellectual property protections, such as blocking illegal downloads. Companies and governments may also block competitor websites or restrict traffic to protect trade secrets \cite{meserve2018google}. 
Although this practice can serve legitimate security purposes, it can also become a rationale to censor or restrict broader sets of content \cite{radsch2023weaponizing,Deibert2008Access}. {Such overblocking of potential legitimate content could affect users' trust in information sources \cite{Deibert2008Access,brown2007internet}.} Regions with strong government-industry ties or heightened espionage concerns tend to enforce such restrictions more aggressively, reflecting the influence of economic and political alliances on censorship patterns \cite{miao2024embedded,ververis2020cross}. 
Additionally, enforcing censorship often leads to unintended non-technical consequences, such as stifled innovation. 

\paragraph{{Managing Public Attitudes as a Form of Control}}
Beyond technical restrictions, governments actively shape public opinions to legitimize censorship. {For example,} state-con\-trolled Chinese media frequently portray the Internet in a negative light, reinforcing public acceptance of Internet restrictions \cite{PewReport2008,wang2015internet}. This {reflects} the concept of \emph{networked authoritarianism} \cite{MacKinnon2011networked}, where authoritarian governments {allow some degree of connectivity but strategically manage and tightly control online discourse} to reinforce state narratives. The example of China demonstrates how deeply rooted cultural norms can lead citizens to regard the government as the sole legitimate authority, thereby making its actions appear more natural and expected \cite{PewReport2008,Burgers2016netauthori}. 
In this context, governments also refer to user studies, stating that the majority of respondents are in favor of Internet censorship and support government control \cite{GIUS2012,guo2012understanding,wang2015internet}. However, this high acceptance rate of Internet censorship is irrespective of the nuances stated by respondents about which types of content should be censored (e.g., pornographic content over online chatting), \cite{wang2015internet,guo2012understanding}. 
Nonetheless, the study by Wang and Mark \cite{wang2015internet} also showed that {over time, even if there is an uprising against censorship at first, normalization reduces resistance, making censorship more effective even without broad technical blocking}. 
Consequently, individuals who adapt and modify their online behavior, impact censorship detection and measurement, as such subtle changes leave fewer visible traces.

\paragraph{{User Behavior and Self-Censorship}}
{As already mentioned, whereas self-censorship is primarily an outcome of censorship and surveillance rather than a deliberate technique by censors,} one psychological motivator {that reinforces it} is fear of isolation: the concern that expressing an unpopular opinion may lead to social exclusion or disapproval. {This fear can lead} to a heightened tendency to withhold opinions, reflecting the willingness to self-censor \cite{Aceto2015,burnett2022self-censoring} (see Sect.~\ref{sect:basics}). Moreover, self-censorship is closely related to the \emph{chilling effect}, where users adapt their online behavior, and the \emph{extended chilling effect}, where such adaptations spill over into offline live \cite{zuboff:age,ExtendedChillingEffect}. Beyond {these individual-level processes,} the \emph{spiral of silence} \cite{noelle-neumann1974spiral,gearhart2025spiral} provides a broader social explanation: users are further discouraged from engaging in discussions when they perceive their opinions to be in the minority, even in the absence of direct censorship \cite{hoffmann2017spiral}. As more people stay silent, the dominant view appears even stronger, which in turn pressures more individuals into self-censorship, reinforcing the spiral of silence and making certain views disappear from public discourse \cite{hayes2007exploring,burnett2022self-censoring,gearhart2025spiral}. {Such social conformity amplify the chilling effect, alter Internet usage, and can mislead passive censorship detection systems that ignore behavioral adaptations \cite{Aceto2015}. Consequently, technical measurements may underestimate censorship levels when they fail to account for human behavior.}

\paragraph{{Challenges in Data Collection and Interpretations}}
{Collecting human-centered data introduces biases as well as ethical and methodological risks. For example, OONI probe usage spikes during censorship events, meaning anomalies may correlate with participant activity rather than censorship prevalence: \emph{users run OONI more often when they are experiencing censorship} \cite{ACSAC24:WorldOfCensorship}.} 
Simultaneously, the use of Volunteer-Operated Devices (VODs) raise \emph{ethical and logistical} challenges \cite{ICLab:SP20}. In highly censored environments, participating in data gathering, {such as} contributing to measurement platforms or reporting blocked content, can invite legal reprisal, political or social risks \cite{Deibert2008Access}. {This produces} self-selection bias, {since only certain groups contribute} \cite{schulz2024warped,wang2015internet}. 
Marwick and Boyd \cite{marwick2018privacy} found that people who have been systematically and structurally marginalized (e.g., LGBTQ communities, people of color) are particularly sensitive to perceived surveillance, leading to heightened privacy concerns and behavioral adaption, such as underreporting \cite{penney2016chilling}. Thus, researchers {bear} responsibility to {protect volunteers, assess political} risks and work {with} trusted organizations \cite{ICLab:SP20}. \\ 
Another challenge, also influencing potential \emph{data misinterpretation}, arises when unskilled users are unable to distinguish between malfunction and censorship. As pointed out by Aceto \ea\cite{Aceto2015}, users might blame a destination host for not delivering a website if no blocking page is shown, although a censor could slow down or block a connection to a particular target. Similarly, users may assume their Internet service provider or destination system is at fault, and thus censorship might be attributed to an end-user's client device. Additionally, some forms of content restriction, such as algorithmic filtering or social media shadowbans, may not be easily detectable. Therefore, to avoid missing the nuanced reasons behind traffic anomalies and to ensure correct data interpretation, local expertise and contextual awareness should be considered \cite{CensoredPlanet2020,khattak2013towards}.\\
Overall, {while technical approaches effectively} detect direct censorship, {they risk} underestimating {suppression when human behavior and societal factors are ignored.} Addressing these challenges {by integrating technical,} psychological and societal {perspectives} will not only improve measurement accuracy, but also support those most affected by restrictions. 


\section{Conclusion}\label{sect:concl}
We surveyed the capabilities of Internet censors. To this end, we introduced a generic taxonomy of censorship systems and provided an in-depth coverage of censorship methodology on the basis of network protocols. We have shown how current censorship is conducted, incorporating sophisticated methods targeting BGP, TLS, and several forms of circumvention tools. 
We provided a detailed explanation of censorship \emph{measurement} on the Internet, transport, and application layers, explained how the measurement of circumvention tool censorship can be conducted and compared available censorship datasets and platforms. 
Finally, we discussed {trends and challenges, including} human aspects.

Overall, several censoring and measurement methods of the Internet, transport and application layers remained consistent for years. However, more {of} the community gained a better understanding of measurement limitations and developed more sophisticated methods that aim to improve precision and accuracy of experiments. Recent techniques incorporated novel network protocols, platforms and technological trends. Another major novelty of the last years is the focus on societal and human aspects (both, from a censorship and from a measurement perspective), which has been a side issue in the past.

\paragraph{Acknowledgments}
We like to thank the students of Ulm University's class on Internet Censorship for the vivid discussions and the feedback gained that influenced parts of this paper.
{We also like to thank Paul Prechtel (Ulm Univ.) for his feedback on the BGP sections of this paper. Finally, we would like to express our thankfulness for the reviewers' feedback, which led to the improvement of the original manuscript.}
\bibliographystyle{elsarticle-harv}
\bibliography{refs}

\section*{Author Biographies}
\nobalance

\subsection*{}
\setlength\intextsep{0pt}
\begin{wrapfigure}{l}{0.16\textwidth}
\centering
\includegraphics[width=0.15\textwidth]{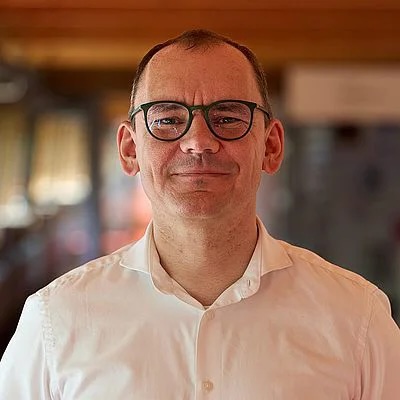}
\end{wrapfigure}
\noindent \textbf{Steffen Wendzel} is a professor at Ulm University, Germany, where he is the chair of the Institute of Information Resource Management (IRM) and the director of the university computing center and university library (kiz). Previously, he was a professor at HS Worms, Germany, a lecturer (\emph{Privatdozent}) at the Dep.\ Mathematics \& Computer Science, Fern\-Universität in Hagen, Germany, and a PostDoc at Fraunhofer FKIE, Bonn, Germany. He received his PhD (2013) and Habilitation (2020) from the University of Hagen, Germany.
Steffen \mbox{(co-)}authored more than 190 publications, of which many appeared in major journals and conferences (e.g., Elsevier COSE, Elsevier FGCS, ACM CSUR, IEEE TDSC, IEEE S\&P Mag., Asia\-CCS, IEEE LCN, Comm.\ ACM, ARES etc.). Website: \url{https://www.wendzel.de} / \url{https://www.uni-ulm.de/en/in/omi/}.

\subsection*{}
\setlength\intextsep{0pt}
\begin{wrapfigure}{l}{0.16\textwidth}
\centering
\includegraphics[width=0.15\textwidth]{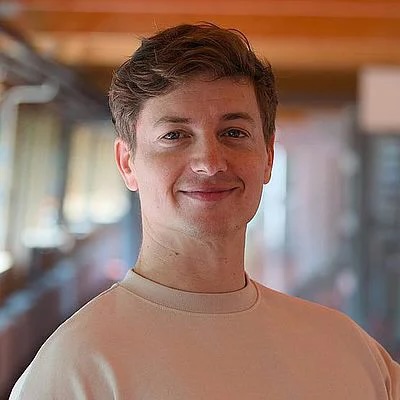}
\end{wrapfigure}
\noindent \textbf{Simon Volpert} is a researcher at the Institute of Information Resource Management (IRM) at the University of Ulm and PhD student at the UzK, Germany. He received his M.Sc. at Ulm University in 2023. His research interests revolve around the challenges of operations at scale, as well as systems benchmarking and data processing. Website: \url{https://www.uni-ulm.de/en/in/omi/}.

\subsection*{}
\setlength\intextsep{0pt}
\begin{wrapfigure}{l}{0.16\textwidth}
\centering
\includegraphics[width=0.15\textwidth]{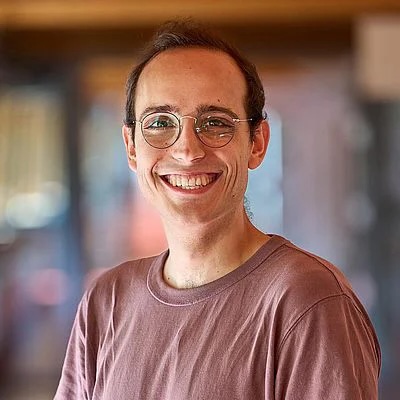}
\end{wrapfigure}
\noindent \textbf{Sebastian Zillien} is a researcher at the Institute of Information Resource Management (IRM) at the University of Ulm and PhD student at the RPTU, Germany. Earlier, he worked for the projects WoDiCoF+ and ATTRIBUT at the University of Applied Sciences in Worms, Germany. He received his M.Sc. in Mobile Computing in 2020 from Hochschule Worms. His research focus includes network and protocol-level security, anomaly detection, covert channels and reliability. Website: \url{https://www.uni-ulm.de/en/in/omi/}.

\subsection*{}
\setlength\intextsep{0pt}
\begin{wrapfigure}{l}{0.16\textwidth}
\centering
\includegraphics[width=0.15\textwidth]{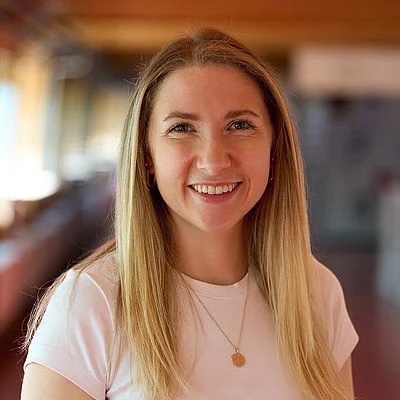}
\end{wrapfigure}
\noindent \textbf{Julia Lenz} is a PhD student at the Institute of Information Resource Management (IRM) at the University of Ulm. She received her B.Sc. from the University of Essex, UK, in Psychology and her M.Sc. from the City University of London, UK, in 2020 with the specialization in Organizational Psychology. Thereafter, she worked on several projects as a scientific researcher at the University of Applied Science in Worms, Germany. With her background in Psychology, her research interests include human factors in cybersecurity and usable security. Website: \url{https://www.uni-ulm.de/en/in/omi/}.

\subsection*{}
\setlength\intextsep{0pt}
\begin{wrapfigure}{l}{0.16\textwidth}
\centering
\includegraphics[width=0.15\textwidth]{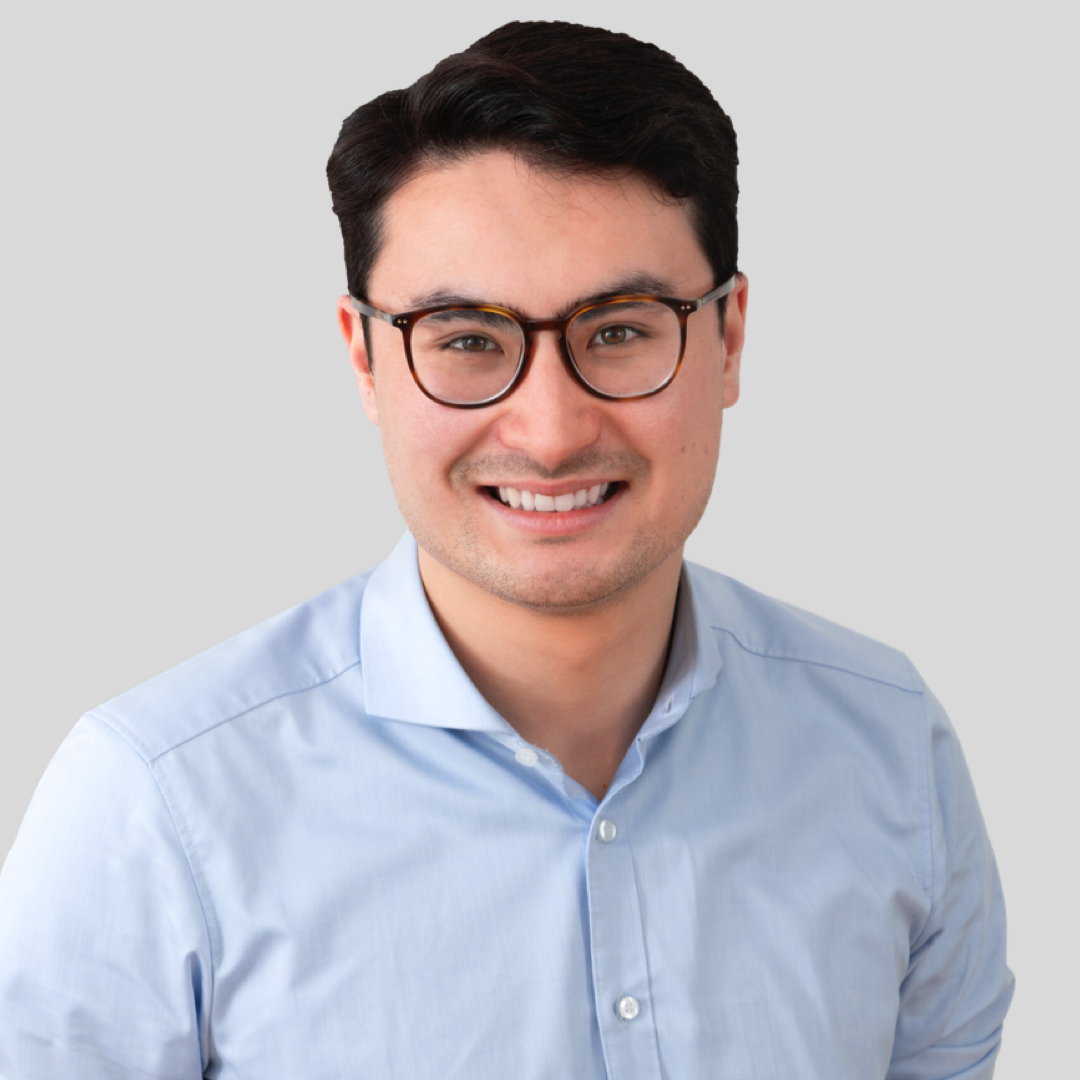}
\end{wrapfigure}
\noindent \textbf{Philip Rünz} has been a Policy Advisor for Security and Defense in the German Bundestag since 2023. He completed a dual study program in Mechatronics at the Baden-Württemberg Cooperative State University (DHBW) in cooperation with the German Armed Forces (B.Eng., 2020) and later specialized in cybersecurity with a degree in Practical Computer Science from the FernUniversität Hagen (M.Sc., 2023). From 2021 to 2023, he worked on a cyber project for the Bundeswehr. Since 2021, he has also been lecturing on signal and control technology at the DHBW Mannheim.

\subsection*{}
\setlength\intextsep{0pt}
\begin{wrapfigure}{l}{0.16\textwidth}
\centering
\includegraphics[width=0.15\textwidth]{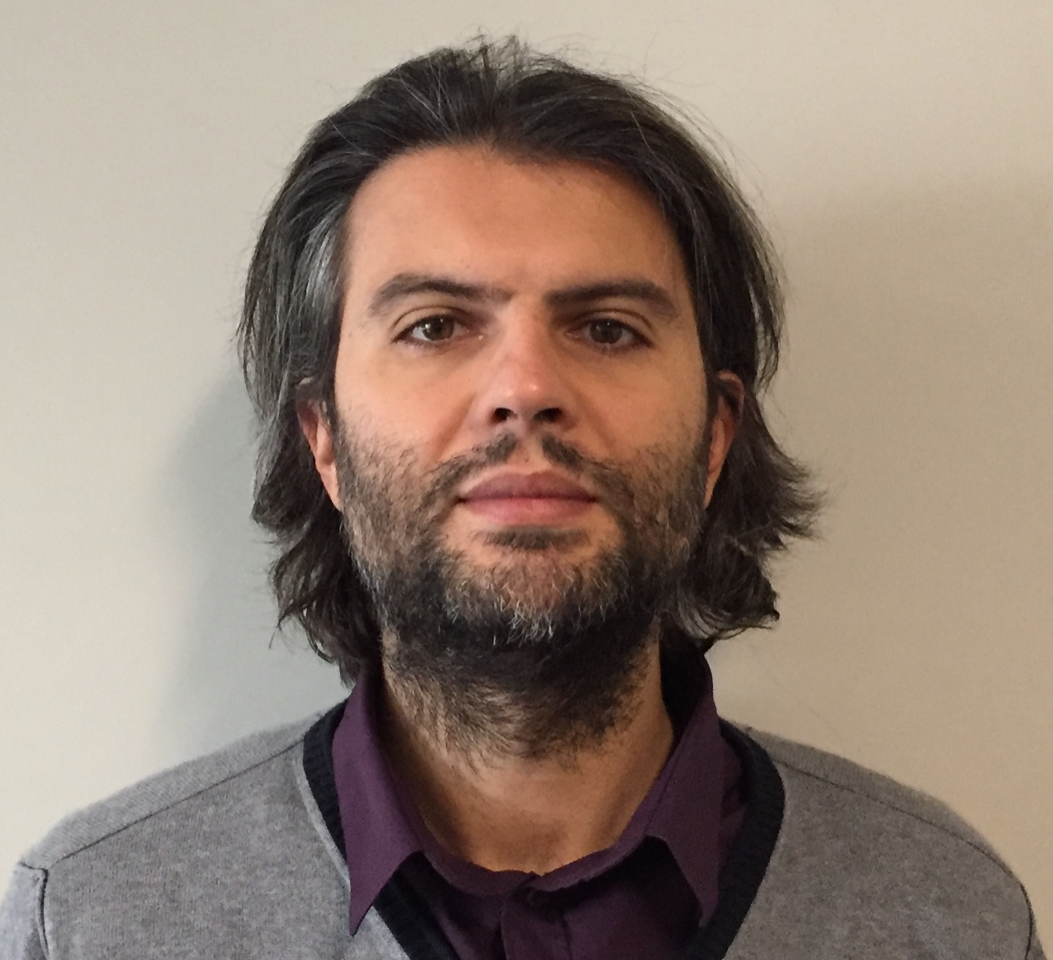}
\end{wrapfigure}
\noindent \textbf{Luca Caviglione} is a Senior Research Scientist at the Institute for Applied Mathematics and Information Technologies of the National Research Council of Italy.
He holds a Ph.D. in Electronic and Computer Engineering from the University of Genoa, Italy.
His research interests include optimization of large-scale computing and network security. He is an author or co-author of more than 200 academic publications, and several patents in the field of p2p and energy-aware computing. He is an associate editor of the IEEE Transactions on Information Forensics \& Security and the head of the IMATI Research Unit of the National Inter-University Consortium for Telecommunications. Website: \url{https://lucacav.github.io}.

\end{document}